\documentclass[aps,prc,preprint,onecolumn,preprintnumbers,superscriptaddress,nofootinbib]{revtex4-1}

\usepackage{amsthm}
\usepackage{amsmath}
\usepackage{graphicx}
\usepackage{slashed}
\usepackage{amssymb}
\usepackage{subcaption}
\usepackage{color}
\usepackage[colorlinks=False, citecolor=blue, urlcolor=blue, linkcolor=blue]{hyperref}

\newcommand*\diff{\mathop{}\!\mathrm{d}}

\newcommand{\nn}{\nonumber}

\newcommand{\be}{\begin{eqnarray}}
\newcommand{\ee}{\end{eqnarray}}
\newcommand{\ma}{\mathrm}
\newcommand{\ml}{\mathcal}
\newcommand{\bs}{\boldsymbol}

\begin{document}

\title{Jet Wake from Linearized Hydrodynamics}

\author{Jorge Casalderrey-Solana}
\email{jorge.casalderrey@ub.edu}
\affiliation{Departament de F\'\i sica Qu\`antica i Astrof\'\i sica \& Institut de Ci\`encies del Cosmos (ICC), 
Universitat de Barcelona, Mart\'{\i}  i Franqu\`es 1, 08028 Barcelona, Spain}

\author{Jos\'e Guilherme Milhano}
\email{gmilhano@lip.pt}
\affiliation{{LIP, Av. Prof. Gama Pinto, 2, P-1649-003 Lisboa , Portugal}}
\affiliation{Instituto Superior T\'ecnico (IST), Universidade de Lisboa, Av. Rovisco Pais 1, 
1049-001, Lisbon, Portugal}

\author{Daniel Pablos}
\email{daniel.pablos@uib.no}
\affiliation{University of Bergen, Postboks 7803, 5020 Bergen, Norway}

\author{Krishna Rajagopal}
\email{krishna@mit.edu}
\affiliation{Center for Theoretical Physics, Massachusetts Institute of Technology, Cambridge, MA 02139 USA}

\author{Xiaojun Yao}
\email{xjyao@mit.edu}
\affiliation{Center for Theoretical Physics, Massachusetts Institute of Technology, Cambridge, MA 02139 USA}

\date{\today}
\preprint{MIT-CTP/5246}

\begin{abstract}
We explore how to improve the hybrid model description of the particles originating from the wake that a jet produced in a heavy ion collision leaves in the droplet of quark-gluon plasma (QGP) through which it propagates, using linearized hydrodynamics on a background Bjorken flow. Jet energy and momentum loss described by the hybrid model become currents sourcing  linearized hydrodynamics. By solving the linearized hydrodynamic equations numerically, we investigate the development of the wake in the dynamically evolving droplet of QGP, study the effect of viscosity, scrutinize energy-momentum conservation, and check the validity of the linear approximation. We find that linearized hydrodynamics works better in the viscous case because diffusive modes damp the energy-momentum perturbation produced by the jet. We calculate the distribution of particles produced from the jet wake by using the Cooper-Frye prescription and find that both the transverse momentum spectrum and the distribution of particles in azimuthal angle are similar in shape in linearized hydrodynamics and in the hybrid model. Their normalizations are different because the momentum-rapidity distribution in the linearized hydrodynamics analysis is more spread out, due to sound modes. Since the Bjorken flow has no transverse expansion, we explore the effect of transverse flow by using local boosts to add it into the Cooper-Frye formula. After including the effects of transverse flow in this way, the transverse momentum spectrum becomes harder: more particles with transverse momenta bigger than $2$ GeV are produced than in the hybrid model. Although we defer implementing this analysis in a jet Monte Carlo, as would be needed to make quantitative comparisons to data, we gain a qualitative sense of how the jet wake may modify jet observables by computing proxies for two example observables: the lost energy recovered in a cone of varying open angle, and the fragmentation function. We find that linearized hydrodynamics with transverse flow effects added improves the description of the jet wake in the hybrid model in just the way that comparison to data indicates is needed. Our study illuminates a path to improving the description of the wake in the hybrid model, highlighting the need to take into account the effects of both transverse flow and the broadening of the energy-momentum perturbation in spacetime rapidity on particle production.
\end{abstract}
\maketitle

\tableofcontents

\section{Introduction}
\label{sect:intro}
Jets are collimated sprays of particles produced in high energy collisions originating from the production of energetic quarks and gluons which in turn each fragments into a shower of quarks and gluons that subsequently hadronize. These objects play a central role in our understanding of collider physics. In the vacuum, as well as in low density hadronic environments such as pp collisions at the LHC, the physics processes that lead to the formation of these QCD structures are under excellent theoretical and experimental control. This knowledge has transformed those objects  into precision tools to understand the physics of the Standard Model and to explore beyond. In dense hadronic environments, like those created in heavy ion collisions at RHIC and the LHC, while our understanding of the properties of these objects are still under intense theoretical investigation~\cite{Gyulassy:1993hr,Wang:1994fx,Baier:1994bd,Baier:1996kr,Zakharov:1996fv,Baier:1996sk,Gyulassy:1999zd,Gyulassy:2000fs,Wiedemann:2000za,Guo:2000nz,Gyulassy:2000er,Wiedemann:2000za,Wang:2001ifa,Arnold:2002ja,Arnold:2002ja,Jeon:2003gi,Majumder:2007zh,MehtarTani:2010ma,CasalderreySolana:2011rz,MehtarTani:2011tz,Ovanesyan:2011xy,MehtarTani:2011gf,MehtarTani:2012cy,Blaizot:2013hx,Blaizot:2013vha,Casalderrey-Solana:2014bpa,Casalderrey-Solana:2015vaa,He:2015pra,Ghiglieri:2015zma,Ghiglieri:2015ala,Casalderrey-Solana:2016jvj,Cao:2017zih,Hulcher:2017cpt,Mehtar-Tani:2017web,Caucal:2018dla,Casalderrey-Solana:2018wrw,He:2018xjv,Casalderrey-Solana:2019ubu}, jets are of particular importance since they act as probes of the quark-gluon plasma (QGP) created in those collisions.

The generic phrase used to describe the modification of jets as they traverse a droplet of QGP is ``jet quenching''. In early studies, the focus was on how the total energy of a jet is degraded; in recent years much attention has been paid to how the shape or structure of the jets are modified in other ways.
When jets traverse the plasma, the energetic partons within a jet shower interact with the plasma constituents and, as a result, lose energy to the medium. Many studies have focused on describing how this parton energy loss mechanism takes place. This by itself is a complicated task, since the interaction between energetic partons and the medium that they probe incorporates physics at multiple scales, from the perturbative dynamics of the short distance splitting processes via which the shower of energetic partons develops to the long distance strongly coupled dynamics of the medium. Since the particular modification of any given jet depends on the parton shower history and varies quite substantially event-by-event, with jets with the same total energy but differing shower structure experiencing very different fractional energy loss and different modifications to their shape, a significant theoretical effort has been devoted to the development of Monte Carlo simulators of jet production events in heavy ion collisions~\cite{Gyulassy:1994ew,Lokhtin:2005px,Renk:2008pp,Lokhtin:2008xi,Armesto:2009fj,Schenke:2009gb,Zapp:2011ya,Zapp:2012ak,Zapp:2013vla,Majumder:2013re,Renk:2013pua,Wang:2013cia,Casalderrey-Solana:2014bpa,Casalderrey-Solana:2015vaa,Casalderrey-Solana:2016jvj,Cao:2017hhk,Cao:2017zih,Hulcher:2017cpt,Cao:2017qpx,Casalderrey-Solana:2018wrw,He:2018xjv,Ke:2018jem,Casalderrey-Solana:2019ubu,Putschke:2019yrg,Dai:2019ddc,Caucal:2019uvr,Caucal:2020xad}.

Complementary to the modification of internal structure of jets, jet quenching also leads to a modification of the medium those objects traverse. This process, which is generically referred to as the backreaction of the medium, is a consequence of the injection of energy and momentum lost by the jet into the plasma. Because momentum is conserved, the wake created in the medium as the jet loses energy and momentum necessarily gives the medium a net momentum in the jet direction, yielding a correlation between the bulk dynamics of the medium and the jet direction. This means that when an experimentalist reconstructs a jet from the hadrons in the final state of the collision, no background subtraction procedure can remove all the particles coming from the hadronization of the medium. What the experimentalist reconstructs as a jet includes hadrons (and energy and momentum) originating from the wake in the medium created by the passing jet as well as hadrons (and energy and momentum) originating from the shower of energetic partons (``the jet itself'') after modification by their passage through the medium. Even though the particle production associated with the jet itself (with its modified internal structure) and that associated with the hadronization of the wake in the medium are quite different in character and origin, it is impossible to separate these two contributions to jet observables at colliders. Therefore, a complete description of collider jet data requires an understanding of the dynamics of the backreaction of the medium.
Since one of the most striking features of the QGP medium is its almost perfect fluid behavior, a natural expectation is that the dynamics of the ``lost'' energy, which is to say the dynamics of the wake in the medium, is also well captured by hydrodynamics. 
These hydrodynamic structures have important consequences for the description of jet observables in heavy ion collisions. 
 Earlier studies applied linearized hydrodynamics with the energy and momentum lost by the jets as external currents, assuming the perturbation caused by the jet energy and momentum deposition into the medium is small~\cite{CasalderreySolana:2004qm,Ruppert:2005uz,CasalderreySolana:2006sq,Neufeld:2008fi,Ayala:2012bv,Ayala:2014sua}. Many subsequent studies followed this line of thinking~\cite{Neufeld:2008dx,Qin:2009uh,Neufeld:2009ep,Yan:2017rku}. More recent studies relaxed the assumption that the perturbation is linear and carried out more complete analyses using full hydrodynamics~\cite{Chaudhuri:2005vc,Betz:2010qh,Floerchinger:2014yqa,Tachibana:2014lja,Tachibana:2020mtb}. A combination of hydrodynamics and weak-coupling partonic transport has also been developed~\cite{Chen:2017zte}. However, these full hydrodynamic analyses are computationally very expensive.
 
 In this paper we will explore the medium backreaction anew, aiming at a simpler description that can be more efficiently implemented into Monte Carlo generators. While our analysis will not depend on the particular model for jet-medium interactions, we will take inspiration from the hybrid strong/weak coupling model that some of us have developed over the last few years ~\cite{Casalderrey-Solana:2014bpa,Casalderrey-Solana:2015vaa,Casalderrey-Solana:2016jvj,Hulcher:2017cpt,Casalderrey-Solana:2018wrw,Casalderrey-Solana:2019ubu}. This model aims at combining a perturbative description of the parton shower at short distances with a non-perturbative, strongly coupled, description of the interaction between partons in the shower and the strongly coupled hydrodynamic medium via the gauge/gravity duality \cite{Maldacena:1997re,Herzog:2006gh,Liu:2006ug,CasalderreySolana:2006rq,Gubser:2006qh,Liu:2006he,CasalderreySolana:2007qw,Chesler:2007an,Gubser:2007ga,Chesler:2007sv,Gubser:2008as,Hatta:2008tx,Dominguez:2008vd,Chesler:2008wd,Chesler:2008uy,Gubser:2009sn,DEramo:2010wup,Arnold:2010ir,Arnold:2011qi,Chesler:2011nc,CasalderreySolana:2011us,Arnold:2012uc,Chesler:2013cqa,DeWolfe:2013cua,Ficnar:2013qxa,Chesler:2014jva,Chesler:2015lsa,Chesler:2015nqz,Casalderrey-Solana:2015tas,Rajagopal:2016uip,Brewer:2017fqy}. In addition, in this hybrid model a very 
 simple prescription for medium backreaction was introduced~\cite{Casalderrey-Solana:2016jvj}. This was based on a small perturbation analysis of the medium response
 on top of an ideal hydrodynamics, together with the assumption that the resulting modification to the final state hadron spectra are small at all $p_T$. Under certain assumptions, which we will review in Section~\ref{sect:spectrum}, a simple prescription solely controlled by energy-momentum conservation was introduced. Though easy to implement numerically, this prescription for the jet wake particle production in the hybrid model is too soft: comparison to experimental measurements~\cite{Chatrchyan:2013kwa,Chatrchyan:2014ava,Khachatryan:2015lha}  shows that it yields too many particles with $p_T<1$ GeV and not enough particles with $p_T\gtrsim 2$ GeV. (See, for example, Figs.~10, 11, 13 and 14 of Ref.~\cite{Casalderrey-Solana:2016jvj}.) 
 
 In this work we will address known oversimplifications of the hybrid model prescription and explore their possible observable consequences. 
 As a first step towards this goal, we apply the framework of linearized hydrodynamics on top of a $1+1$D boost invariant hydrodynamics (the Bjorken fluid), by assuming the energy and momentum deposited by the high energy partons in the jet shower are small. By solving the linearized hydrodynamic equations numerically, we demonstrate the time evolution of the jet wake and scrutinize the energy and momentum conservation in the evolution. Furthermore, we investigate the effect of viscosity on the bulk dynamics of the jet wake perturbation and on the assumed linear approximation. Then, we convert the jet wake into particle production by using the Cooper-Frye formula. We find that the particle distribution from the linearized hydrodynamics is similar to that from the hybrid model, except that the momentum-rapidity distribution in the former is more spread out. 
 While our hydrodynamic analysis itself does not include transverse expansion, we can nevertheless also explore the effect of transverse flow. We do so by introducing a transverse flow profile, extracted from a $2+1$D viscous hydrodynamic simulator ``VISHNU"~\cite{Song:2007ux,Shen:2014vra} and calibrated with experimental observables of light particles at low transverse momentum~\cite{Bernhard:2016tnd}, to appropriately boost the jet-induced fluid disturbances and explore the effects of transverse flow on the hadronization of the wake via the Cooper-Frye formula.
  We find that the transverse flow
  makes the particle production from the wake harder than in the hybrid model, in line with indications from experimental measurements. 
We will also explore how this improvement in the backreaction description could influence jet observables by studying the recovery of energy induced by particles from the wake in conical regions around the direction of the wake (and jet) as well as the momentum distribution of those particles, which affects the fragmentation functions for medium-modified jets.

This paper is organized as follows: In Section~\ref{sect:linear_hydro}, we introduce our framework of linearized hydrodynamics on a background Bjorken flow. We also explain how to convert the energy loss rate into external currents of energy and momentum in the hydrodynamic equations. Then in Section~\ref{sect:numerics}, we explain the numerical recipes to solve the linearized hydrodynamics in detail. In this section we also discuss the jet wake development, tests of energy-momentum conservation, and tests of the validity of the linear approximation. We study distributions of particles produced from the jet wake in Section~\ref{sect:spectrum}. We also provide comparisons between the results from the linearized hydrodynamics, with and without the effects of transverse flow, and the hybrid model prescription. And, we discuss the dependence of the effects of transverse flow on hadron mass. Next, in Section~\ref{sect:observable} we consider the potential effects from the jet wake on jet observables. Finally, we draw conclusions and look ahead in Section~\ref{sect:conclusion}.

\section{Linearized Hydrodynamics on Bjorken Flow}
\label{sect:linear_hydro}

One of the most striking properties of the QGP as produced in high-energy colliders is that it behaves as an almost perfect fluid. This means that the collective properties of bulk particle production, with transverse momentum smaller than a few GeV are correctly captured by a hydrodynamical description of the medium in which all dynamics are reduced to the conservation of the energy momentum tensor, which is approximated by a gradient expansion around equilibrium. However, harder particles may not be assumed to be part of the fluid; as a extreme example, high-energy jets contain fragments of several tens of GeV, which do not have time to arrive close to thermal equilibrium and may be viewed as external probes of the medium. Certainly, the distinction between these two different sets of components is not sharp; a complete description of the jet-medium interaction should be able to account for the dynamics at all energy scales and only below some scale those dynamics would be well approximated by hydrodynamics. However, since currently such description is not available, in this paper we will simply assume that such separation between soft and hard modes exists and that the soft modes are well described via hydrodynamics. 

From the point of view expressed above, the energy loss that jets experience in the medium may be viewed as a transfer of energy from hard modes to soft modes. While the overall energy-momentum tensor of the system is conserved, from the point of view of the energy-momentum tensor of soft, bulk particles, $T^{\mu \nu}$ this energy-momentum transfer may be viewed as source in the conservation equation, which we may write as

\be
\nabla_\mu T^{\mu\nu} = J^\nu\,,
\ee
where $J^\nu$ depends on the distribution and dynamics (e.g. energy loss) of hard modes which, in the case of jet-medium interactions, depend on the jet properties. 

In the absence of a source, since, as already stated, the dynamics are captured by hydrodynamics, the energy-momentum of soft bulk matter can be written in a gradient expansion. (See Refs.~\cite{Jeon:2015dfa,Romatschke:2017ejr} for recent reviews.) The building blocks of this approximation are the (local) energy density field $\varepsilon$ and  the fluid velocity field $u^\mu$, normalized by $u^2 \equiv u_\mu u^\mu=1$. The energy-momentum tensor is then constructed using symmetry properties and a power-counting prescription based on the total order of derivatives. To linear order in gradients and neglecting the bulk viscosity for simplicity, we have
\be
\label{eqn:Tmunu}
T^{\mu\nu} = (\varepsilon + P)u^\mu u^\nu - P g^{\mu\nu} + 2\eta \nabla^{\langle\mu} u^{\nu\rangle} \,,
\ee
 Here  $P$ denotes the equilibrium pressure, which is a function of $\varepsilon$, $g_{\mu\nu}$ stands for the metric tensor and $\eta$ is the shear viscosity. The shear term can be written as
\be
2\nabla^{\langle\mu} u^{\nu\rangle} = \Delta^{\mu\rho} \nabla_\rho u^\nu + 
\Delta^{\nu\rho} \nabla_\rho u^\mu -
\frac{2}{3} \Delta^{\mu\nu} \nabla_\rho u^\rho\,,
\ee
where $\Delta^{\mu\nu} \equiv g^{\mu\nu} - u^\mu u^\nu$. 

The main assumption in the hydrodynamic description of the medium backreaction is that the disturbance of the bulk motion of the hydrodynamic fluid (the wake) introduced by the jet can also be described via hydrodynamics. This implies that, also in the presence of the source, the energy-momentum tensor may be approximated in a gradient expansion as in Eq.~(\ref{eqn:Tmunu}). Certainly, this will be the case sufficiently far away from the jet, when the disturbance has relaxed and is characterised by long-wavelength modes. However, close to the jet the gradients of the flow field will be larger, which may challenge the hydrodynamic assumption. Nevertheless, explicit numerical analysis of the out-of-equilibrium dynamics of both strongly coupled and weakly coupled gauge theories (see the pioneering works~\cite{Chesler:2010bi, Heller:2011ju, Kurkela:2015qoa, Chesler:2015bba, Chesler:2016ceu} and see Refs.~\cite{ CasalderreySolana:2011us, Chesler:2015lsa, Busza:2018rrf, Florkowski:2017olj, Romatschke:2017ejr} for recent reviews) as well as the recent studies of hydrodynamic attractors \cite{Heller:2015dha,Strickland:2017kux,Blaizot:2019scw,Jaiswal:2019cju,Kurkela:2019set} suggest that hydrodynamics may be applicable even when gradients are large and the matter is away from local thermal equilibrium, although a non-hydrodynamic explanation of the attractor phenomenon in terms of a pre-hydrodynamic epoch during which the dynamics is dominated by one or a few low-lying modes in an effective Hamiltonian would delay the onset of hydrodynamization~\cite{Brewer:2019oha}.
Furthermore, explicit analysis of the medium response to an energetic colored object in strongly-coupled gauge theories~\cite{Chesler:2007an,Gubser:2007ga,Chesler:2007sv,Chesler:2008wd,Chesler:2008uy,Gubser:2009sn,Chesler:2011nc,CasalderreySolana:2011us} demonstrates that, at least in the strongly coupled fluid of ${\cal N}=4$ supersymmetric Yang-Mills theory, the energy-momentum tensor disturbance induced by these energetic particles behave hydrodynamically even at distances as short as $1/\pi T$ away from the jet. These theoretical advances motivate us to explore the  observable consequences of a hydrodynamic treatment of the backreaction of the medium to high-energy jets in heavy ion collisions. Similar explorations have been performed in Refs.~\cite{CasalderreySolana:2004qm,CasalderreySolana:2006sq,Neufeld:2008fi,Neufeld:2008dx,Ruppert:2005uz,Qin:2009uh,Neufeld:2009ep,Yan:2017rku,Chaudhuri:2005vc,Betz:2010qh,Floerchinger:2014yqa,Tachibana:2014lja,Tachibana:2020mtb}.

In addition to assuming the validity of hydrodynamics, in this work we will assume that the disturbance introduced by the external current $J^\mu$ is small and leads to a perturbation of the stress-energy tensor that can be well-approximated as linear. This assumption allows us to decompose the total stress-energy tensor into two parts: $T_{(0)}^{\mu\nu} + \delta T^{\mu\nu}$, where $T_{(0)}^{\mu\nu}$ denotes the stress-energy tensor of the background fluid in the absence of the external current and $\delta T^{\mu\nu}$ is the stress-energy tensor of the perturbation induced by the external current. These two components evolve according to
\be
\nabla_\mu T_{(0)}^{\mu\nu} &=& 0 \\
\nabla_\mu \delta T^{\mu\nu} &=& J^\nu \,.
\ee
The first of these two equations controls the collective flow of matter in the absence of the jet. The second equation determines the backreaction of the soft modes to the passage of the jet. 
We want to emphasize that these two equations are not independent from each other. The perturbation piece $\delta T^{\mu\nu}$ generally depends on the details on the background fluid expansion, encoded by
 $T_{(0)}^{\mu\nu}$. However, the backreaction on the background fluid $T_{(0)}^{\mu\nu}$ from the perturbation $\delta T^{\mu\nu}$ can be neglected since the perturbation is assumed to be small.

Because of the interconnection between these two components, to proceed further we need to specify the flow of the background fluid. In a realistic simulation, this bulk flow 
depends on the particular properties of each event, such as the geometry introduced by the impact parameter of the collision and the fluctuations coming from the nucleonic and subnucleonic structure of the colliding nuclei. While detailed hydrodynamic simulations of the bulk dynamics for such configurations are publicly available, in this exploratory work we will simplify the description of this flow and postpone the analysis of the interplay between the backreaction and the variation of the flow properties in different events to future work. For this reason, in this work we consider the simplest model for the background expansion: the Bjorken (or boost invariant) flow, which assumes homogeneity in the transverse plane and, therefore, has no transverse expansion. We leave analysis of linearized hydrodynamic perturbations on top of a radially expanding background flow to future studies. 

It is easy to define the Bjorken flow in the Milne $(\tau,x,y,\eta_s)$ coordinate system, where $\eta_s$ denotes the spacetime-rapidity and the subscript ``s" is used to distinguish it from the shear viscosity. In this coordinate system, the metric tensor is given by $g_{\mu\nu} = \rm{diag} (1, -1, -1, -\tau^2)$ and the covariant derivative on a (contravariant) vector field can be written as $\nabla_\nu u^\mu = \partial_\nu u^\mu + \Gamma^\mu_{~\rho\nu} u^\rho$. Here $\partial_\nu$ denotes the standard derivative and the nonvanishing Christoffel symbols are given by
\be
\Gamma^\tau_{~\eta\eta} = \tau \,, \ \ \ \ \ 
\Gamma^\eta_{~\eta\tau}=\Gamma^\eta_{~\tau\eta} = \frac{1}{\tau}\,.
\ee
The velocity field in the Bjorken flow is given in these coordinates by
\be
u^\mu_0 \equiv (u^\tau_0, u^x_0, u^y_0, u^{\eta_s}_0) = (1,0,0,0) \,.
\ee
Then one can work out the shear term in the stress-energy tensor: $2\nabla_{\phantom{1}}^{\langle\mu} u^{\nu \rangle}_0 = \rm{diag}(0, \frac{2}{3\tau}, \frac{2}{3\tau}, \frac{-4}{3\tau^3})$. The hydrodynamic equation $\nabla_\mu T^{\mu\nu}_{(0)}=0$ for the background Bjorken flow is given by
\be
\label{eqn:bjorken}
\frac{\partial \varepsilon_0}{\partial\tau}  + \frac{\varepsilon_0 + P_0 }{\tau} - \frac{4\eta}{3\tau^2} = 0\,.
\ee
To solve the hydrodynamic equation \eqref{eqn:bjorken}, we need the equation of state that relates $P_0$ to $\varepsilon_0$. We will use the textbook result for a non-interacting gas of gluons and $N_f=2$ flavors of massless quarks, which is not far from the results of lattice QCD calculations of the QCD equation of state for QGP with $N_f=2+1$ (2 light and 1 strange quarks) at high temperature~\cite{Bazavov:2014pvz}. For the background energy density and pressure, we have
\be
\label{eqn:eos1}
\varepsilon_0+P_0 &=& Ts_0 = \frac{4g}{\pi^2}T^4 \,,
\ee
where $s_0$ is the entropy density and $g=40$ is the number of massless degrees of freedom in non-interacting $N_f=2$ QGP.
This equation of state leads to a speed of sound
\be
\label{eqn:eos2}
c_s^2 \equiv \frac{P_0}{\varepsilon_0} = \frac{1}{3}\,,
\ee
which also controls the dynamics of the fluid perturbation.

To solve the system, we also need to specify the value of the viscosity. In this paper, we will consider and compare two characteristic values of this transport coefficient: the ideal case, $\eta=0$; and the value inferred from strongly coupled computations $\eta= \frac{s}{4\pi}$, where $s$ denotes the entropy density.

To explicitly write out the hydrodynamic equation $\nabla_\mu \delta T^{\mu\nu} = J^\nu$ for the perturbation, we must work out the explicit expressions for the perturbed stress-energy tensor $\delta T^{\mu\nu}$ and the external current $J^\nu$, which is what we are going to do in the next two subsections.

\subsection{Linear Perturbation of Bjorken Flow}
\label{sect:bjorken}

As discussed earlier, we assume the perturbation caused by the jet energy and momentum deposition leads to a small perturbation. So, the perturbed velocity field can be written as
\be
u^\mu = u^\mu_0 + \delta u^\mu = (1, \delta u^x, \delta u^y, \delta u^{\eta_s}) \,.
\ee
Since the perturbation $\delta u^\mu$ is small, the fluid velocity is still properly normalized if we neglect higher order terms beyond the linear order: $(u^\mu_0 + \delta u^\mu)^2 = 1 + \ml{O}((\delta u)^2)$.

The total stress-energy tensor $T^{\mu\nu}$ with the perturbation can be calculated by plugging $u^\mu_0+\delta u^\mu$ into Eq.~(\ref{eqn:Tmunu}) and decomposing $\varepsilon=\varepsilon_0+\delta \varepsilon$ and $P=P_0+\delta P$. Since $\varepsilon+P = Ts$ and $\eta=\frac{s}{4\pi}$, we also decompose the shear viscosity 
\be
\eta = \eta_0+\delta \eta = \frac{s_0}{4\pi}+\frac{\delta s}{4\pi} = \frac{s_0}{4\pi}+ \gamma_\eta\, \delta \varepsilon\ ,
\ee 
where $\delta\varepsilon = T \delta s$ and we have defined
\be
\label{eqn:gamma-eta}
\gamma_\eta \equiv \frac{\eta_0}{\varepsilon_0+P_0} = \frac{\eta_0}{Ts_0}\,.
\ee
The physical significance of $\gamma_\eta$ is that it controls the relaxation time of hydrodynamic excitations of the fluid. Expanding the expression \eqref{eqn:Tmunu} for $T^{\mu\nu}$ to linear order in the perturbation and subtracting the unperturbed background $T^{\mu\nu}_{(0)}$, we find
\be
\delta T^{\mu\nu} &=& \begin{pmatrix} 
\delta\varepsilon &&&\\
 & \delta P &&\\
 & & \delta P &\\
 &&  &  \frac{\delta P}{\tau^2} \\ \end{pmatrix} + (\varepsilon_0 + P_0) \begin{pmatrix} 
 & \delta u^x & \delta u^y & \delta u^{\eta_s} \\
\delta u^x &  &&\\
\delta u^y & &  &\\
\delta u^{\eta_s} &&  & 
\end{pmatrix} + \delta\eta\begin{pmatrix} 
0&&&\\
&\frac{2}{3\tau}&&\\
&&\frac{2}{3\tau}&\\
&&&-\frac{4}{3\tau^3}
\end{pmatrix} \nn\\
&+& \eta_0 \begin{pmatrix} 
0 & \frac{2}{3\tau}\delta u^x & \frac{2}{3\tau}\delta u^y & 
-\frac{4}{3\tau}\delta u^{\eta_s} \\
\frac{2}{3\tau}\delta u^x  & A^{xx} & -\partial_x \delta u^y - \partial_y \delta u^x &  -\partial_x \delta u^{\eta_s} -\frac{1}{\tau^2} \partial_{\eta_s} \delta u^x\\
\frac{2}{3\tau}\delta u^y  & -\partial_y \delta u^x -\partial_x \delta u^y & A^{yy} &  -\partial_y \delta u^{\eta_s} -\frac{1}{\tau^2} \partial_{\eta_s} \delta u^y\\
 -\frac{4}{3\tau}\delta u^{\eta_s}  & -\frac{1}{\tau^2} \partial_{\eta_s} \delta u^x -\partial_x \delta u^{\eta_s} & -\frac{1}{\tau^2} \partial_{\eta_s} \delta u^y -\partial_y \delta u^{\eta_s}&  A^{{\eta_s}{\eta_s}} \end{pmatrix} ,\ \ \ \
\ee
where
\be
A^{xx} &=&-\frac{4}{3}\partial_x \delta u^x + \frac{2}{3}\partial_y \delta u^y + \frac{2}{3}\partial_{\eta_s} \delta u^{\eta_s} \\
A^{yy} &=&-\frac{4}{3}\partial_y \delta u^y + \frac{2}{3}\partial_x \delta u^x + \frac{2}{3}\partial_{\eta_s} \delta u^{\eta_s}\\
A^{{\eta_s}{\eta_s}} &=& -\frac{4}{3\tau^2}\partial_{\eta_s} \delta u^{\eta_s} + \frac{2}{3\tau^2}\partial_x \delta u^x + \frac{2}{3\tau^2}\partial_y \delta u^y \,.
\ee
We use the short-hand notation $\partial_x=\frac{\partial}{\partial x}$ and similarly for $y$, $\eta_s$ and $\tau$.

The linearized hydrodynamic equation $\nabla_\mu \delta T^{\mu\nu} = \partial_\mu \delta T^{\mu\nu} +\Gamma^\mu_{~\rho\mu} \delta T^{\rho\nu} + \Gamma^\nu_{~\rho\mu} \delta T^{\mu\rho} = J^\nu $ can then be written as
\be
\partial_\tau \delta\varepsilon + \Big(1-\frac{\gamma_\eta}{\tau}\Big)\frac{\delta\varepsilon + \delta P}{\tau}
+ \partial_x \Big( (\varepsilon_0 + P_0)\delta u^x + \frac{4\eta_0}{3\tau}\delta u^x \Big) && \nn\\
+ \partial_y \Big( (\varepsilon_0 + P_0)\delta u^y + \frac{4\eta_0}{3\tau}\delta u^y \Big) + \partial_{\eta_s} \Big( (\varepsilon_0 + P_0)\delta u^{\eta_s} - \frac{8\eta_0}{3\tau}\delta u^{\eta_s} \Big)&=& J^\tau \\
\Big( \partial_\tau + \frac{1}{\tau} \Big) \Big( (\varepsilon_0 + P_0) \delta {\bs u}^\perp + \frac{2\eta_0}{3\tau} \delta {\bs u}^\perp \Big) + {\bs \partial}^\perp \delta P  + \frac{2\gamma_\eta}{3\tau}{\bs \partial}^\perp  \delta\varepsilon   && \nn\\
- \eta_0 \Big( \partial^{\perp2} + \frac{\partial_{\eta_s}^2}{\tau^2} \Big) \delta {\bs u}^\perp - \frac{1}{3}\eta_0  {\bs \partial}^\perp  \Big({\bs \partial}^\perp \cdot \delta {\bs u}^\perp + \partial_{\eta_s} \delta u^{\eta_s} \Big) &=& {\bs J}^\perp \\
\Big( \partial_\tau + \frac{3}{\tau} \Big) \Big( (\varepsilon_0 + P_0) \delta u^{\eta_s} - \frac{4\eta_0}{3\tau} \delta u^{\eta_s} \Big) + \frac{1}{\tau^2}\partial_{\eta_s} \delta P - \frac{4\gamma_\eta}{3\tau^3}\partial_{\eta_s} \delta\varepsilon  &&\nn\\
- \eta_0 \Big( \partial_x^2+\partial_y^2 + \frac{\partial_{\eta_s}^2}{\tau^2} \Big) \delta u^{\eta_s} - \frac{1}{3\tau^2} \eta_0 \partial_{\eta_s} \Big( {\bs \partial}^\perp \cdot \delta {\bs u}^\perp + \partial_{\eta_s} \delta u^{\eta_s} \Big) &=& J^{\eta_s} \,,
\ee
where the bold symbols denote Euclidean 2-vectors and the dot product between two Euclidean vectors is given by ${\bs a}\cdot{\bs b} = a^xb^x + a^yb^y$.

We solve the equations of linearized hydrodynamics in momentum space. For simplicity, we define
\be
\label{eqn:define_gperp}
{\bs g}^\perp(\tau, {\bs x}^\perp, \eta_s) &\equiv & (\varepsilon_0 + P_0) \delta {\bs u}^\perp \\
\label{eqn:define_geta}
g^{\eta} (\tau, {\bs x}^\perp, \eta_s) &\equiv & (\varepsilon_0 + P_0) \delta u^{\eta_s} \,.
\ee
The Fourier transformed quantities in momentum space, which we denote by $\delta \tilde{\varepsilon}$,  $\tilde{\bs g}^\perp$, $\tilde{g}^\eta$, and $\tilde{J}^\mu$ are then defined via
\be
\delta \varepsilon(\tau, {\bs x}^\perp, {\eta_s}) & = & \int \frac{\diff k^{\eta} \diff^2k^\perp }{(2\pi)^3} e^{i{\bs k}^\perp \cdot {\bs x}^\perp + i k^{\eta} {\eta_s}} \delta \tilde{\varepsilon}(\tau, {\bs k}^\perp, k^{\eta}) \\
{\bs g}^\perp (\tau, {\bs x}^\perp, {\eta_s}) & = & \int \frac{\diff k^{\eta} \diff^2k^\perp }{(2\pi)^3} e^{i{\bs k}^\perp \cdot {\bs x}^\perp + i k^{\eta} {\eta_s}} \tilde{\bs g}^\perp (\tau, {\bs k}^\perp, k^{\eta}) \\
g^{\eta} (\tau, {\bs x}^\perp, {\eta_s}) & = & \int \frac{\diff k^{\eta} \diff^2k^\perp }{(2\pi)^3} e^{i{\bs k}^\perp \cdot {\bs x}^\perp + i k^{\eta} {\eta_s}} \tilde{g}^{\eta} (\tau, {\bs k}^\perp, k^{\eta}) \\
J^\mu(\tau, {\bs x}^\perp, {\eta_s}) & = & \int \frac{\diff k^{\eta} \diff^2k^\perp }{(2\pi)^3} e^{i{\bs k}^\perp \cdot {\bs x}^\perp + i k^{\eta} {\eta_s}} \tilde{J}^{\mu} (\tau, {\bs k}^\perp, k^{\eta})
\ee
where we are slightly abusing the  notation: we use $g^\eta$ instead of $g^{\eta_s}$ in the definitions and the momentum conjugate to the spacetime-rapidity $\eta_s$ is labeled $k^\eta$ rather than $k^{\eta_s}$. We believe the physical meanings of $g^\eta$ and $k^\eta$ are clear; they have nothing to do with the shear viscosity $\eta$.

Plugging the Fourier transforms into the linearized hydrodynamic equations and using the definition of the speed of sound, we obtain
the equations for the excitation in Fourier space. In these equations, we will perform one additional approximation: we neglect all terms that are suppressed by a factor of order $\gamma_\eta / \tau$ relative to terms that are of order unity, with $\gamma_\eta$ as defined in \eqref{eqn:gamma-eta}. 
This approximation is motivated by the origin of the hydrodynamic approximation as a gradient expansion and by observing that due to the boost invariant symmetry assumed in Bjorken flow, the gradients of the velocity field scale as $1/\tau$.
In essence this approximation amounts to considering the flow field in regions where $\frac{\gamma_\eta}{\tau} \ll 1$, where the strict gradient expansion is valid. With this approximation, the equations for the perturbations are
\be
\label{eqn:li_hydro1}
\Big(\partial_\tau + \frac{1+c_s^2}{\tau} \Big) \delta \tilde{\varepsilon} + i{\bs k}^\perp \cdot \tilde{\bs g}^\perp  + i k^\eta \tilde{g}^\eta &=& \tilde{J}^\tau\\
\label{eqn:li_hydro2}
\Big(\partial_\tau + \frac{1}{\tau} \Big) \tilde{\bs g}^\perp + ic_s^2 {\bs k}^\perp \delta\tilde{\varepsilon} + \gamma_\eta \Big( {k^\perp}^2 + \frac{{k^\eta}^2}{\tau^2} \Big) \tilde{\bs g}^\perp + \frac{1}{3}\gamma_\eta {\bs k}^\perp ({\bs k}^\perp \cdot \tilde{\bs g}^\perp + k^\eta \tilde{g}^\eta) &=& \tilde{\bs J}^\perp \\
\label{eqn:li_hydro3}
\Big(\partial_\tau + \frac{3}{\tau} \Big) \tilde{g}^\eta + \frac{ic_s^2 k^\eta }{\tau^2}  \delta\tilde{\varepsilon} + \gamma_\eta \Big( {k^\perp}^2 + \frac{{k^\eta}^2}{\tau^2} \Big) \tilde{g}^\eta + \frac{1}{3\tau^2}\gamma_\eta k^\eta ({\bs k}^\perp \cdot \tilde{\bs g}^\perp + k^\eta \tilde{g}^\eta) &=& \tilde{J}^{\eta_s} \,.
\ee
For later convenience, we decompose $\tilde{\bs g}^\perp$ into the longitudinal $\tilde{g}_L$ and transverse $\tilde{g}_T$ modes:
\be
\tilde{g}_L &\equiv& \frac{1}{|{\bs k}^\perp|}{\bs k}^\perp \cdot \tilde{\bs g}^\perp \\
\tilde{g}_T &\equiv& \tilde{\bs g}^\perp - \frac{{\bs k}^\perp}{|{\bs k}^\perp|} \tilde{g}_L \,.
\ee
Then from \eqref{eqn:li_hydro2} we obtain the hydrodynamic equations for $\tilde{g}_L$ and $\tilde{g}_T$:
\be
\label{eqn:gL}
\Big(\partial_\tau + \frac{1}{\tau} \Big) \tilde{g}_L + ic_s^2 |{\bs k}^\perp| \delta\tilde{\varepsilon} + \gamma_\eta \Big( {k^\perp}^2 + \frac{{k^\eta}^2}{\tau^2} \Big) \tilde{g}_L + \frac{1}{3}\gamma_\eta |{\bs k}^\perp| (|{\bs k}^\perp|\tilde{g}_L  + k^\eta \tilde{g}^\eta) = \frac{{\bs k}^\perp \cdot\tilde{\bs J}^\perp }{|{\bs k}^\perp|} \\
\label{eqn:gT}
\Big(\partial_\tau + \frac{1}{\tau} \Big) \tilde{g}_T + \gamma_\eta \Big( {k^\perp}^2 + \frac{{k^\eta}^2}{\tau^2} \Big) \tilde{g}_T  = \tilde{\bs J}^\perp - \frac{{\bs k}^\perp {\bs k}^\perp \cdot\tilde{\bs J}^\perp }{|{\bs k}^\perp|^2} \,.
\ee
These equations illustrate the convenience of the decomposition of the momentum flux we have just introduced, since we see that the hydrodynamic equation for the transverse mode decouples from the other equations. This is the manifestation of the diffusive mode of a homogeneous plasma in this expanding system. In contrast, the longitudinal mode couples with the energy perturbation $\delta \tilde{\varepsilon}$ and the $\eta_s$-momentum perturbation $\tilde{g}^\eta$. This coupled channel describes the sound excitations of the system. These two channels are sourced by different components of the current $J
^\mu$. To be able to solve the dynamics, we need to specify the functional form of this source.

\subsection{External Current from Hybrid Model}
\label{sect:Jmu}

In this subsection we describe how we treat the energy deposition by the jet into the plasma and how we use this knowledge to model the source term that feeds the hydrodynamic response of the medium. While detailed calculations of the energy loss rate of hard partons propagating plasma are available in the literature, the hydrodynamization processes via which this energy becomes a modification to (e.g. a wake in) the hydrodynamic fluid and the manifestation of these deposition and hydrodynamization processes via this type of source terms is not under good theoretical control. For this reason, the specification of this source will require additional assumptions. Whenever possible, we will draw inspiration from explicit computations of the jet-medium interaction in both the weak and strong coupling limits.

Although some of our consideration will be general, to make the analysis concrete we will focus on the strong-coupling limit of jet-medium interaction and use knowledge from the $\mathcal{N}=4$ SYM description via a gravity dual. Following the hybrid strong/weak coupling model perspective, we will employ the following energy loss rate for a single hard particle traversing the fluid~\cite{Chesler:2014jva,Chesler:2015nqz}
\be
\label{eqn:sc}
\frac{\diff E}{\diff \tau} &=& \frac{4}{\pi}E_{\ma{in}} \frac{\tau^2}{\ell^2_\ma{stop}} \frac{1}{\sqrt{\ell^2_\ma{stop} - \tau^2}} \\
\ell_\ma{stop} &=& \frac{1}{2\kappa_{sc}}\frac{E_{\ma{in}}^{1/3}}{T^{4/3}} \,,
\ee
where $E_{\ma{in}}$ is the ``incoming'' energy of the parton, before it loses any energy to the hydrodynamic medium, and $T$ is the local temperature of the QGP medium.
In strongly coupled $\mathcal{N}=4$ SYM, this energy loss rate is fixed and the parameter $\kappa_{sc}=1.05 \lambda^{1/6}$, with $\lambda$ the 't Hooft coupling. In the hybrid strong/weak coupling model it is assumed~\cite{Casalderrey-Solana:2014bpa} that the functional form of this rate remains the same in strongly coupled QCD as in strongly coupled $\mathcal{N}=4$ SYM gauge theory and that all differences between these two theories can be condensed into the value of this parameter, which becomes a fitting parameter of the model. Taking the phenomenological analysis of Ref.~\cite{Casalderrey-Solana:2018wrw} as a guide, we shall 
choose the value $\kappa_{sc} = 0.4$ throughout this paper.

In a future Monte Carlo study, we would apply the rate of energy loss  \eqref{eqn:sc} and the entire analysis of the present paper to each energetic parton in each jet shower in an ensemble of events in which in each event jets are produced at a different location and propagate in different directions (typically two or more jets, which may have interesting consequences~\cite{Yan:2017rku,Pablos:2019ngg}) within an expanding, cooling, droplet of QGP. In the present study, we shall simplify things by considering only a single energetic parton traversing the (boost invariant; Bjorken) hydrodynamic fluid. 

To fix the total energy-momentum injection from the hard parton into the plasma we also need to determine the rate of momentum loss. We can easily relate this rate to the rate of energy loss by assuming that the parton moves at the speed of light with a fixed spacetime rapidity $\eta_{s\, {\rm parton}}$ 
and transverse direction $\hat{n}^\perp_{\rm parton}$,
\be
\label{eqn:dP}
\diff P^i &=& v^i \diff E  \\
\label{eqn:vperp}
v^\perp &=& \frac{\hat{n}^\perp_{\rm parton}}{\cosh(\eta_{s\, {\rm parton}})} \\
\label{eqn:vz}
v^z &=& \tanh{(\eta_{s\, {\rm parton}})} \,.
\ee
These rates then constrain the functional form of the current $J^\mu$ that describes the injection of energy and momentum into the medium but do not by themselves fix it, as we shall see. 

Working in Milne coordinates and assuming that the disturbances induced by the jet fall sufficiently fast at spacetime rapidities away from the jet, we can begin by relating $J^\mu$ to the rates of energy and momentum loss  by integrating over the space-like fixed-$\tau$ hypersurface. By means of Stokes' theorem
\be
\label{eqn:conserve_ef}
\int \tau \diff x \diff y \diff \eta_s  \,J^t &=& \frac{\diff}{\diff\tau} \int \tau \diff x \diff y \diff \eta_s\, \delta T^{t \tau}  =\frac{\diff E}{\diff\tau}
\\
\label{eqn:conserve_pf}
\int \tau \diff x \diff y \diff \eta_s  \,J^i &=& \frac{\diff}{\diff\tau} \int \tau \diff x \diff y \diff \eta_s\, \delta T^{i \tau}  =\frac{\diff P^i}{\diff\tau}\,,
\ee
where $t, \, i=x,\, y,\, z$ are Minkowski spacetime coordinates and we have used the fact that the energy and momentum fluxes through the fixed $\tau$-hypersurface are the changes to the energy and momentum stored in soft modes of the system, e.g. in the hydrodynamic fluid. Details of the construction of conservation laws from the Killing vector and the stress-energy tensor are provided in Appendix~\ref{sect:killing}, as are the conservation laws for the background Bjorken flow. Eqs.~(\ref{eqn:killing_J}, \ref{eqn:gausslaw3}, \ref{eqn:gausslaw4}) in Appendix~\ref{sect:killing} provide the explanation for why it is correct that in \eqref{eqn:conserve_ef} we see a relationship between $J^t$, with $t$ the Minkowski time, and ${\rm d}E/{\rm d}\tau$.

The integral constraints \eqref{eqn:conserve_ef} and \eqref{eqn:conserve_pf} will serve to fix the magnitude of $J^\mu$ but they do not fix the full functional form of the current. 
Following physical reasoning, the simplest assumption is that most of the energy deposition occurs in some small region around the local position of the hard parton while it is traversing the plasma. Assuming a Gaussian parametrization for the deposition process of energy and momentum in the rest frame of the fluid cell, we write the current (in Milne coordinates) as
\be
\label{eqn:Jmu}
&&J^{\mu} (\tau,x,y,\eta_s) \nn\\
&=& \frac{C^\mu(\tau)}{(2\pi)^{3/2}\sigma^2 \sigma_{\eta_s}} \exp\bigg(-\frac{(x-x_{\rm parton}(\tau))^2+(y-y_{\rm parton}(\tau))^2}{2\sigma^2} - \frac{(\eta_s -\eta_{s\, {\rm parton}})^2}{2\sigma_{\eta_s}^{\ 2}}\bigg)  \,,
\ee
where $(x_{\rm parton}(\tau),y_{\rm parton}(\tau),\eta_{s\, {\rm parton}})$ is the transverse position at time $\tau$, and the spacetime rapidity, of the energetic parton propagating through --- and losing energy and momentum to --- the medium, and where
$\sigma$ and $\sigma_{\eta_s}$ are the Gaussian widths in the transverse and spacetime-rapidity directions, respectively. With this parametrization and for fixed values of the width parameters, the $\tau$-dependent functions $C^\mu(\tau)$ are fixed by the rates of energy and momentum loss through the integral constraints \eqref{eqn:conserve_ef} and \eqref{eqn:conserve_pf}. Using the relations between the Minkowski $t$ and $z$ components of $J^\mu$ and the Milne $\tau$ and $\eta_s$ components, namely
\be
 J^{t} &=& (\cosh\eta_s) J^{\tau} + \tau(\sinh\eta_s)  J^{\eta} \\
J^{z} &=& (\sinh\eta_s)  J^{\tau} + \tau(\cosh\eta_s)  J^{\eta} \,,
\ee
as well as Eqs.~(\ref{eqn:conserve_ef}, \ref{eqn:conserve_pf}), we obtain
\be
C^\tau(\tau) &=& \bigg(\cosh(\eta_{s\, {\rm parton}}) - \frac{\sinh^2(\eta_{s\, {\rm parton}})}{\cosh(\eta_{s\, {\rm parton}})} \bigg) \frac{1}{\tau}\frac{\diff E}{\diff \tau}\exp\Big(-\frac{\sigma_{\eta_s}^2}{2}\Big) \\
C^\perp(\tau) &=& \frac{\hat{n}^\perp}{\cosh(\eta_{s\, {\rm parton}})} \frac{1}{\tau}\frac{\diff E}{\diff \tau} \\
\label{eqn:C_eta}
C^{\eta_s}(\tau) &=& 0 \ .
\ee
After fixing these coefficients we have fully specified the source term and we are now ready to solve the  linearized hydrodynamics equations, (\ref{eqn:li_hydro1}, \ref{eqn:li_hydro3}, \ref{eqn:gL}, \ref{eqn:gT}, \ref{eqn:sc}, \ref{eqn:Jmu}), to study the jet wake. In the next Section, we will explain the numerical procedure that we use to solve these equations and to generate the medium response to a high-energy parton.

\section{Numerical Solution}
\label{sect:numerics}
\subsection{Numerical Setup}

In the previous Section, we have completely specified the equilibrium and transport properties of the system we simulate as well as the rates of energy and momentum loss that govern the energy and momentum injected into the medium and the equations that govern the medium response. Prior to solving these equations and describing the backreaction of the medium numerically,  we also need to describe the geometry and parameters that specify the events we simulate.

As we have said, in this paper we shall choose a particularly simple hydrodynamic solution for the bulk fluid, treating it as a longitudinally expanding boost invariant fluid with no transverse expansion. 
 We take the initial temperature of the fluid to be $T_0=400$ MeV at an initial time that we choose to be $\tau_0=0.6$~fm$/c$.
The bulk matter then evolves hydrodynamically, expanding longitudinally and cooling, in a way that is determined by the equation of state and shear viscosities specified above. We shall assume that the background flow freezes out, which is to say hadronizes, when the local temperature drops to a freezeout temperature that we choose to be $T_f=154$ MeV. The freezeout time is $\tau_f\approx 10.5$ fm$/c$ for the ideal case in which we choose $\eta=0$ and is $\tau_f\approx 11.7$ fm$/c$ when we choose a fluid with nonzero shear viscosity, specifically $\eta_0=s_0/(4\pi)$. When the fluid viscosity is nonzero, heat can be generated as the fluid expands; this is why it takes a little longer for the fluid to cool to $T_f$. While the first order hydrodynamic equation (\ref{eqn:bjorken}) is known to lead to unphysical behavior due to the acausal nature of the relativistic Navier-Stokes equations, for the values of $\eta_0/s_0$, $\tau_0$ and $T_0$ that we use in this work --- which satisfy $\tau_0 T_0 \gg \eta_0/s_0$ ---  these effects are negligible and the solution of the equation (\ref{eqn:bjorken}) provides a good description of the evolution of the system obtained with causal completions of hydrodynamics \cite{Baier:2006um}.

On top of this fluid, we now wish to compute the linearized hydrodynamic response to an energetic probe traversing the medium and losing energy at the rate~(\ref{eqn:sc}), with the strong coupling parameter taken to be $\kappa_{sc}=0.4$. We shall choose the Gaussian widths in Eq.~(\ref{eqn:Jmu}) as $\sigma = \frac{1}{\pi T}$ and $\sigma_{\eta_s} = \frac{1}{\pi}$. We will study a high energy parton with an initial energy $E_{\ma{in}}=100$ GeV that starts at a point $(x_0,y_0)$ in the transverse plane and travels along the $+x$-direction in the transverse plane and $\eta_{s\, {\rm parton}}=0$. We assume the jet direction does not change during the in-medium evolution, so $x_{\rm parton}(\tau)=x_0+\frac{1}{\cosh(\eta_{s\, {\rm parton}})}(\tau-0.6)$ fm, $y_{\rm parton}(\tau)=y_0$. The shape of the solution is independent of the choice of $x_0$ and $y_0$, since our fluid is translation-invariant in the transverse directions. In a future Monte Carlo study we would choose the probability distribution for $x_0$ and $y_0$ for an ensemble of jets using a Glauber model and would choose their direction of propagation at random. In the present calculation, with a medium that is homogeneous in the transverse directions, any choice of $x_0$ and $y_0$ is as good as any other, as long as we make sure that the wake generated is contained within the box that we use for our numerical computation. 

Later, in Section~\ref{sect:spectrum}, we shall model the effects of transverse, radial, flow by using the velocity field extracted from a longitudinally boost invariant $2+1$D viscous hydrodynamic simulation of a central heavy ion collision to boost the hadrons that freezeout from each point on a freezeout surface. Doing so will mean that each point in the medium is no longer equivalent to others, and in particular will require choosing a point in the transverse plane that is the origin for the radial flow, which need not be the same as the point where the high energy parton starts. This also means that, in Section~\ref{sect:spectrum}, it will be necessary for us to make sure that (most of) our wake falls within the transverse extent of the freezeout surface extracted from the $2+1$D hydrodynamic simulation.

We introduce an additional parameter $\tau_e$, after which the jet energy loss is switched off. This parameter is necessary since our system is infinite in extent and homogeneous in the transverse plane and, unlike in a real collision, jets never leave the plasma. 
To mimic the finite size of a real collision, we impose that for $\tau > \tau_e$, the right hand sides of (\ref{eqn:li_hydro1}, \ref{eqn:li_hydro2}, \ref{eqn:li_hydro3}) vanish. We can change the value of $ \tau_e$ to change how much energy the jet loses during the in-medium evolution.

Once all parameters are specified we are ready to simulate the medium response numerically. We do the calculation in momentum space, solving the Fourier transformed versions of the linearized hydrodynamics equations, namely Eqs.~(\ref{eqn:li_hydro1}, \ref{eqn:li_hydro2}, \ref{eqn:li_hydro3}). We discretize momentum in a grid of size $N_x\times N_y\times N_{\eta}$, where $N_i$ is the number of grid points in the $i$-direction in momentum space. We choose the momentum ranges $k^x\in[-10,10]$ GeV, $k^y\in[-10,10]$ GeV, $k^\eta\in[-60,60]$. We solve the linearized hydrodynamics equations at each momentum grid point by using the fourth-order Runge-Kutta method 
and then do the Fourier transform to go back to position space. 
Contributions outside the momentum ranges specified above are suppressed exponentially because of our Gaussian parametrization (\ref{eqn:Jmu}) of the source. In practice, we only need to solve the linearized hydrodynamics equations for points in momentum space with $k^\eta \in[-20,20]$ since it turns out that contributions from $|k^\eta|\in[20,60]$ are already exponentially suppressed. 

Our choice of the momentum range for $k^x$ gives a resolution of $\Delta x = \frac{\pi}{(k^x)_{\ma{max}}} \approx 0.06$ fm in the transverse plane with $(k^x)_{\ma{max}}=10$ GeV, and similarly for the $y$-direction. For the $\eta_s$-direction, the resolution power is 
$\Delta \eta_s = \frac{\pi}{60}\approx 0.05$. The range of the $x, y, \eta_s$ space included in our calculation is a box of size $\frac{\pi N_x}{(k^x)_{\ma{max}}} \times \frac{\pi N_y}{(k^y)_{\ma{max}}}  \times \frac{\pi N_\eta}{(k^\eta)_{\ma{max}}} $ centered at the origin. That is, we establish as our convention that $x_0=y_0=0$ corresponds to the center of our calculational box. We must then choose $N_x$, $N_y$ and $N_\eta$ large enough such that our calculational box encompasses the whole jet wake in the transverse plane and spacetime rapidity. The conditions can be approximately written as
\be
x_0+ \sigma + (\tau_e-\tau_0) + c_s(\tau_f-\tau_e) &<& \frac{\pi N_x}{2(k^x)_{\ma{max}}} \\
-x_0+\sigma+c_s(\tau_f-\tau_0) &<& \frac{\pi N_x}{2(k^x)_{\ma{max}}}\\
\sigma + c_s (\tau_f-\tau_0) &<& \frac{\pi N_y}{2(k^y)_{\ma{max}}} \\
\sigma_{\eta_s} + c_s (\tau_f-\tau_0) &<& \frac{\pi N_{\eta}}{2(k^{\eta})_{\ma{max}}} \,.
\ee
where the left-hand-side of the first line gives the farthest distance the jet wake can reach along $+x$ direction while left-hand-side of the second line gives that along $-x$ direction. In the viscous case, the size of the jet wake can be somewhat 
bigger than the above estimate because perturbations to the hydrodynamic fluid can diffuse in addition to spreading at the speed of sound. Since the farthest distances along the $+x$ and $-x$ directions are different, we can optimize the value of $x_0$ to minimize the required value for $N_x$, which further reduces the computing resources needed to solve the linearized hydrodynamics. After fixing $N_x$, we choose $N_y=N_x$ by symmetry.

We shall present results from two calculations in this paper:
\begin{enumerate}

	\item The high energy parton traverses an ideal Bjorken fluid starting from $\tau_0=0.6$ fm$/c$ and losing energy and momentum until $\tau_e=4.6$ fm$/c$.
	We choose $x_0=-1.5$ fm, $y_0=0$ for the position of the energetic parton at $\tau_0$, whose direction of motion we choose to be in the positive $x$ direction. This choice of the direction and point of origin of the high energy parton ensures that the wake that the parton excites in the fluid remains within our calculational box until freezeout at $\tau_f=10.5$~fm$/c$, as we
	specify the size of our calculational box by choosing $N_x=N_y=275$ and $N_\eta=751$.
	
	\item Same as Case 1, but the fluid is a viscous Bjorken fluid with $\eta_0/s_0=1/(4\pi)$, which freezes out at $\tau_f=11.7$~fm$/c$.
	We choose $x_0$ and $y_0$ as in Case 1 but take
	$N_x=N_y=321$ and $N_\eta=751.$ (We need a slightly larger calculational box for this Case because of the later freezeout and because when the fluid has a nonzero viscosity the wake diffuses and so spreads a little farther.)
	
	
\end{enumerate}
In a future Monte Carlo calculation, we would need a shower of energetic partons not just one, and would need to choose an ensemble of events with varying points of origin for the energetic partons and varying directions of motion.

\subsection{Development of the Wake}

\begin{figure}
    \begin{subfigure}[t]{0.49\textwidth}
        \centering
        \includegraphics[height=2.4in]{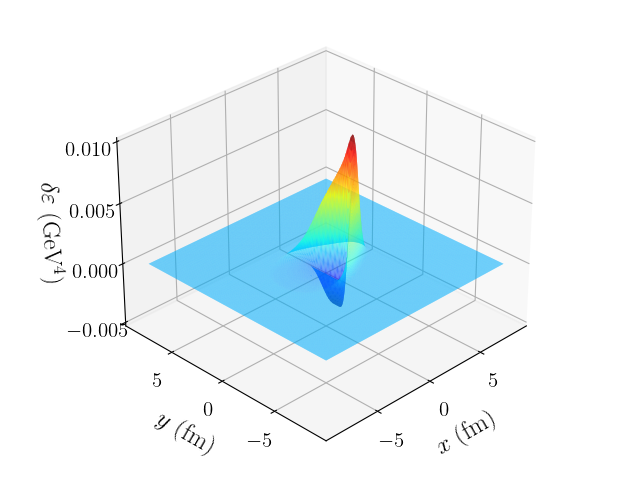}
        \caption{Case 1 (ideal), $\tau=4.9$ fm$/c$.}
    \end{subfigure}%
    ~
    \begin{subfigure}[t]{0.49\textwidth}
        \centering
        \includegraphics[height=2.4in]{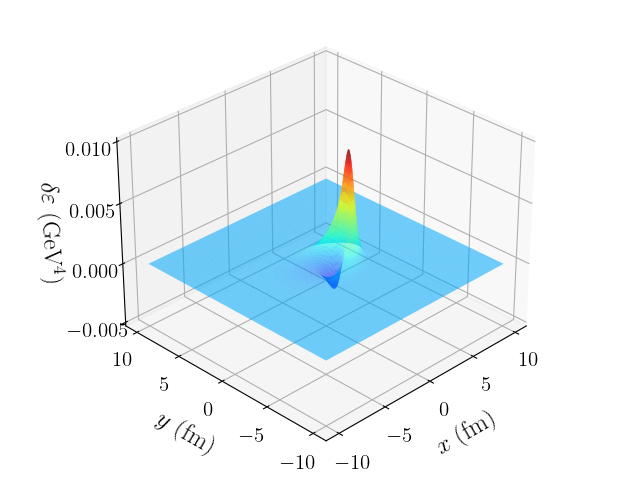}
        \caption{Case 2 (viscous), $\tau=4.9$ fm$/c$.}
    \end{subfigure}%
    
    \begin{subfigure}[t]{0.49\textwidth}
        \centering
        \includegraphics[height=2.4in]{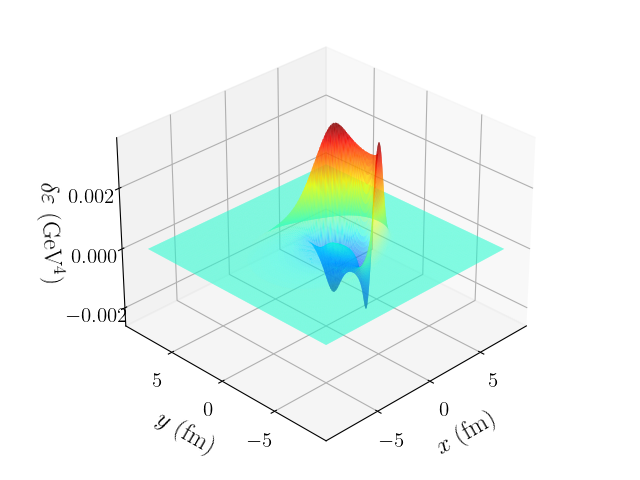}
        \caption{Case 1 (ideal), $\tau=7.7$ fm$/c$.}
    \end{subfigure}%
    ~
    \begin{subfigure}[t]{0.49\textwidth}
        \centering
        \includegraphics[height=2.4in]{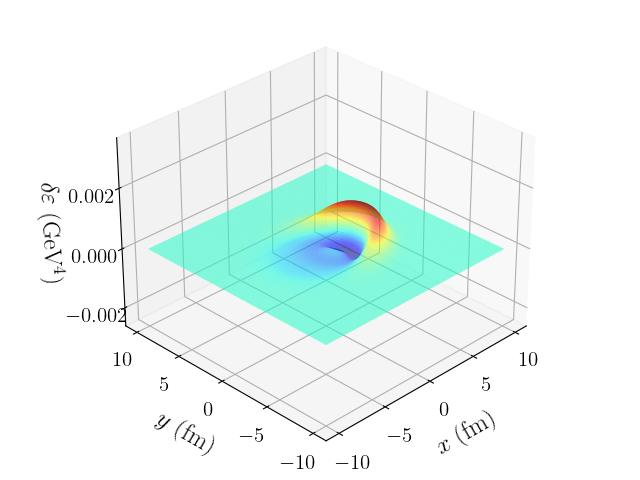}
        \caption{Case 2 (viscous), $\tau=8.3$ fm$/c$.}
    \end{subfigure}%
    
    \begin{subfigure}[t]{0.49\textwidth}
        \centering
        \includegraphics[height=2.4in]{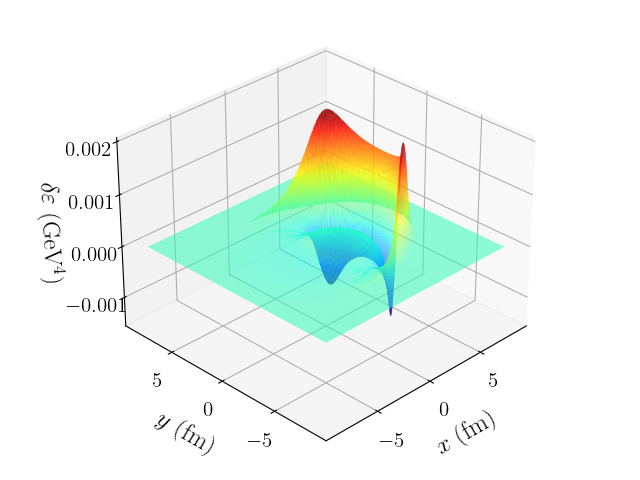}
        \caption{Case 1 (ideal), $\tau=10.5$ fm$/c$ (freezeout).}
    \end{subfigure}%
    ~
    \begin{subfigure}[t]{0.49\textwidth}
        \centering
        \includegraphics[height=2.4in]{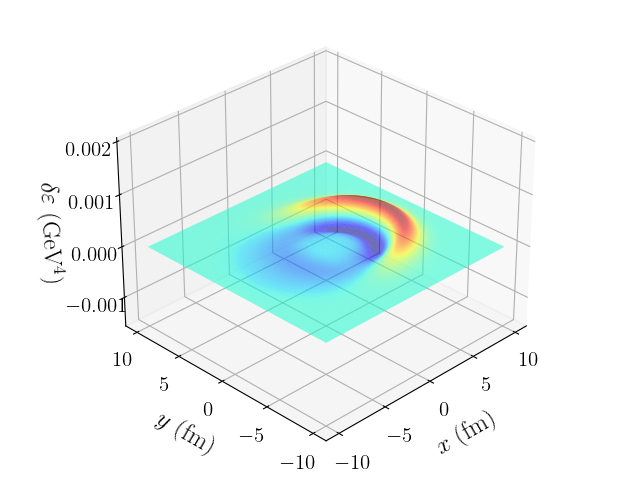}
        \caption{Case 2 (viscous), $\tau=11.7$ fm$/c$ (freezeout).}
    \end{subfigure}%
\caption{Plots of $\delta \varepsilon(\eta_s=0)$ as a function of $x$ and $y$ at three different proper times $\tau$ for Case 1 (ideal fluid; left panels) and 2 (viscous fluid; right panels).}
\label{fig:3D_epsilon}
\end{figure}

With our calculational framework, setup and parameters now fully specified, we can now study how the jet wake develops during the in-medium evolution. In Fig.~\ref{fig:3D_epsilon}, the energy perturbation $\delta\varepsilon(x,y,\eta_s=0)$ is plotted at three different times for both Case 1 and 2. For both Cases, the first time plotted is $\tau=4.9~{\rm fm}/c$, just after energy deposition by the hard parton is complete. In both Cases the third time plotted is at freezeout, when the temperature of the background medium has dropped to $T_f=154$~MeV, and the second time plotted is at the midpoint in time between the first and third.
The wake moves along the $+x$-direction, with energy accumulation on a wavefront that is apparent in the Figures and that follows the energetic parton, and energy depletion just behind this wavefront. In the case of an ideal fluid, Case 1, only the propagating sound mode exists while in the viscous fluid, Case 2, there is also a diffusive mode. This is why the energy perturbation in the viscous fluid, Case 2, spreads out more in the transverse plane,  with its peak value dropping more rapidly with time. Note that at each time plotted the left and right panels in Fig.~\ref{fig:3D_epsilon} have the same vertical axes, but the vertical axes for plots at different times differ.
We see that the magnitude of the perturbation is initially similar in the two cases, but indeed it drops much more rapidly with time in the viscous fluid.

\begin{figure}
    \begin{subfigure}[t]{0.49\textwidth}
        \centering
        \includegraphics[height=2.4in]{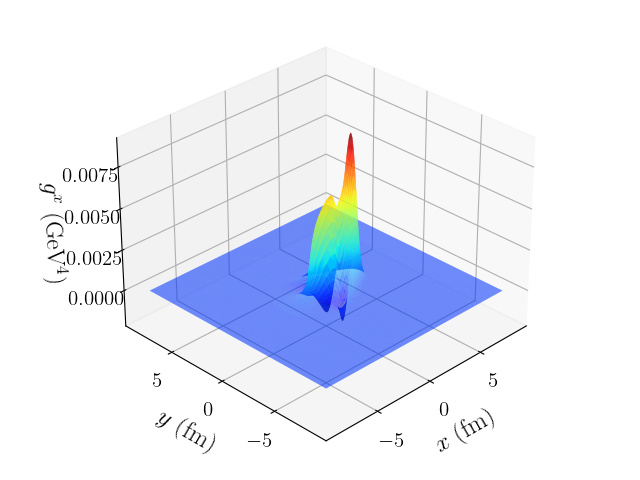}
        \caption{Case 1 (ideal), $\tau=4.9$ fm$/c$.}
    \end{subfigure}%
    ~
    \begin{subfigure}[t]{0.49\textwidth}
        \centering
        \includegraphics[height=2.4in]{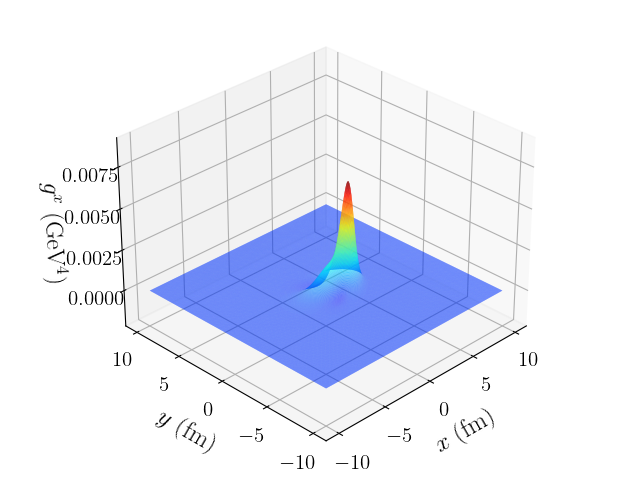}
        \caption{Case 2 (viscous), $\tau=4.9$ fm$/c$.}
    \end{subfigure}%
    
    \begin{subfigure}[t]{0.49\textwidth}
        \centering
        \includegraphics[height=2.4in]{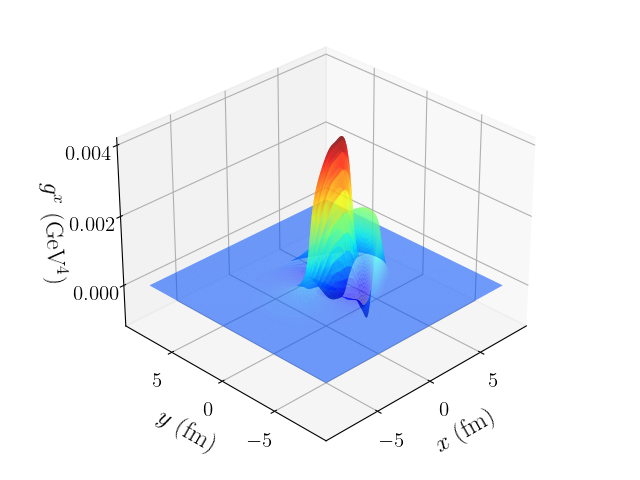}
        \caption{Case 1 (ideal), $\tau=7.7$ fm$/c$.}
    \end{subfigure}%
    ~
    \begin{subfigure}[t]{0.49\textwidth}
        \centering
        \includegraphics[height=2.4in]{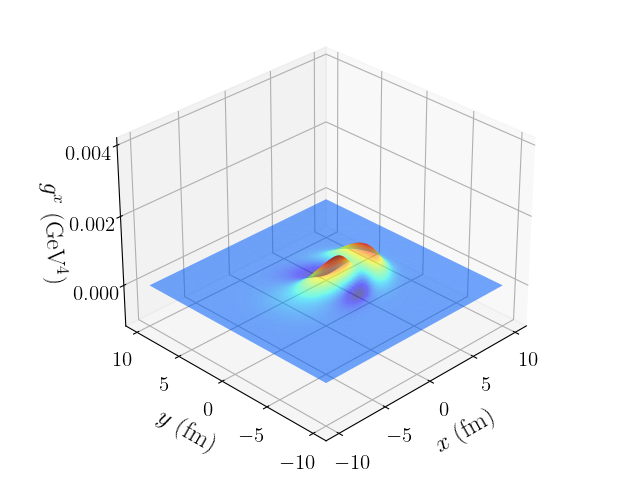}
        \caption{Case 2 (viscous), $\tau=8.3$ fm$/c$.}
    \end{subfigure}%
    
    \begin{subfigure}[t]{0.49\textwidth}
        \centering
        \includegraphics[height=2.4in]{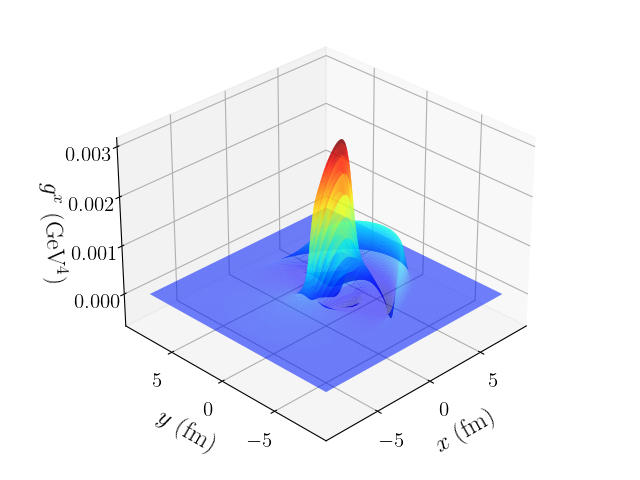}
        \caption{Case 1 (ideal), $\tau=10.5$ fm$/c$ (freezeout).}
    \end{subfigure}%
    ~
    \begin{subfigure}[t]{0.49\textwidth}
        \centering
        \includegraphics[height=2.4in]{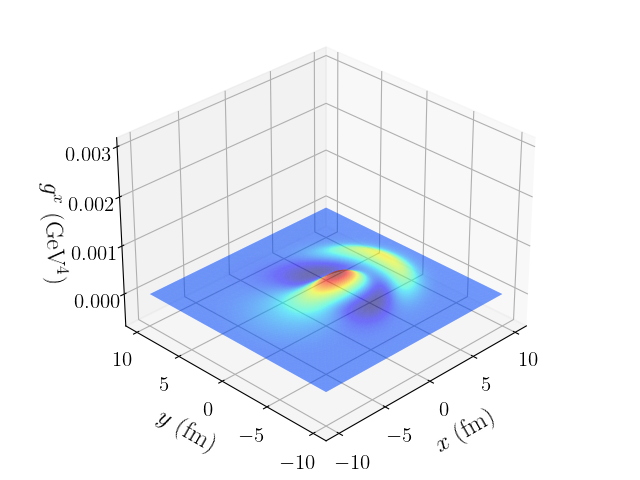}
        \caption{Case 2 (viscous), $\tau=11.7$ fm$/c$ (freezeout).}
    \end{subfigure}%
\caption{Plots of $g^x(\eta_s=0)$ as functions of $x$ and $y$ at three different proper times $\tau$ for Cases 1 (ideal fluid; left panels) and 2 (viscous fluid; right panels).}
\label{fig:3D_gx}
\end{figure}

Next we plot the momentum perturbation $g^x(x,y,\eta_s=0)$ in Fig.~\ref{fig:3D_gx}. Since the jet moves along the $+x$-direction and only deposits momentum along the $+x$-direction, we will not show plots for $g^y$ here. In the ideal case, a wing-shaped peak structure in $g^x$ is left behind by the energetic parton propagating along the $x$-axis, and this moving wake of fluid remains quite prominent at late times. This structure arises from the transverse mode $g_T$. The transverse mode does not diffuse in the ideal case (see Eq.~(\ref{eqn:gT})). 
In addition to sustaining the wing-shaped peak along the $x$-axis, a Mach cone structure is also seen in a wavefront spreading out from the place where the energy density has its positive peak in Fig.~\ref{fig:3D_epsilon}. This structure originates from the coupling between the longitudinal mode and the energy perturbation (see Eq.~(\ref{eqn:gL})). In the viscous case, the transverse mode $g_T$ diffuses and spreads so we see much less accumulation of momentum perturbation along the $x$-axis, a much less prominent wing-shaped structure along the $+x$-direction, less wake. For the same reason, the Mach cone structure coming from the longitudinal mode is also spread out. Note again that at each time plotted the left and right panels of 
Fig.~\ref{fig:3D_gx} have the same vertical axes (whereas we have chosen different vertical axes at different times). We see that, as in Fig.~\ref{fig:3D_epsilon}, the magnitude of the perturbation drops much more rapidly with time in the viscous fluid.

\subsection{Energy-Momentum Conservation}
\label{sect:conservation}

\begin{figure}
    \begin{subfigure}[t]{0.48\textwidth}
        \centering
        \includegraphics[height=2.3in]{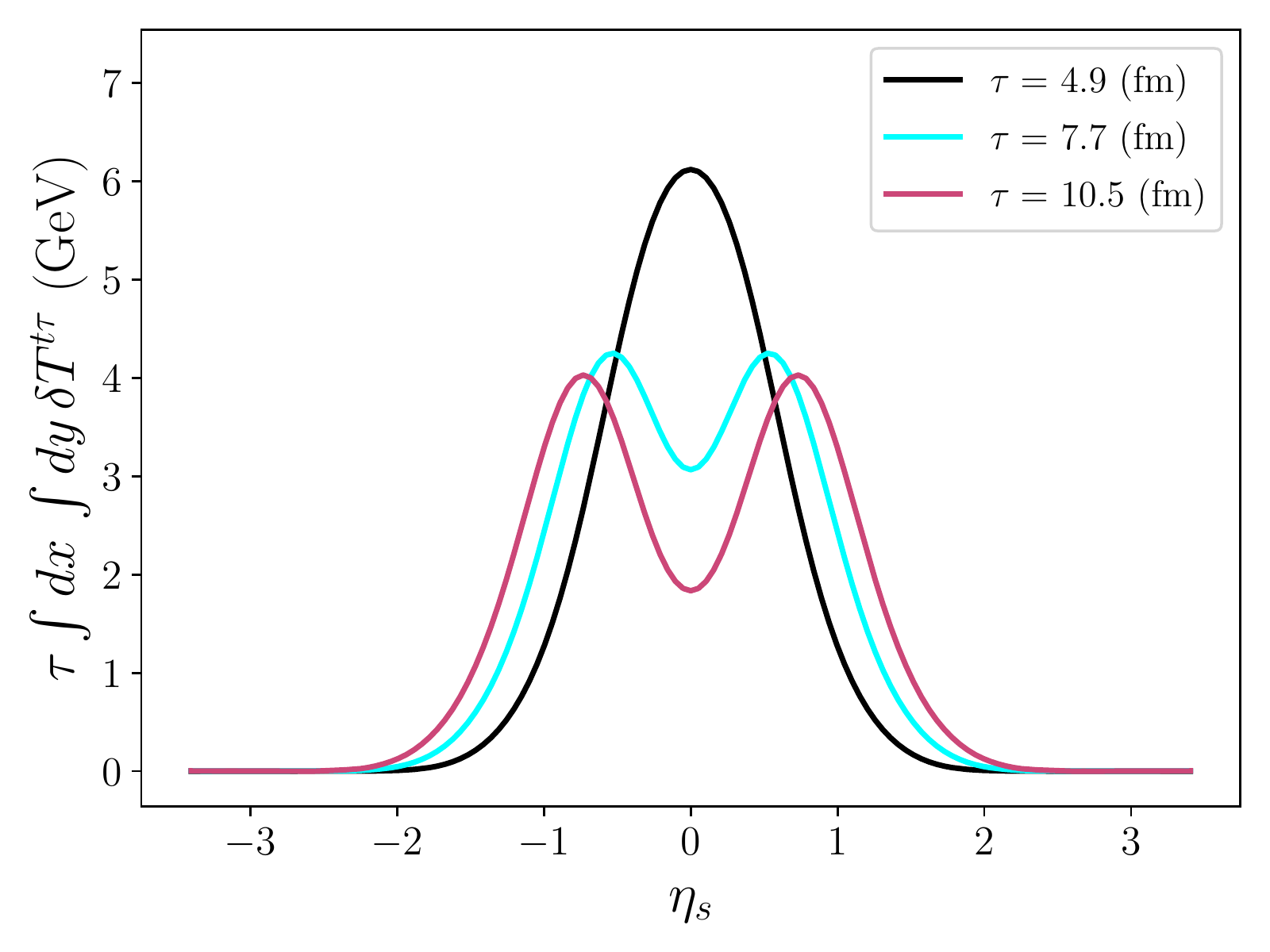}
        \caption{Integrated $\delta T^{t\tau}$ in Case 1 (ideal).}
    \end{subfigure}%
    ~
    \begin{subfigure}[t]{0.48\textwidth}
        \centering
        \includegraphics[height=2.3in]{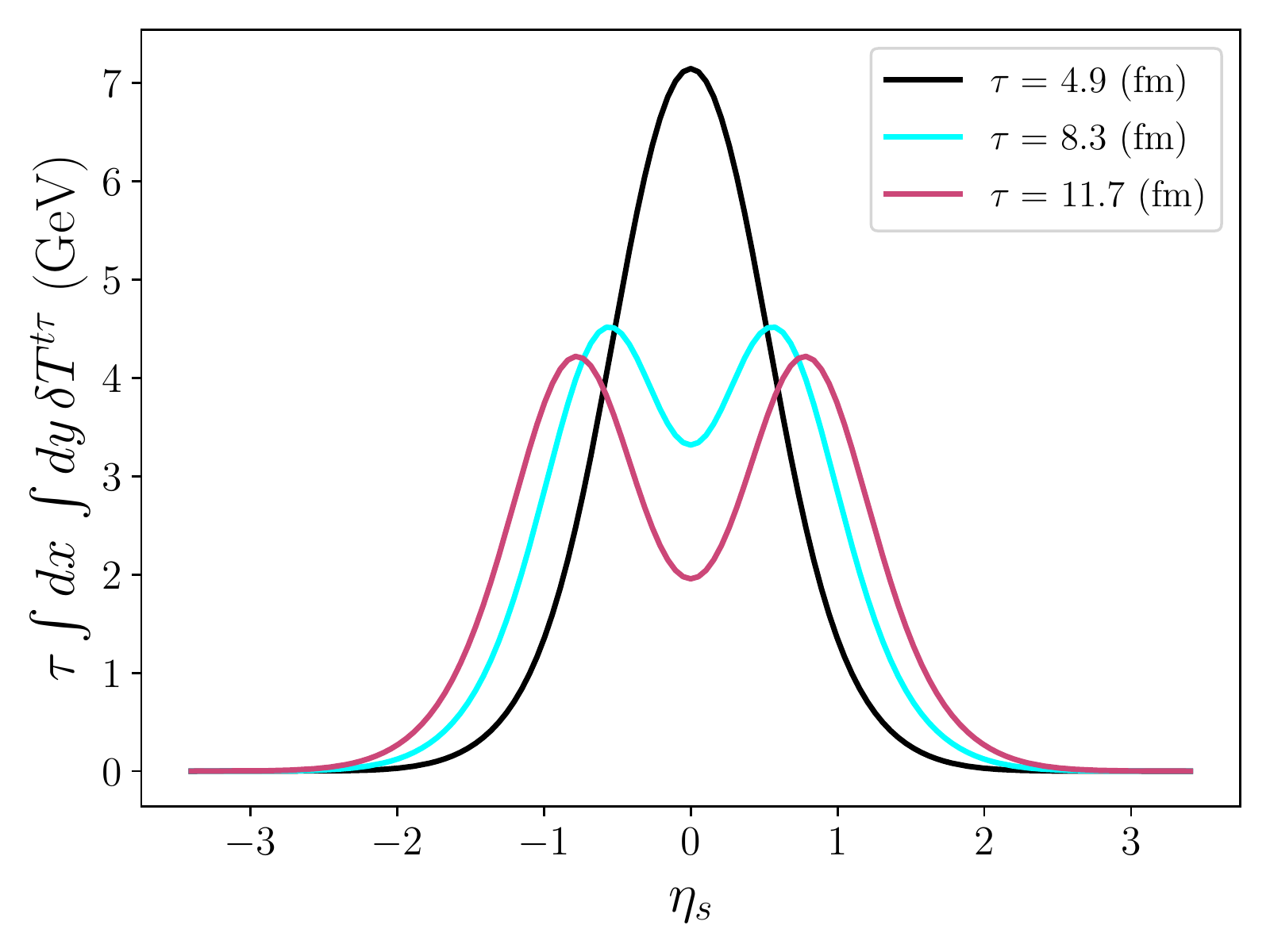}
        \caption{Integrated $\delta T^{t\tau}$ in Case 2 (viscous).}
    \end{subfigure}%
\caption{$\delta T^{t\tau}$ integrated over $x$ and $y$ and multiplied by $\tau$ in Case 1 and 2, in each panel at three different values of $\tau$. Integrating the curves over $\eta_s$ gives the total energy deposited by the high energy parton.}
\label{fig:f_eta}
\end{figure}

As a test of our numerical solver, we check that the energy and momentum injected into soft modes by the passing high energy parton are indeed 
stored in the hydrodynamic fields at the (much later) time of freezeout. As we discussed in Section~\ref{sect:Jmu} and Appendix~\ref{sect:killing}, the integration of $\delta T^{\tau\tau} = \delta\varepsilon$ and $\delta T^{\eta_s\tau} = g^\eta$ at some $\tau>\tau_e$ do not reproduce the energy and the $z$-momentum deposited by the high energy parton. Instead, the conserved energy and momentum are related to the integrals of the stress tensor components as 
\be
\int\tau \diff x \diff y \diff \eta_s\, \delta T^{t\tau}(\tau,x,y,\eta_s) &=& \Delta E_{\ma{tot}} \\
\int\tau \diff x \diff y \diff \eta_s\, \delta T^{\perp\tau}(\tau,x,y,\eta_s) &=& \Delta P^\perp_{\ma{tot}} \\
\int\tau \diff x \diff y \diff \eta_s\, \delta T^{z\tau}(\tau,x,y,\eta_s) &=& \Delta P^z_{\ma{tot}} \,,
\ee
where $\delta T^{t\tau} = (\cosh\eta_s)\delta T^{\tau\tau} + \tau(\sinh\eta_s)\delta T^{\eta_s\tau}$ and $\delta T^{z\tau} = (\sinh\eta_s)\delta T^{\tau\tau} + \tau(\cosh\eta_s)\delta T^{\eta_s\tau}$. In Fig.~\ref{fig:f_eta}, $\delta T^{t\tau}$ integrated over $x$ and $y$, multiplied by $\tau$ is shown as a function of $\eta_s$ for Cases 1 and 2. We have plotted this quantity because integration of each curve over $\eta_s$ yields the total energy in the perturbation. In each panel of Fig.~\ref{fig:f_eta}, energy conservation corresponds to the constancy of the integral of the curves at three different times. (Note that more energy is deposited in the viscous fluid, Case 2, because parton energy loss and hence energy deposition is larger when the fluid is hotter, and the viscosity slows down the drop in the temperature of the fluid.) Due to the sound modes in Eq.~(\ref{eqn:li_hydro3}), the energy deposition propagates along $\pm \eta_s$ directions and extends to a quite large rapidity region ($|\eta_s| \lesssim 2$).

To complete our check of energy and momentum conservation, we calculate the total energy and momentum lost by the high energy parton from the strong coupling formula (\ref{eqn:sc}) together with (\ref{eqn:dP}):
\be
\Delta E_{\ma{tot}} &=& \int_{\tau_0}^{\tau_e} \diff\tau \frac{\diff E}{\diff \tau} \\
P^i_{\ma{tot}} &=& \int_{\tau_0}^{\tau_e} \diff\tau v^i \frac{\diff E}{\diff \tau} \,,
\ee
in which the velocity $v^i$ is given in Eqs.~(\ref{eqn:vperp}, \ref{eqn:vz}). The quantities $\Delta E_{\ma{tot}}$ and $P^i_{\ma{tot}}$ represent the total energy and momentum lost by, and hence deposited into the medium by, the passing high energy parton.

For each of the two Cases considered here, we calculate the total deposition of energy and momentum into the fluid by the energetic parton 
and compute the energy and momentum in the hydrodynamic wake at freezeout from the linearized hydrodynamics. Since in both our calculations we have a high energy parton moving along the $+x$-direction with constant $\eta_s=0$, we have $v^x=1$, $v^y=0$ and $v^z=0$. Therefore, all the momentum deposited is along the $+x$-direction. In Table~\ref{tab:conservation} we show our results for the energy and momentum in the $x$-direction stored in the hydrodynamic fields at freezeout and compare them with the energy and momentum deposited into the plasma by the high energy parton. We observe that indeed energy and momentum conservation both work well. 
We also observe that momentum conservation is more accurate than energy conservation. We believe this is due to the $\cosh\eta_s$ and $\sinh\eta_s$ terms in $T^{t\tau}$, which can lead to numerical instability at large $\eta_s$. Our test of energy conservation confirms that we have this well under control.
We have also tested that momenta along the two directions perpendicular to the jet are very small in magnitude ($<10^{-15}$), consistent with zero, and very well conserved.

\begin{table}
\centering
\begin{tabular}{ | c | c | c | c | c| } 
\hline
Case & $\Delta E_{\ma{tot}}$ Deposited &   $\Delta E_{\ma{tot}}$ Recovered & $\Delta P^x_{\ma{tot}}$ Deposited & $\Delta P^x_{\ma{tot}}$ Recovered \\ 
\hline
1(ideal)   & 8.768   &8.775 & 8.768 & 8.768 \\
\hline
2(viscous) & 9.820   & 9.868 & 9.820 & 9.819\\
\hline
\end{tabular}
\caption{Tests of energy-momentum conservation. All magnitudes are in GeV.}
\label{tab:conservation}
\end{table}

\subsection{Validity of Linear Approximation}
\label{sect:u}

\begin{figure}
    \begin{subfigure}[t]{0.49\textwidth}
        \centering
        \includegraphics[height=2.2in]{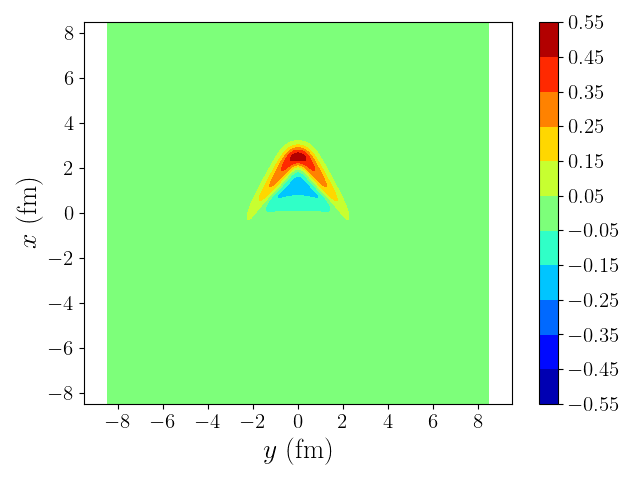}
        \caption{Case 1 (ideal), $\tau=4.9$ fm$/c$.}
    \end{subfigure}%
    ~
    \begin{subfigure}[t]{0.49\textwidth}
        \centering
        \includegraphics[height=2.2in]{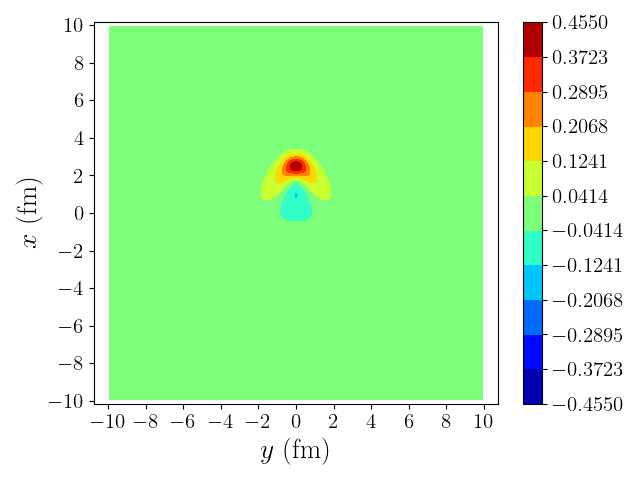}
        \caption{Case 2 (viscous), $\tau=4.9$ fm$/c$.}
    \end{subfigure}%

    \begin{subfigure}[t]{0.49\textwidth}
        \centering
        \includegraphics[height=2.2in]{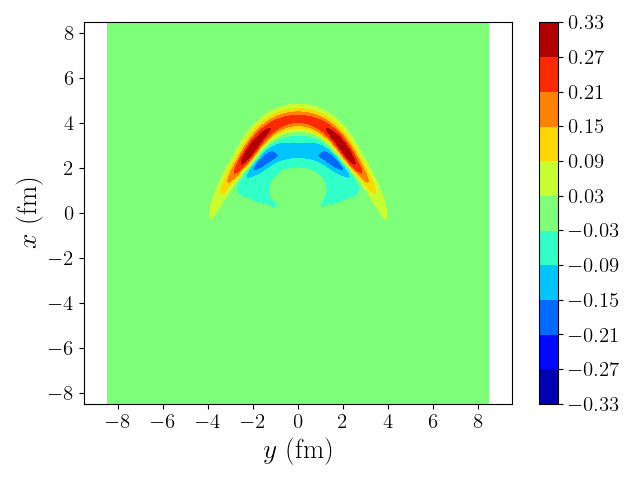}
        \caption{Case 1 (ideal), $\tau=7.7$ fm$/c$.}
    \end{subfigure}%
    ~
    \begin{subfigure}[t]{0.49\textwidth}
        \centering
        \includegraphics[height=2.2in]{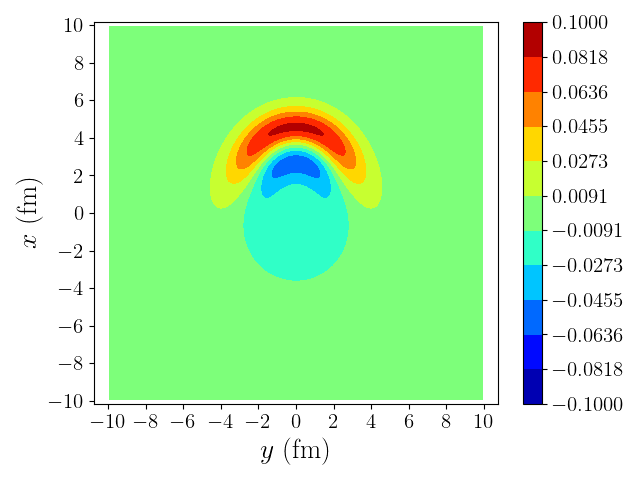}
        \caption{Case 2 (viscous), $\tau=8.3$ fm$/c$.}
    \end{subfigure}%

    \begin{subfigure}[t]{0.49\textwidth}
        \centering
        \includegraphics[height=2.2in]{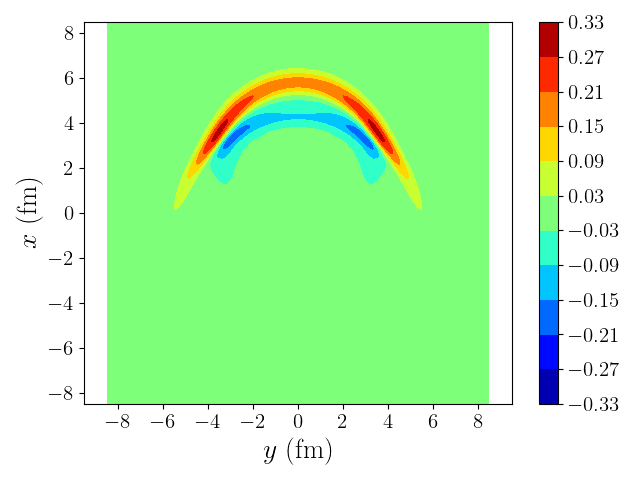}
        \caption{Case 1 (ideal), $\tau=10.5$ fm$/c$.}
    \end{subfigure}%
    ~
    \begin{subfigure}[t]{0.49\textwidth}
        \centering
        \includegraphics[height=2.2in]{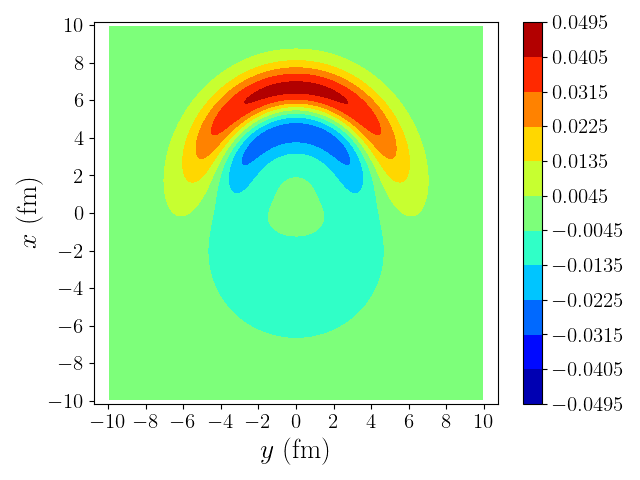}
        \caption{Case 2 (viscous), $\tau=11.7$ fm$/c$.}
    \end{subfigure}%
\caption{Plots of $\frac{\delta\varepsilon}{\varepsilon_0}(\eta_s=0)$ as functions of $x$ and $y$ at three different times $\tau$ for Case 1 (ideal fluid; left panels) and 2 (viscous fluid; right panels). Note that we have used different color bars in different panels; assessing the strength of the perturbations (in this and the next two Figures also) requires looking at the color bars to see the magnitudes corresponding to the reddest and bluest colors.}
\label{fig:de_over_e}
\end{figure}

\begin{figure}
    \begin{subfigure}[t]{0.49\textwidth}
        \centering
        \includegraphics[height=2.2in]{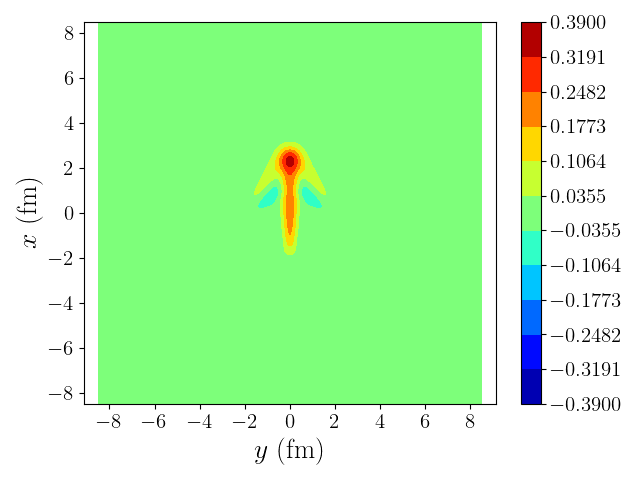}
        \caption{Case 1 (ideal), $\tau=4.9$ fm$/c$.}
    \end{subfigure}%
    ~
    \begin{subfigure}[t]{0.49\textwidth}
        \centering
        \includegraphics[height=2.2in]{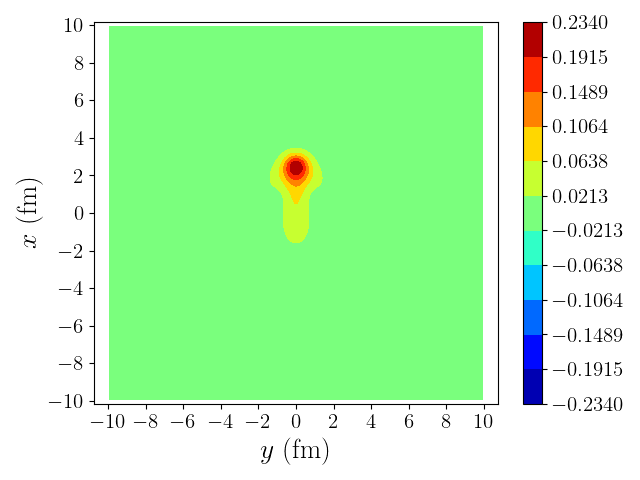}
        \caption{Case 2 (viscous), $\tau=4.9$ fm$/c$.}
    \end{subfigure}%

    \begin{subfigure}[t]{0.49\textwidth}
        \centering
        \includegraphics[height=2.2in]{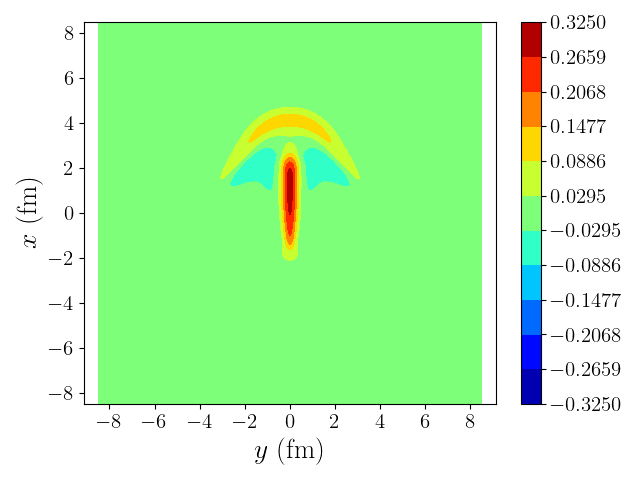}
        \caption{Case 1 (ideal), $\tau=7.7$ fm$/c$.}
    \end{subfigure}%
    ~
    \begin{subfigure}[t]{0.49\textwidth}
        \centering
        \includegraphics[height=2.2in]{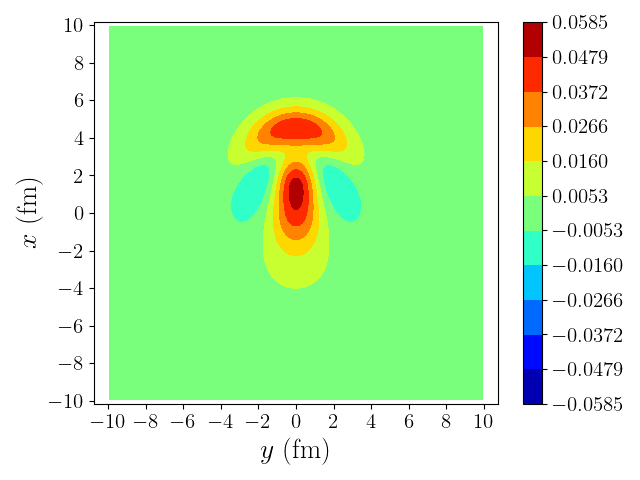}
        \caption{Case 2 (viscous), $\tau=8.3$ fm$/c$.}
    \end{subfigure}%
    
    \begin{subfigure}[t]{0.49\textwidth}
        \centering
        \includegraphics[height=2.2in]{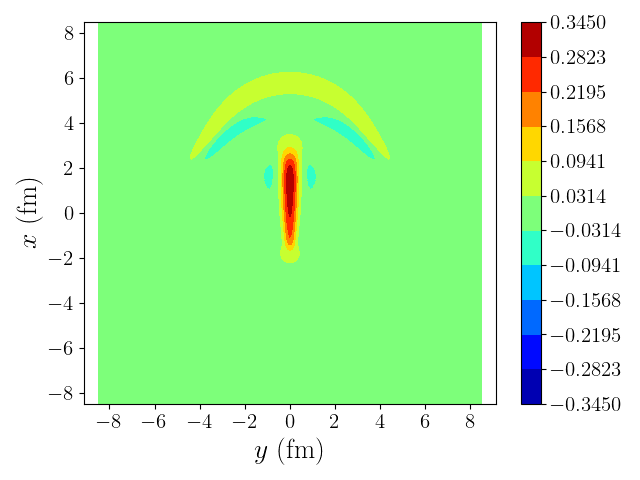}
        \caption{Case 1 (ideal), $\tau=10.5$ fm$/c$.}
    \end{subfigure}%
    ~
    \begin{subfigure}[t]{0.49\textwidth}
        \centering
        \includegraphics[height=2.2in]{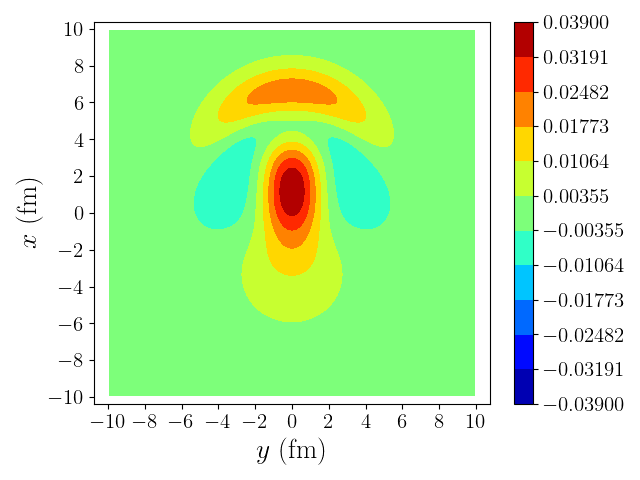}
        \caption{Case 2 (viscous), $\tau=11.7$ fm$/c$.}
    \end{subfigure}%
\caption{Plots of $\delta u^x(\eta_s=0)$ as functions of $x$ and $y$ at three different times $\tau$ for Case 1 (ideal fluid; left panels) and 2 (viscous fluid; right panels).}
\label{fig:du_x}
\end{figure}

To close this Section, we check the validity of our use of linearized hydrodynamics, which is based on the assumption that $\delta\varepsilon/\varepsilon_0$ (and hence $\delta P/P_0$), $\delta u^x$, $\delta u^y$ and $\delta u^{\eta_s}$ are all small, meaning that a truncation to linear order in these quantities is a good approximation. To this end, we show the contour plots of $\delta\varepsilon/\varepsilon_0$ and $\delta u^x$ at different times in Figs.~\ref{fig:de_over_e} and~\ref{fig:du_x} respectively. 
These two quantities provide an estimate for the magnitude of the perturbation, since the jet loses momentum along the $x$-direction and $\delta u^y$ and $\delta u^{\eta_s}$ are zero right after the deposition. They become nonzero by dynamical coupling with $\delta\varepsilon$ and $\delta u^x$, and thus are smaller than $\delta u^x$, as we have confirmed. Both $\delta\varepsilon/\varepsilon_0$ and $\delta u^x$ are shown at $\eta_s=0$, which is the value of the spacetime rapidity at which they attain their maximum value. The strength of the perturbations shown in the different panels of Figs.~\ref{fig:de_over_e} and~\ref{fig:du_x} are quite different; the color bars beside each panel provide the magnitudes corresponding to the reddest and bluest colors.

In both Cases 1 and 2, we see that the perturbations right after the deposition of energy and momentum into the fluid is completed are significant in magnitude, in small regions of space. As time goes on, in Case 2 where the fluid has a nonzero viscosity the perturbations soon become quite small in magnitude, and significantly smaller than in Case 1 where the fluid is ideal. This happens because the perturbations in the viscous case spread out further in the transverse plane via diffusion, and thus become smaller in magnitude. 
These observations highlight the fact that the use of the linearized approximation to hydrodynamics is under much better control when we consider energy deposition in a fluid with nonzero viscosity, even with a viscosity as small as $\eta_0/s_0 = 1/(4\pi)$ as in our Case 2, than when we consider energy deposition in an ideal fluid. In the viscous fluid, Case 2, the quantitative reliability of the linearized approximation is in doubt in small regions of space at early times but the approximation rapidly becomes quantitatively reliable. In the ideal fluid, Case 1, the small regions of space in which the reliability of the approximation is in doubt persist. These observations also highlight the importance of incorporating viscous effects into the dynamics of the wake, since turning on a viscosity as small as that in Case 2 introduces quite substantial changes to the magnitude of the wake by the time of freezeout. We have reported results for Case 1, the ideal fluid, because it is a standard calculational benchmark. In the next Section, however, when we consider particle production from the perturbed fluid we shall focus entirely on Case 2, where the fluid has a nonzero viscosity and the linearized approximation that we employ throughout is under good control long before freezeout.

\begin{figure}[t]
    \begin{subfigure}[t]{0.49\textwidth}
        \centering
        \includegraphics[height=2.2in]{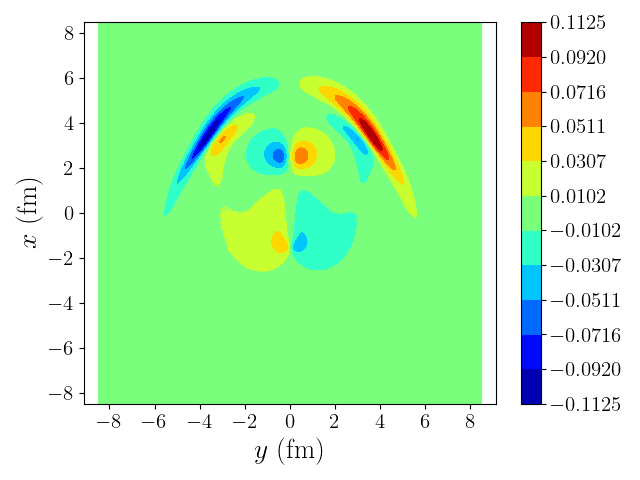}
        \caption{$\delta u^y(\eta_s=0)$ in Case 1 (ideal).}
    \end{subfigure}%
    ~
    \begin{subfigure}[t]{0.49\textwidth}
        \centering
        \includegraphics[height=2.2in]{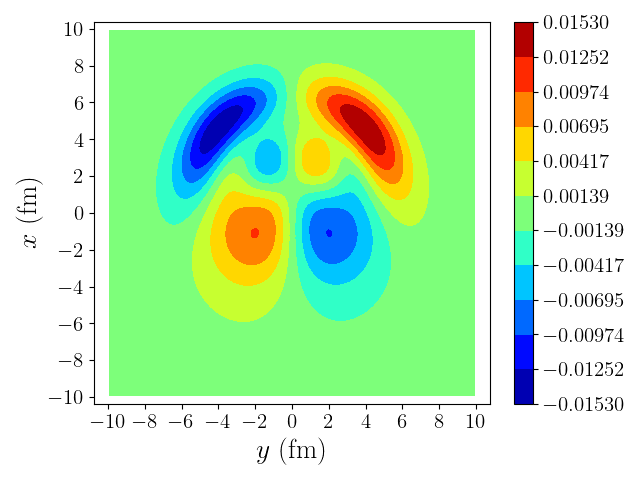}
        \caption{$\delta u^y(\eta_s=0)$ in Case 2 (viscous).}
    \end{subfigure}%
    
    \begin{subfigure}[t]{0.49\textwidth}
        \centering
        \includegraphics[height=2.2in]{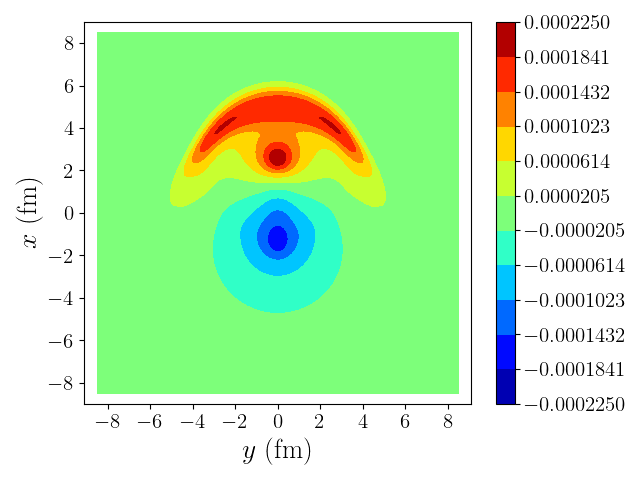}
        \caption{$\delta u^{\eta_s}(\eta_s\approx0.52)$ in Case 1 (ideal).}
    \end{subfigure}%
    ~
    \begin{subfigure}[t]{0.49\textwidth}
        \centering
        \includegraphics[height=2.2in]{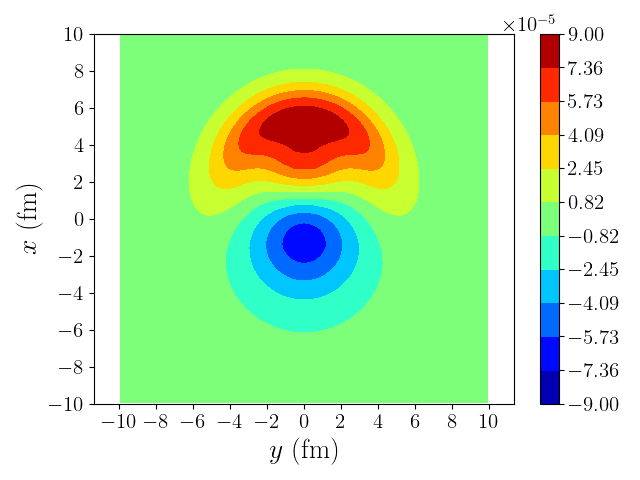}
        \caption{$\delta u^{\eta_s}(\eta_s\approx0.52)$ in Case 2 (viscous).}
    \end{subfigure}%
\caption{Plots of $\delta u^y(\eta_s=0)$ and $\delta u^{\eta_s}(\eta_s=0.52)$ as functions of $x$ and $y$ at freezeout for Case 1 (ideal fluid; left panels; $\tau_f=10.5$~fm$/c$) and Case 2 (viscous fluid; right panels; $\tau_f=11.7$~fm$/c$).}
\label{fig:du_yeta}
\end{figure}

For later reference, we plot $\delta u^y(\eta_s=0)$ and $\delta u^{\eta_s}$ at freezeout in Fig.~\ref{fig:du_yeta}. Since $\delta u^{\eta_s}=0$ at zero rapidity because of symmetry, we show its value at $\eta_s\approx 0.52$. As those plots clearly show, $\delta u^y$ and $\delta u^{\eta_s}$ are much smaller than $\delta\varepsilon/\varepsilon_0$ and $\delta u^x$, consistent with the discussion above.

\begin{figure}[t]
    \begin{subfigure}[t]{0.49\textwidth}
        \centering
        \includegraphics[height=2.2in]{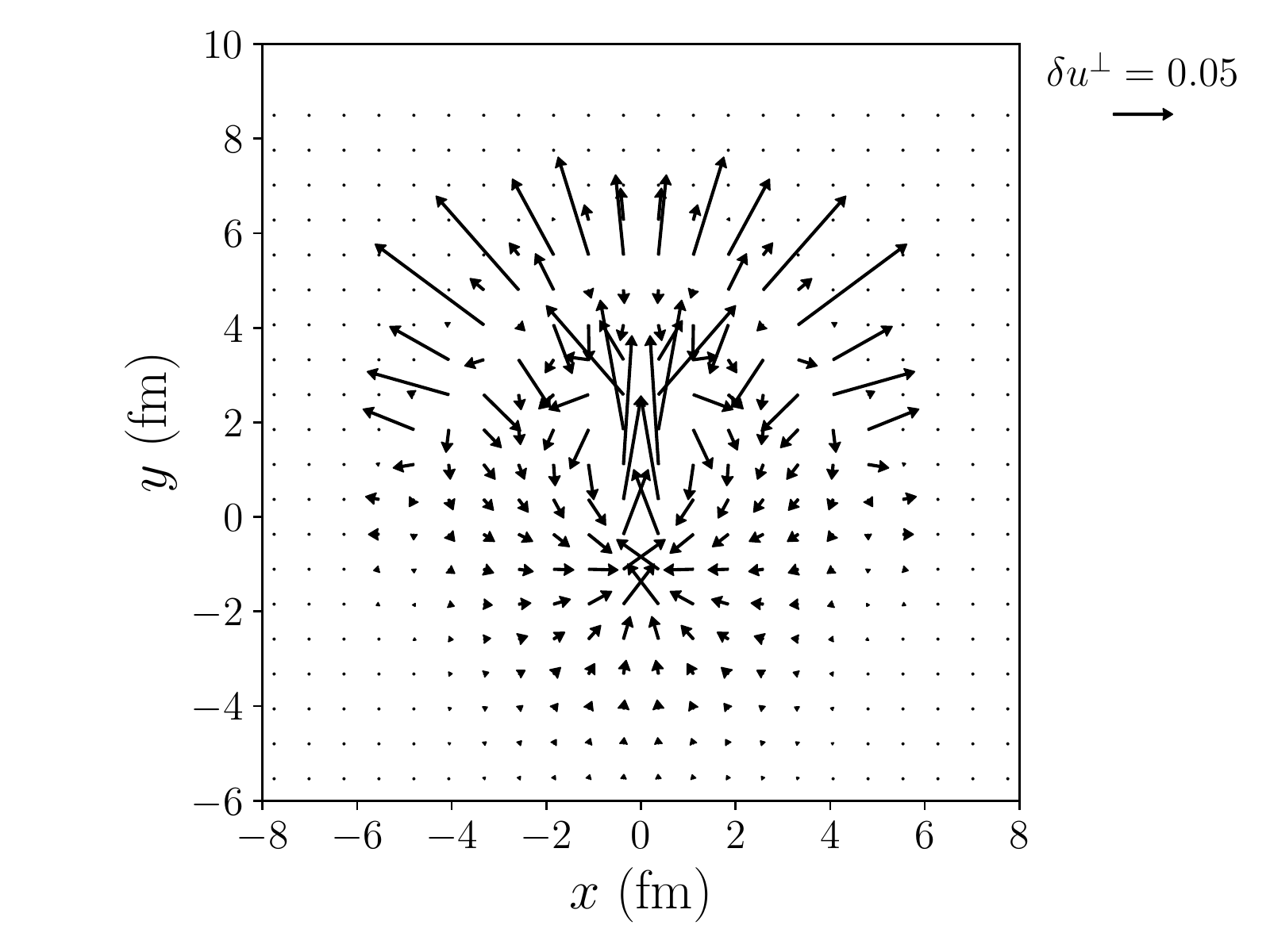}
        \caption{$\delta {\bs u}^\perp(\eta_s=0)$ in Case 1 (ideal).}
    \end{subfigure}%
    ~
    \begin{subfigure}[t]{0.49\textwidth}
        \centering
        \includegraphics[height=2.2in]{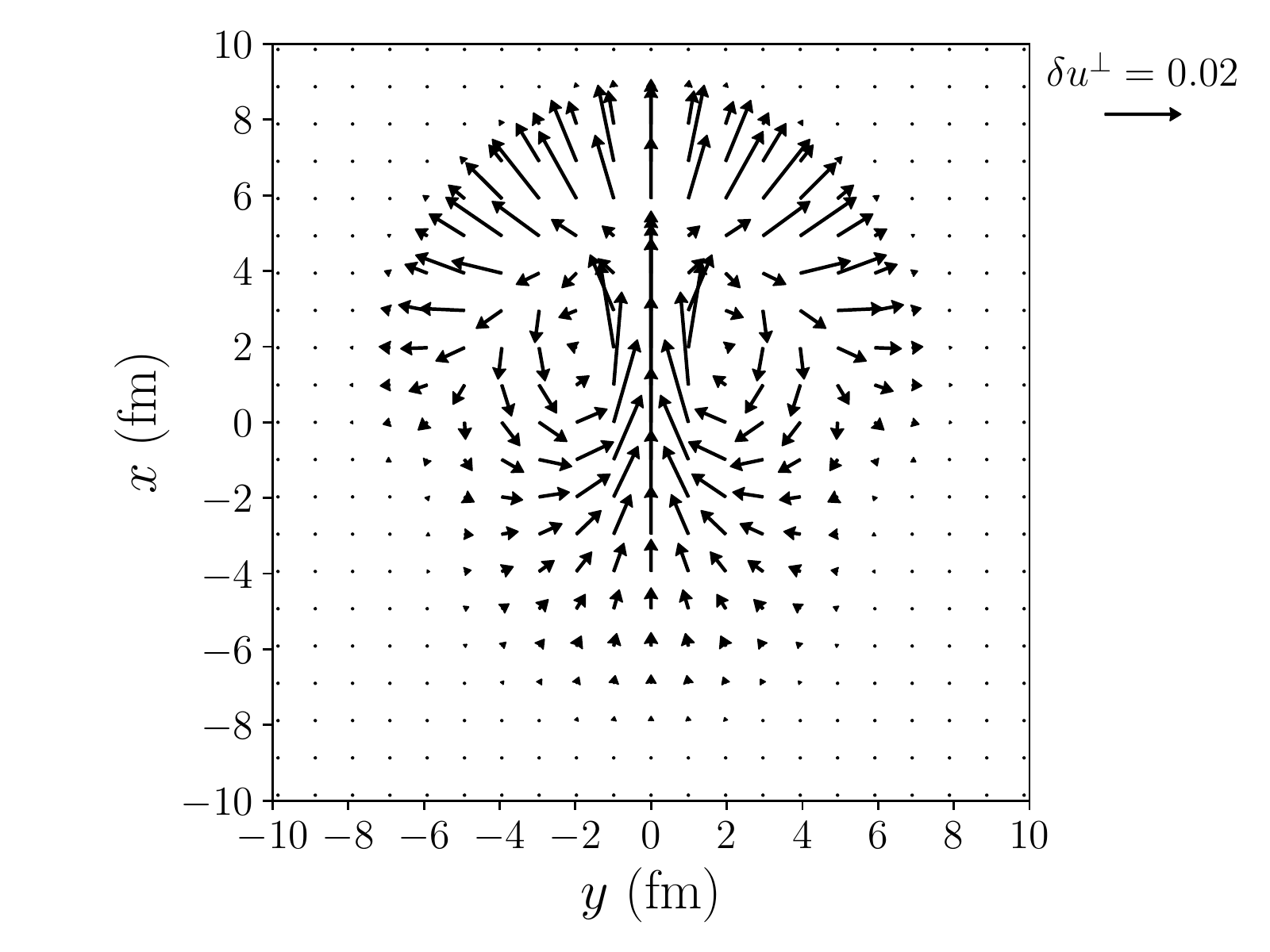}
        \caption{$\delta {\bs u}^\perp(\eta_s=0)$ in Case 2 (viscous).}
    \end{subfigure}%
\caption{Plots of $\delta {\bs u}^\perp(\eta_s=0)$ as functions of $x$ and $y$ at freezeout for Case 1 (ideal fluid; left panels; $\tau_f=10.5$~fm$/c$) and Case 2 (viscous fluid; right panels; $\tau_f=11.7$~fm$/c$).}
\label{fig:du_perp}
\end{figure}

Finally, in Fig.~\ref{fig:du_perp} we plot the vector field $\delta {\bs u}^\perp = (\delta u^x, \delta u^y)$ at freezeout in the transverse plane at $\eta_s=0$. We see a vortical flow pattern in the wake left behind by the energetic parton in both the ideal and viscous cases. As discussed in Ref.~\cite{Serenone:2021zef}, the jet creates a vortex ring  around the jet direction in its wake. (Because of the longitudinal expansion, there is no azimuthal symmetry under rotation about the jet direction meaning that the ring is not circular.) We see a cross-section of this smoke-ring-like pattern in Fig.~\ref{fig:du_perp}, indicating that it persists until freezeout even in the presence of viscosity.

\section{Particle Spectrum from Jet Wake}
\label{sect:spectrum}

\subsection{Cooper-Frye Formula}
\label{sect:CooperFrye}

After the system has cooled down sufficiently, the fluid breaks apart into hadrons that fly out of the collision zone and are ultimately detected in the experimentalists' detector. While a full microscopic description of this freezeout process is complicated, in particular because as it freezes out the hydrodynamic fluid falls out of equilibrium, the standard procedure for treating freezeout from a hydrodynamic fluid in local thermal equilibrium is to use the Cooper-Frye prescription~\cite{Cooper:1974mv} in which particle production at freezeout is modeled by assuming that the particles formed from each cell in the hydrodynamic fluid
are produced according to the equilibrium distribution with temperature $T_f$ in the local fluid rest frame defined by the velocity of the fluid cell.\footnote{We will not introduce viscous corrections to the particle distribution at freezeout.} Integrating over all the fluid cells on a specified freezeout hypersurface that we shall denote by $\sigma^\mu$ yields the spectrum for hadrons with a given mass $m$ produced at freezeout:   
\be
\label{eq:cfgen}
E\frac{\diff N}{\diff^3 p} = \frac{1}{(2\pi)^3} \int \diff \sigma^\mu p_\mu \, f\Big(\frac{u^\mu p_\mu}{T}\Big)\,,
\ee
where $f(x)$ is the Bose-Einstein distribution $(e^x-1)^{-1}$ for bosons with mass $m$ and the Fermi-Dirac distribution $(e^x+1)^{-1}$ for fermions with mass $m$.

The freezeout hypersurface is typically taken to be the isothermal hypersurface at which the fluid temperature drops to a specified $T_f$. 
Within our assumption of homogeneous Bjorken flow, for the unperturbed background fluid the isothermal freezeout surface is also isochronous, since all fluid cells cool down to the freezeout temperature $T=T_f$ at the same freezeout time $\tau_f$. For the bulk particles, taking the velocity field as 
 $u^\mu_0=(u^\tau_0, u^x_0, u^y_0, u^{\eta_s}_0) = (1,0,0,0)$, we find
\be
\diff\sigma^\mu p_\mu &=& \tau_f \diff x \diff y \diff \eta_s\, m_T\cosh(\ma{y}-\eta_s)\\[2pt]
\label{eqn:u0_dot_p}
\frac{u_0^\mu p_\mu}{T} &=& \frac{m_T \cosh(\ma{y}-\eta_s)}{T_f} \,,
\ee
where $m_T\equiv\sqrt{p_T^2+m^2}$ is the transverse mass ($p_T$ is the transverse momentum and $m$ denotes the mass of a specific particle produced from the wake) and $\ma{y}$ is the momentum rapidity.

When we consider the perturbed fluid, including the wake that originates from the deposition of energy and momentum by the passing high energy parton and that has evolved hydrodynamically until the freezeout time as we have analyzed in previous Sections, the isochronous and isothermal surfaces no longer coincide because the energy perturbation $\delta \varepsilon$ changes the
value of the temperature at the freezeout time of the background fluid. Nevertheless, in the case of a viscous fluid (even with a viscosity as small as $\eta_0/s_0=1/(4\pi)$ as in Case 2), at the freezeout time $\tau_f$ the perturbations from the wake are small, 
which means that the consequent change in
the temperature of the fluid in the wake relative to the background temperature is also small and so are the differences between the isochronous and isothermal surfaces. For this reason, throughout our analysis we shall compute particle production at the fixed $\tau_f$, isochronous, hypersurface. 
The perturbations to the fluid at the $\tau=\tau_f$ hypersurface due to the presence of the wake perturb the final-state spectra of hadrons, as we shall compute.

At the time $\tau=\tau_f$, 
 the flow velocity is $u^\mu = (1,\delta u^x, \delta u^y, \delta u^{\eta_s})$, where the velocity perturbations are related to $g^{\perp}$ and $g^{\eta}$ via Eqs. (\ref{eqn:define_gperp}, \ref{eqn:define_geta}), and $g^{\perp}$ and $g^{\eta}$ can be extracted from our linearized hydrodynamic calculation at $\tau=\tau_f$. Also, the local temperature at freezeout is modified by $\delta \varepsilon$, see Eq.~\eqref{eqn:eos1}. From Eqs. (\ref{eqn:eos1}, \ref{eqn:eos2}, \ref{eqn:define_gperp}, \ref{eqn:define_geta}) we then find
\be
\label{eqn:u_dot_p}
u^\mu p_\mu &=& m_T \cosh(\ma{y}-\eta_s) - \frac{{\bs g}^\perp \cdot {\bs p}_\perp}{\varepsilon_0+P_0} - \frac{\tau_f g^\eta m_T\sinh(\ma{y}-\eta_s)}{\varepsilon_0+P_0} \\
\label{eqn:T_deltaT}
T &=& T_f + \delta T = \Big[\frac{\pi^2}{3g}\big( \varepsilon_0+\delta\varepsilon \big) \Big]^{1/4} \,,
\ee
where $g=40$ is the number of degrees of freedom in the plasma phase.
By taking the difference between the particle distribution from the fluid with the perturbation and that without, we obtain the distribution of particles originating from the jet wake in the linearized hydrodynamic approach:
\be
\label{eqn:spectrum}
\frac{\diff \Delta N}{p_T \diff p_T \diff\phi \diff \ma{y}} = \frac{1}{(2\pi)^3} \int \tau \diff x \diff y  \diff\eta_s \,
m_T \cosh(\ma{y}-\eta_s) \Big[ f \Big(\frac{u^\mu p_\mu}{T_f+\delta T}\Big) - f \Big(\frac{u_0^\mu p_\mu}{T_f}\Big) \Big]\,,
\ee
where $\phi$ is the azimuthal angle of the transverse momentum of the produced particle. 
The determination of the spectrum and the distribution in azimuthal angle and momentum rapidity $\ma{y}$ from the expression \eqref{eqn:spectrum} requires knowledge of the hydrodynamic perturbations, which in the linearized approximation to hydrodynamics we obtain via the numerical results from the previous Section. 

In the hybrid strong/weak coupling model of Ref.~\cite{Casalderrey-Solana:2016jvj}, a series of further assumptions and simplifications were employed, which reduced \eqref{eqn:spectrum} to a closed-form expression depending only on the amount of energy and momentum deposited in the fluid and not on knowledge of the detailed form of the resulting fluid perturbation.
The analysis of Ref.~\cite{Casalderrey-Solana:2016jvj} was, first of all, based on ideal hydrodynamics where the comoving entropy $\tau s$ ($s$ denotes the entropy density) is conserved. Second, 
in the Cooper-Frye analysis of freezeout \eqref{eq:cfgen}, the distinction between fermions and bosons was ignored in Ref.~\cite{Casalderrey-Solana:2016jvj}, with the distribution $f$ taken as the Boltzmann distribution. Third, in Ref.~\cite{Casalderrey-Solana:2016jvj}
the Boltzmann exponential was expanded to linear order in the velocity perturbation, which is to say that it was assumed that small perturbations to the hydrodynamic fluid yield small perturbations to the hadron spectra at all $p_T$, an assumption that is only valid at low $p_T$.
We shall not need to make any of these simplifying assumptions in our treatment of the wake.
The fourth assumption made in Ref.~\cite{Casalderrey-Solana:2016jvj}, which we shall also assume, is that the direction of the high energy parton, in azimuth and in spacetime rapidity, does not change as the high energy parton propagates through the fluid. In particular, this means that the high energy parton does not deposit any momentum in the $\eta_s$ direction into the fluid, see (\ref{eqn:C_eta}). 
Fifth, it was also assumed in Ref.~\cite{Casalderrey-Solana:2016jvj} that the energy and transverse momentum lost by a jet with momentum rapidity $\ma{y}_J$ turns into perturbations of the fluid at spacetime rapidity $\eta_s=\ma{y}_J$ throughout, i.e., $\delta\varepsilon(\tau, \eta_s)\propto \delta(\eta_s-\ma{y}_J)$ and $\delta P^\perp(\tau, \eta_s)\propto \delta(\eta_s-\ma{y}_J)$ for all $\tau$, including at $\tau=\tau_f$, the freezeout time. 
In our linearized hydrodynamics calculation, we shall not need this assumption either.
Making these five assumptions and simplifications, as in Ref.~\cite{Casalderrey-Solana:2016jvj},  yields the following hybrid model expression for the excess of particles generated by the wake in the fluid at freezeout:
\be
\label{eqn:spectrum_hb}
\frac{\diff \Delta N_{\ma{hybrid}}}{p_T \diff p_T \diff\phi \diff \ma{y}} &=& \frac{1}{32\pi}\frac{m_T}{T_f^5}\cosh(\ma{y}-\ma{y}_J) \exp\Big[ -\frac{m_T\cosh(\ma{y}-\ma{y}_J)}{T_f} \Big] \nn\\
\label{eqn:hb}
&\times&\bigg[ p_T \Delta  P^\perp \cos(\phi-\phi_J) + \frac{\Delta E}{3\cosh \ma{y}_J}m_T\cosh(\ma{y}-\ma{y}_J) \bigg]\,
\ee
where $\ma{y}_J$ and $\phi_J$ are the momentum rapidity and transverse azimuthal angle of the ``jet'', e.g. the high energy parton. 

In this work we will test how particle production described by the oversimplified hybrid model expression \eqref{eqn:hb} compares to particle production described by the full linearized hydrodynamics result, obtained by integrating Eq.~(\ref{eqn:spectrum}).



\subsection{Effects of Transverse Flow}
\label{sect:TransverseFlow}

As we have stressed, 
our framework of linearized hydrodynamics is based on a Bjorken flow background, which has no flow in the transverse plane. While this assumption has allowed us to have analytical control over the background flow, in the hydrodynamic description of ultra-relativistic heavy ion collisions transverse flow is an important ingredient for capturing the momentum distribution of particles. In most central collisions at the LHC, where the radial flow velocity can be as large as $\beta=0.7$, this effect makes the spectrum of soft particles significantly harder than our background spectrum. It is therefore important to estimate the effect of the transverse flow on jet observables. (For complementary studies see Refs.~\cite{Yan:2017rku,Tachibana:2020mtb}.)

Transverse flow alters the medium backreaction in at least two important ways. The first effect is the alteration of the perturbed hydrodynamic fields: as the background fluid expands, the perturbation induced by the jet moves with the flow, leading to a distortion of the field profile. In our calculation to this point, we only have longitudinal expansion; incorporating transverse flow in our setup from the beginning would modify the ``shape'' of the wake that we have calculated in Section III. Investigating this would require repeating our linear hydrodynamic analysis for perturbations to a realistic flow profile, incorporating transverse flow and deviations from boost invariance in the longitudinal direction in the unperturbed solution to hydrodynamics. This problem is a well-posed application of the linear hydrodynamic formalism, but the reduction in the symmetry of the hydrodynamic flow of the background fluid would make the numerical calculation more challenging. As we are not doing a phenomenological analysis anyway (e.g. as we are considering only a single energetic parton rather than an ensemble of jet showers) we leave repeating our calculation with a more realistic background flow profile to future work. We note also that including the effects of transverse flow on the ``shape'' of the wake need not modify the magnitude of the perturbations to the hydrodynamic background from the wake in a significant way. 

The second effect of transverse flow is more important. 
It occurs at hadronization and is due to the transverse boost of fluid cells at freezeout. In Section~\ref{sect:CooperFrye}, we used the Cooper-Frye prescription to compute the distribution of particles coming from the freezeout of each unit cell of the fluid on the freezeout hypersurface --- using the equilibrium distribution in the local fluid rest frame defined by the {\it longitudinal} velocity of each fluid cell on the freezeout hypersurface.
If we modify the velocity of a fluid cell so as to take into account its
transverse, e.g. radial, flow this additional component of the fluid velocity means that the distribution of particles produced at freezeout must be computed in a new boosted frame. And, this modification affects the production of {\it all} particles: those coming from the background hydrodynamic fluid and those coming from the wake perturbation. Since the particle
distribution depends exponentially on this background flow velocity, see Eqs.~\eqref{eq:cfgen} and \eqref{eqn:spectrum}, this effect on the particle spectrum can be very significant and is certainly much more important than the first effect above. In this Section, we will estimate the modification of the spectrum and the azimuthal and momentum rapidity distributions of the hadrons originating at freezeout due to this second effect.

According to the previous discussion, in our estimate of the effect of transverse flow we will assume that the flow profiles that we have computed in the previous Section remain unchanged, as does the freezeout hypersurface. However, we will imagine that these profiles, which is to say the wake that we have computed, freezes out from a hypersurface at which the radial, transverse flow
is significant, meaning that when the particle production needs to be calculated in the local rest frame the boost to this frame now includes a transverse boost.
When the fluid breaks into particles, the momentum of a particle produced at each cell in the local rest frame, $p^\mu_{\ma{cell}}$, whose distribution is determined via the integrand of equation~(\ref{eq:cfgen}), differs from the momentum observed in the laboratory frame, $p^\nu_{\ma{lab}}$ by a local transverse boost
\be
\label{eqn:plab}
p^\mu_{\ma{cell}} = \Lambda^\mu_{\ \nu}({\bs {v}}_{\ma{cell}}) p^\nu_{\ma{lab}}\,,
\ee
where $\Lambda^\mu_{\ \nu}({\bs v}_{\ma{cell}})$ is a Lorentz transformation with ${\bs v}_{\ma{cell}}={\bs v}_{\ma{cell}}(\tau,x,y,\eta_s)$ which is the transverse velocity of the fluid cell seen in the laboratory frame. The longitudinal boost has already been accounted for when we write $u_0=(1,0,0,0)$ in the $(\tau,x,y,\eta_s)$ coordinate system, so we only need to include the transverse boost in Eq.~(\ref{eqn:plab}).
This modification is incorporated into the distribution of particles produced from a given fluid cell at freezeout by replacing the $u_0^\mu p_\mu$ and $u^\mu p_\mu$ terms in the distribution \eqref{eqn:spectrum}, respectively, by their boosted counterparts:
\be
\label{eqn:u0_dot_p2}
u_0^\mu p_\mu &\to& u_0^\mu \Lambda_{\mu\nu}({\bs {v}}_{\ma{cell}}) p^\nu\\
\label{eqn:u_dot_p2}
u^\mu p_\mu &\to& u^\mu \Lambda_{\mu\nu}({\bs {v}}_{\ma{cell}}) p^\nu \,.
\ee
We then need to obtain ${\bs v}_{\ma{cell}}$ for each fluid cell on the freezeout surface from a longitudinally boost invariant $2+1$D hydrodynamic calculation.

\begin{figure}
    \begin{subfigure}[t]{0.49\textwidth}
        \centering
        \includegraphics[height=2.3in]{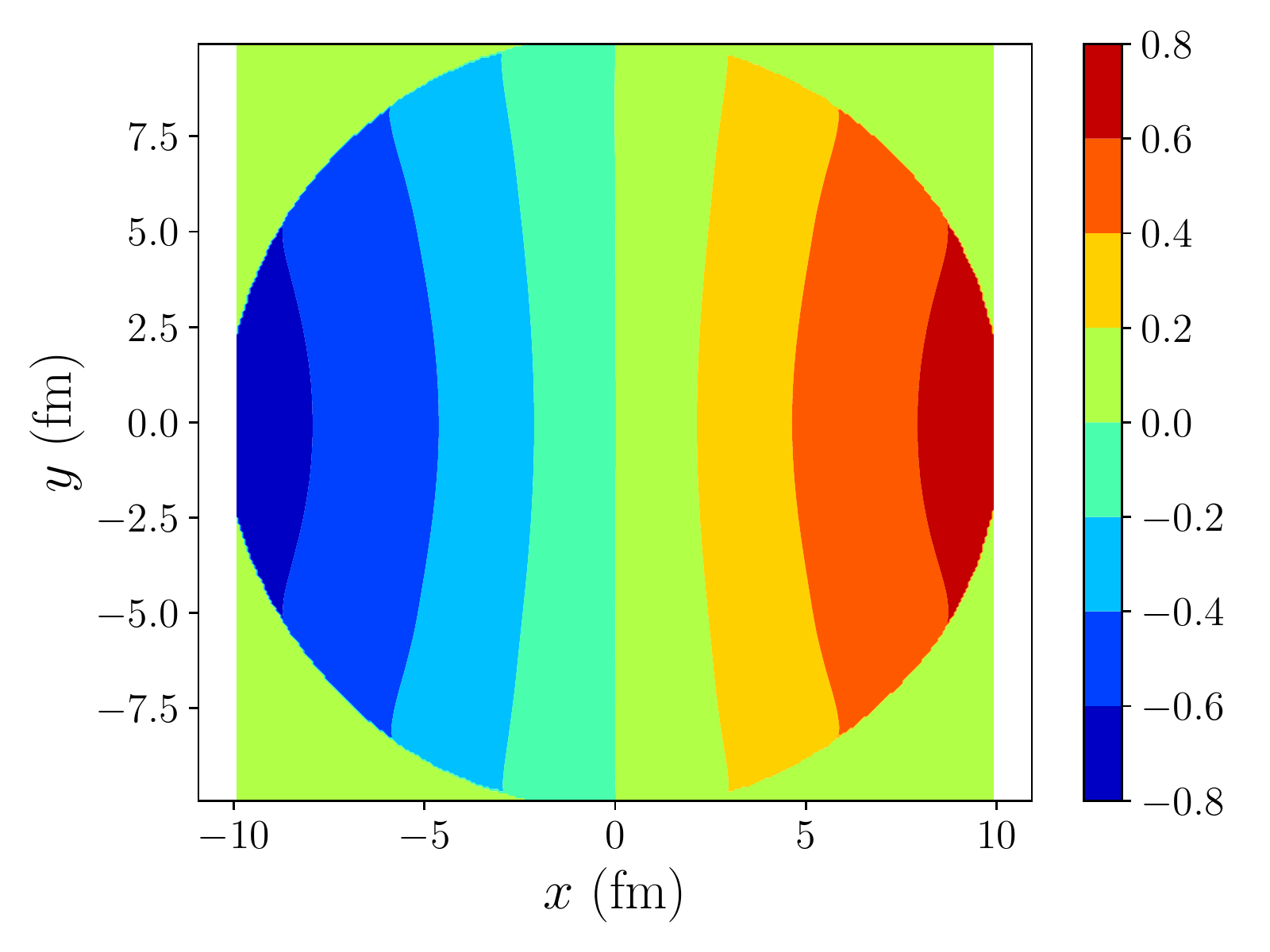}
        \caption{$v_{\ma{cell},x}$ profile.}
    \end{subfigure}%
    ~
    \begin{subfigure}[t]{0.49\textwidth}
        \centering
        \includegraphics[height=2.3in]{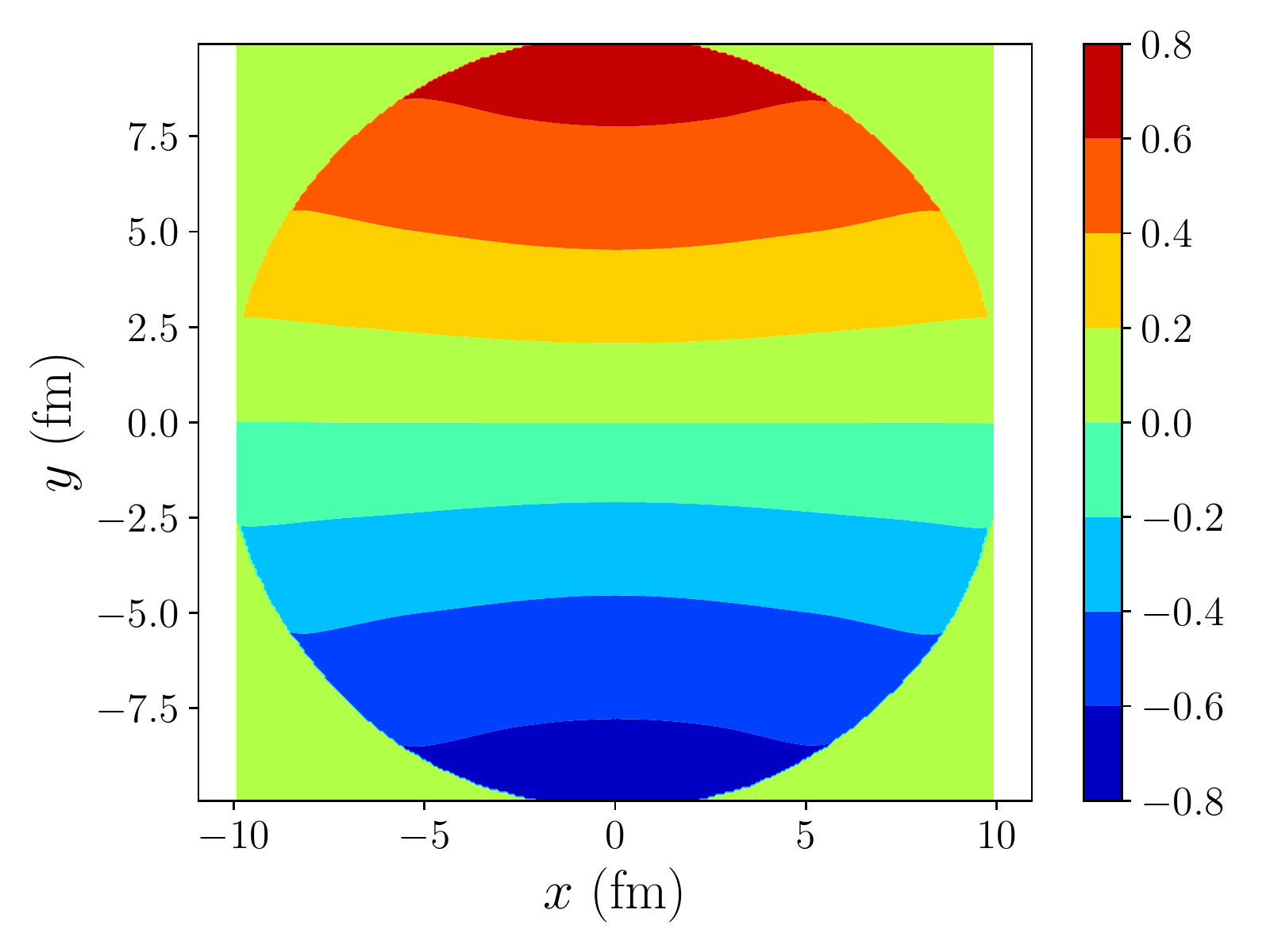}
        \caption{$v_{\ma{cell},y}$ profile.}
    \end{subfigure}%
\caption{Transverse flow velocities ${\bs v}_{\ma{cell}}$ at freezeout, $T_f=154$ MeV, extracted from the calibrated $2+1$D viscous hydrodynamics calculation of Ref.~\cite{Bernhard:2016tnd}. The transverse flow velocities are set to zero outside the freezeout hypersurface, which is of finite size in the transverse plane. For a central collision, the transverse extent of the freezeout hypersurface is a circle, outside of which we set ${\bs v}_{\ma{cell}}=0$, light green in the Figure.}
\label{fig:flow}
\end{figure}

For our estimates, we extract the spacetime dependence of the transverse flow from a hydrodynamical simulation that incorporates radial transverse flow obtained
using the ``VISHNU" package~\cite{Song:2007ux,Shen:2014vra}, which solves longitudinally boost invariant $2+1$D viscous hydrodynamics for azimuthally symmetric, central, heavy ion collisions. We shall use results from the calculations calibrated with experimental measurements on low transverse momentum observables found in Ref.~\cite{Bernhard:2016tnd}. The extracted transverse velocity at mid-rapidity 
${\bs v}_{\ma{cell}} = (v_x,v_y,0)$ at the freezeout hypersurface determined by $T_f=154$ MeV is depicted in Fig.~\ref{fig:flow} for central collisions (centrality $0$-$5\%$). When the transverse flow is taken into account, the freezeout hypersurface is finite in extent in the transverse plane. Outside its boundary, we simply set the transverse flow velocity to be zero. 
To make sure that this oversimplification has little influence on our results, we must ensure that the wake in the hydrodynamic fluid is located, at the time of freezeout, almost entirely within the colored circles in Fig.~\ref{fig:flow}. We choose the centers of these circles, e.g. the point of origin of the radial flow at the center of the collision, to lie at the center of our calculational box. And, the choices that we made in Section~\ref{sect:numerics} for the point of origin of the high energy parton and the direction of its motion, relative to our calculational box, were made with malice aforethought: they ensure that the wake that we have calculated lies within the colored circles in Fig.~\ref{fig:flow}. (See the bottom-right panels in Figs.~\ref{fig:de_over_e} and~\ref{fig:du_x} and the right panels of Fig.~\ref{fig:du_yeta}.) 

\subsection{Results, Including the Effects of Radial Flow}

\begin{figure}[t]
        \centering
        \includegraphics[height=3.5in]{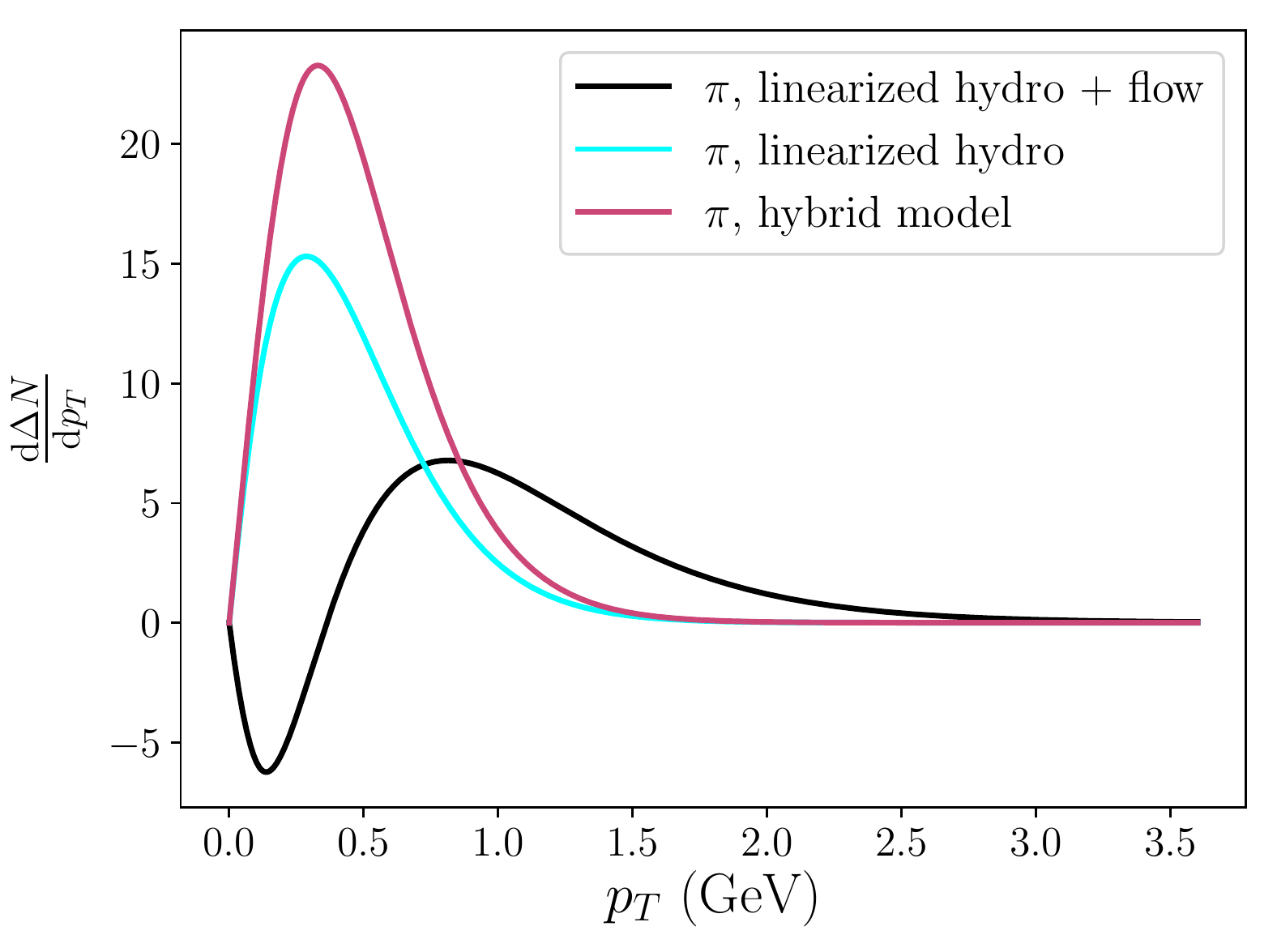}
\caption{$p_T$ spectrum of pions produced from the jet wake. Results from our linearized hydrodynamics calculations, with and without the effects of transverse flow effects added, are compared with the spectrum of the wake implemented in the hybrid model in Ref.~\cite{Casalderrey-Solana:2016jvj}.}
\label{fig:spectrum}
\end{figure}

\begin{figure}
    \begin{subfigure}[t]{0.49\textwidth}
        \centering
        \includegraphics[height=2.4in]{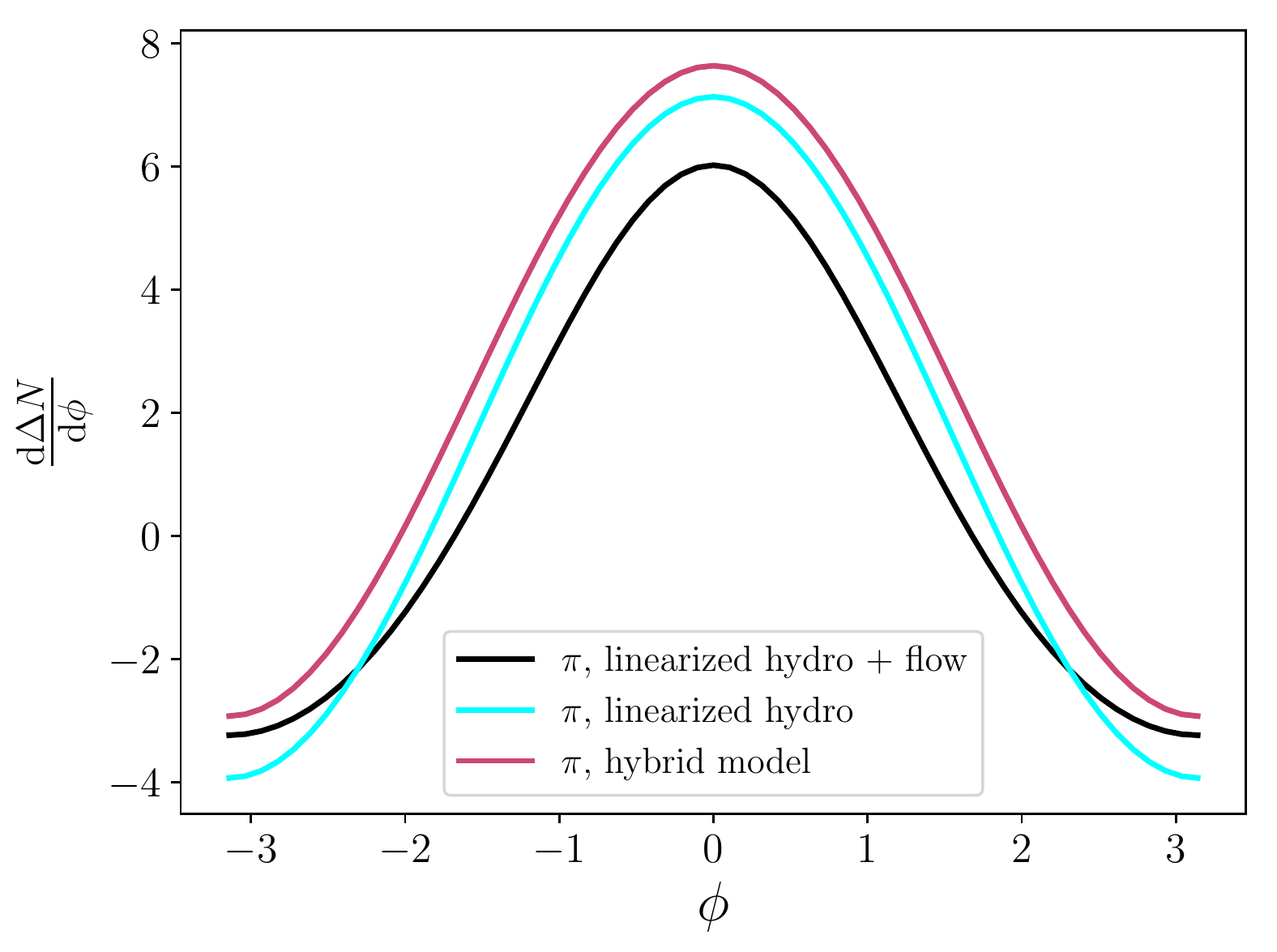}
        \caption{$\phi$ distribution.}
        \label{fig:dist_phi}
    \end{subfigure}%
    ~
    \begin{subfigure}[t]{0.49\textwidth}
        \centering
        \includegraphics[height=2.4in]{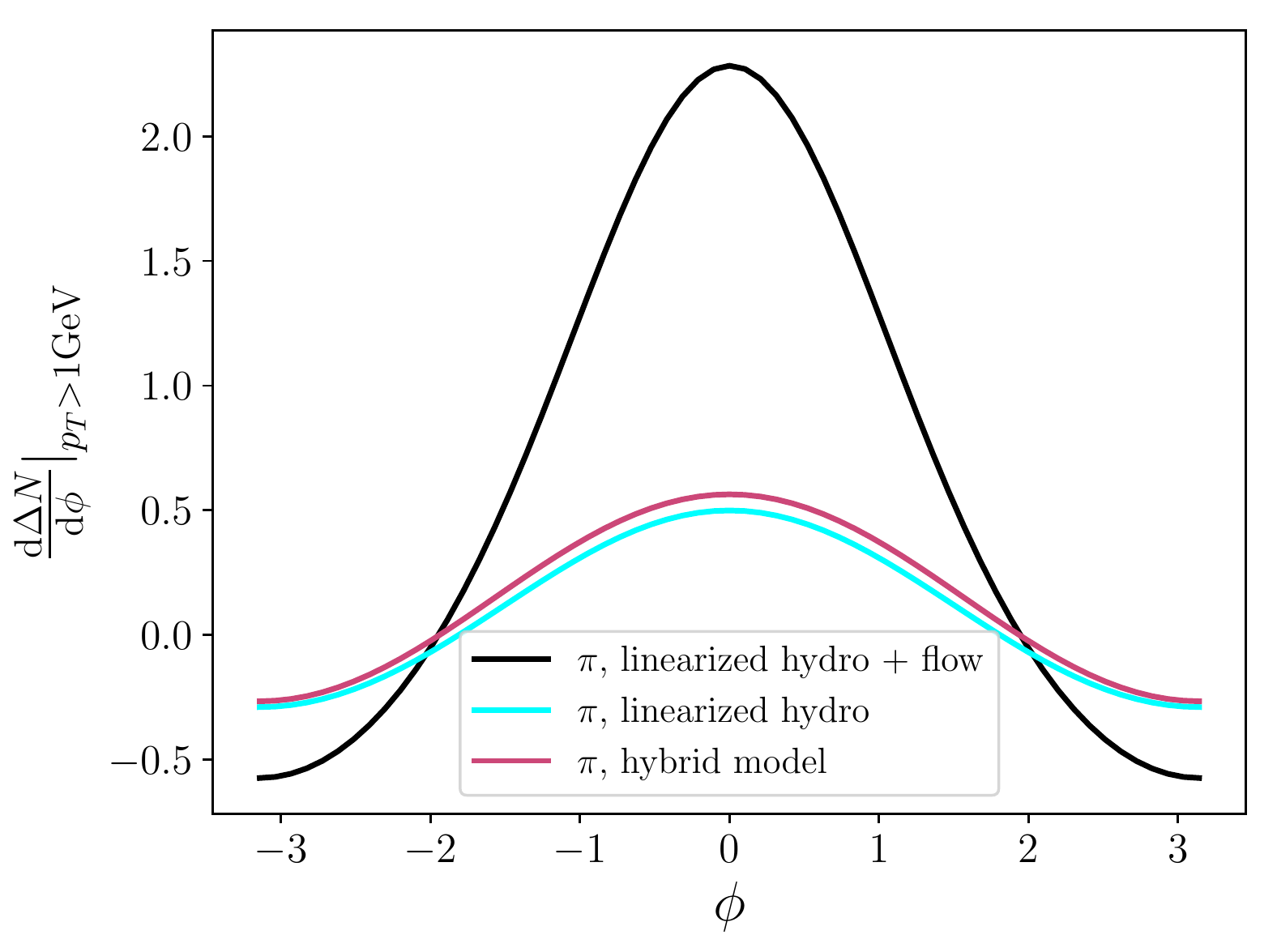}
        \caption{$\phi$ distribution for $p_T>1$ GeV.}
        \label{fig:dist_phi_pT>1}
    \end{subfigure}%
    
    \begin{subfigure}[t]{0.49\textwidth}
        \centering
        \includegraphics[height=2.4in]{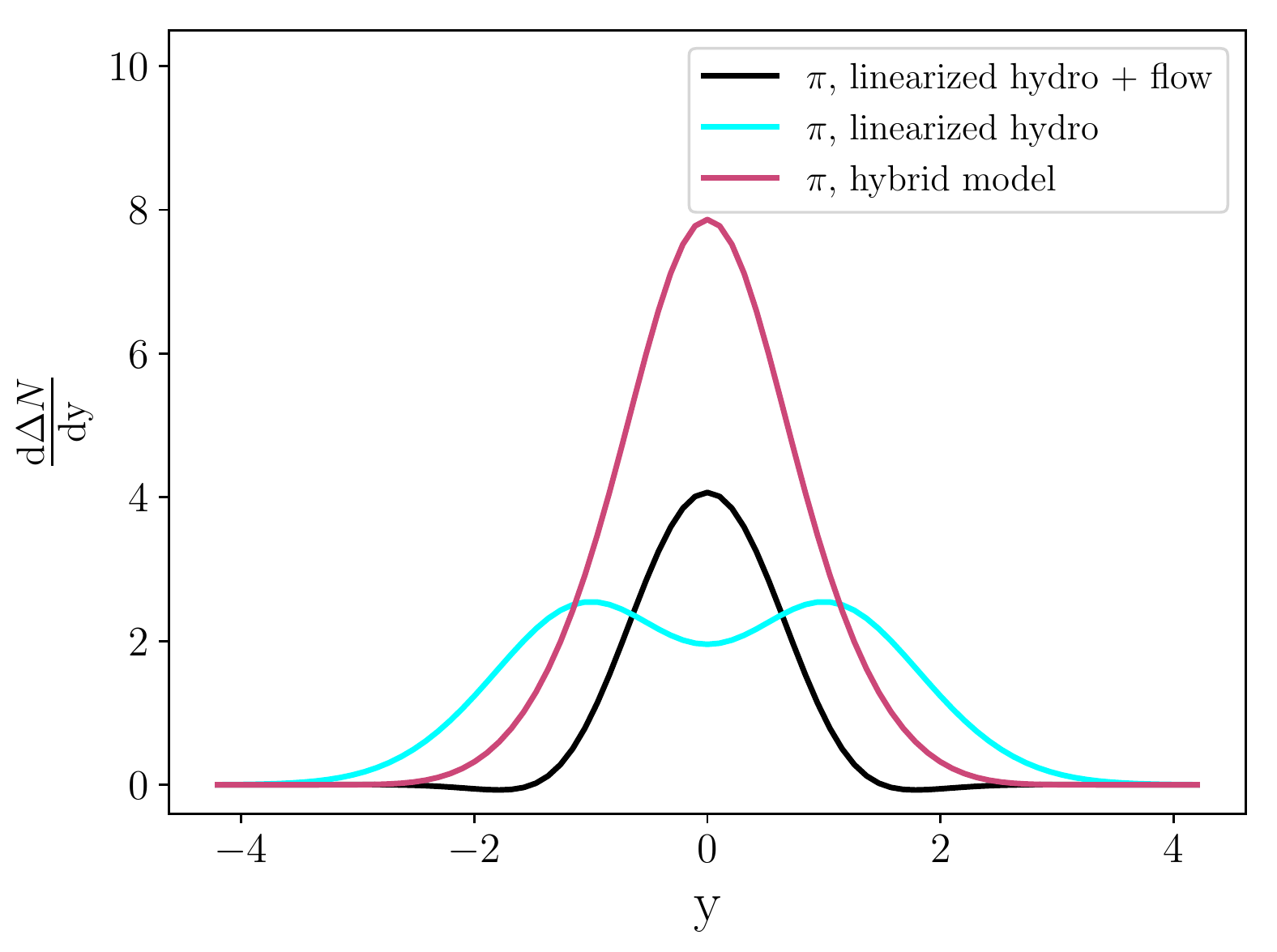}
        \caption{$\ma{y}$ distribution.}
          \label{fig:dist_y}
    \end{subfigure}%
    ~
    \begin{subfigure}[t]{0.49\textwidth}
        \centering
        \includegraphics[height=2.4in]{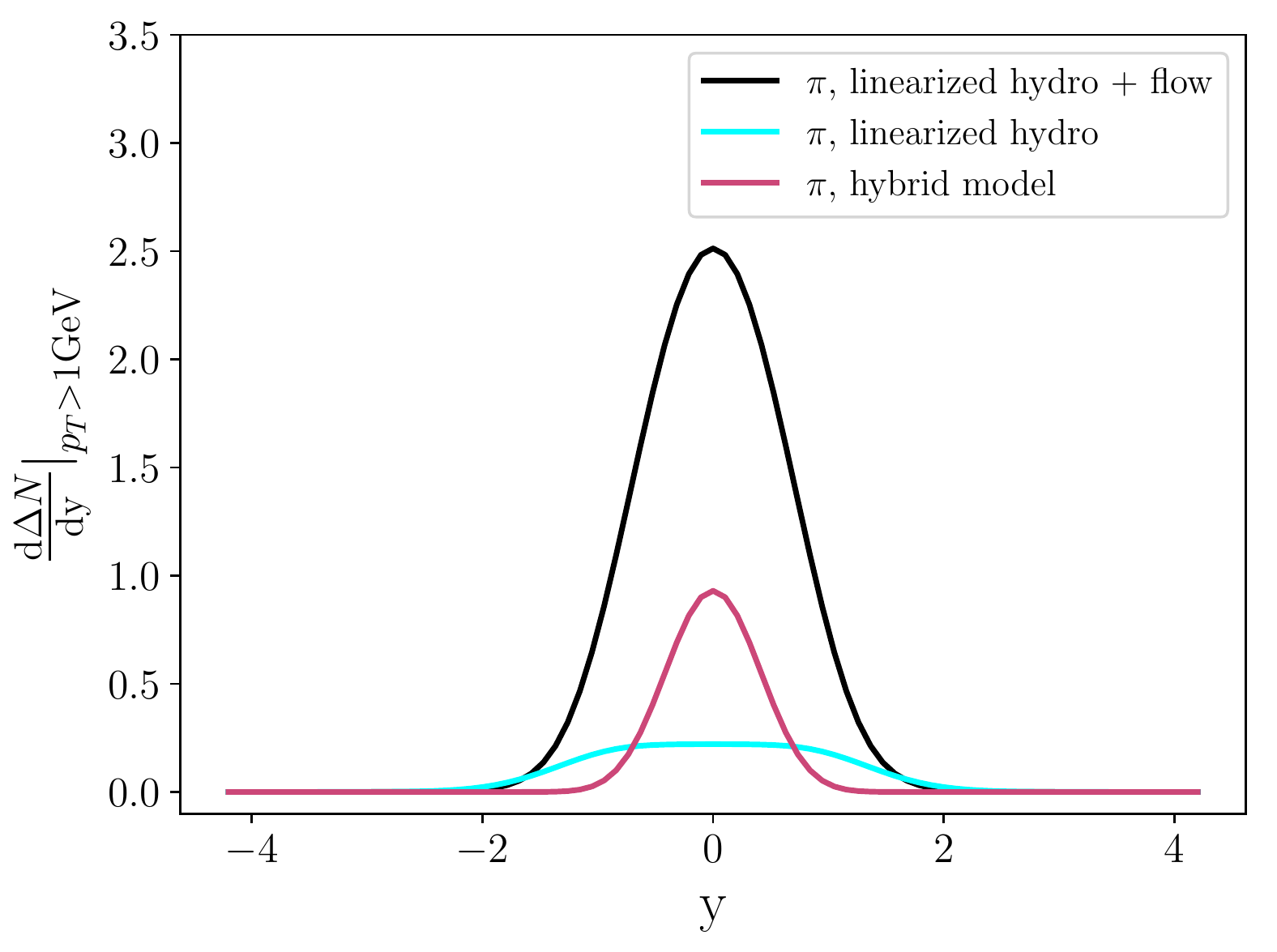}
        \caption{$\ma{y}$ distribution for $p_T>1$ GeV.}
        \label{fig:dist_y_pT>1}
    \end{subfigure}%
\caption{Distributions in azimuthal angle (upper panels) and momentum rapidity (lower panels) for the pions produced from the wake excited in the hydrodynamic fluid by a high energy parton. In each panel, results from the linearized hydrodynamics with and without the effects of transverse flow added are compared with each other and with the distributions obtained in the simplified treatment of the hybrid model. In the left panels, we include all pions whereas in the right panels we include only those pions that have $p_T>1$~GeV.}
\label{fig:distribution}
\end{figure}

We are now ready to look at the results of our calculations of the distribution of particles coming from the freezeout of the perturbed hydrodynamic fluid. We only consider what we referred to as Case 2 in Section~\ref{sect:numerics}, which is to say the case where the perturbation is the wake that a high energy parton leaves in a fluid with a small but nonzero viscosity. For the present, we assume that all the particles produced at freezeout are pions with  mass $m_\pi=138$ MeV. We begin by looking at the spectrum of pions produced from the wake, shown in 
Fig.~\ref{fig:spectrum}. Then, in Fig.~\ref{fig:distribution} we look at the distribution of pions in azimuthal angle and in momentum rapidity. In both these Figures, we compare three calculations: 
the oversimplified calculation from the hybrid model of Ref.~\cite{Casalderrey-Solana:2016jvj};
our linearized hydrodynamics calculation with no transverse expansion --- see Eqs.~(\ref{eqn:u0_dot_p}, \ref{eqn:u_dot_p}, \ref{eqn:T_deltaT}, \ref{eqn:spectrum}) in Section~\ref{sect:CooperFrye}; our linearized hydrodynamics with the effects of radial transverse flow added --- see Eqs.~(\ref{eqn:T_deltaT}, \ref{eqn:spectrum}, \ref{eqn:u0_dot_p2}, \ref{eqn:u_dot_p2}) in Section~\ref{sect:TransverseFlow}. For the linearized hydrodynamic calculations with and without the effects of transverse flow, we normalize the distributions calculated from the Cooper-Frye formula by requiring the total energy of all the particles (pions) produced to be equal to the total energy lost by the high energy parton
\be
\label{eqn:normalize}
\Delta E_{\ma{tot}} = \int \diff p_T \diff\phi \diff\ma{y}\, m_T\cosh\ma{y} \frac{\diff\Delta N}{\diff p_T \diff\phi \diff\ma{y}} \,.
\ee

First, we compare the $p_T$ spectra shown in Fig.~\ref{fig:spectrum}. The linearized hydrodynamic calculation with no effects of transverse flow yields a similar $p_T$ spectrum as that obtained in the simplified hybrid model calculation, although the normalization is different 
because the momentum rapidity distribution in our linearized hydrodynamic calculation is significantly broader (see below). The similar $p_T$ shape is expected since the particle spectrum in both cases must be close to a thermal spectrum at $T_f=154$ MeV when the perturbation is small, as is the case  when the fluid has a nonzero viscosity which allows for the perturbation to spread diffusively, as we have seen in Section~\ref{sect:u}, 
 We also see in Fig.~\ref{fig:spectrum} that the inclusion of radial transverse flow leads to a blue-shift in the effective temperature, which makes the $p_T$ spectrum harder, as compared with the other two cases.
 In addition, 
 we notice that the modifications introduced by radial flow make the spectrum of the wake negative at low $p_T$. What we are computing --- see \eqref{eqn:spectrum} --- is the difference between two spectra: one for pions produced at freezeout from the fluid perturbed by the wake, and the other for pions produced at freezeout from the unperturbed fluid. 
 In kinematic regions where this quantity is negative, there are fewer pions after we include the perturbation induced by the wake left behind by the high energy parton. What we see from Fig.~\ref{fig:spectrum} is that this does not arise without radial flow, but the boost created by radial flow hardens the spectrum of the wake so much that in addition to pushing it up significantly at $p_T\gtrsim 1$~GeV it makes it negative at the lowest $p_T$.

Next we analyze the distribution in the azimuthal angle, shown in the upper panels of Fig.~\ref{fig:distribution}. The $\phi$ distribution in the hybrid model (\ref{eqn:hb}) is approximately sinusoidal, by which we mean approximately of the form $a+b\cos\phi$. Similarly sinusoidal behavior is also seen in the linearized hydrodynamic calculation without effects of transverse flow. As we have shown, the perturbation in the fluid velocity is small in the viscous linearized hydrodynamics. So expanding the exponential in the Cooper-Frye prescription to linear order in the velocity perturbation works reasonably well. Then the $\phi$-distribution in the linearized hydrodynamic calculation without the flow effect added can also be approximated in the form of $a+b\cos\phi$.
The shape of the inclusive azimuthal distribution remains also similar when transverse flow effects are included. 
However, since the $p_T$-distribution is severely modified, it is interesting to compare the different distributions in azimuthal angle for semi-hard, $p_T>1$ GeV, particles as this is the kinematic regime in which radial flow acts to push the spectrum up (see Fig.~\ref{fig:spectrum}). Indeed, in this region of $p_T$ the effect of radial flow becomes apparent also in the azimuthal distribution. As can be observed from Fig.~\ref{fig:dist_phi_pT>1}, when flow effects are introduced the production of semi-hard particles becomes more strongly correlated with the direction of the high energy parton. Relative to the cases without flow, the yield close to the jet increases and number of particles produced at large angles from the jet decreases. 
This stronger correlation with the jet direction may be due to the fact that in the configuration that we have studied the background radial flow is aligned with the direction of propagation of the jet at the time of freezeout (regardless of the fact that at the initial time $\tau_0$ they pointed in opposite directions). The alignment of the direction of the momentum of the wake (which is the direction of the high energy parton) and the direction of the background radial flow at freezeout enhances the production of particles along the source direction. (To see this, consider Eqs.~\eqref{eqn:u0_dot_p2} and \eqref{eqn:u_dot_p2}, substituted into \eqref{eqn:spectrum}.)  In a future realistic simulation, however, jets would be produced at all starting points and in all directions, and although the parton direction and the flow direction will tend to be aligned by the time of freezeout there will certainly be jets in the ensemble for which this is not the case. Therefore, when all such configurations are averaged over in a future more realistic Monte Carlo study, the magnitude of the observed correlation (e.g. the degree to which the azimuthal distribution becomes more peaked due to the effects of radial flow) is likely to be somewhat reduced.

Next, we discuss our results for the distribution of particles from the wake in momentum rapidity, depicted in the second row of Fig.~\ref{fig:distribution}. For this distribution, the linearized hydrodynamic calculation with no effects of radial flow and the hybrid model approximation differ significantly. 
The linearized hydrodynamic calculation, without transverse flow, leads to a much broader distribution in $y$ than the hybrid model. This originates from the broad distribution of the energy and momentum of the wake in spacetime rapidity $\eta_s$ in our linearized hydrodynamic calculation. Once the energy and momentum are deposited, they propagate along the $\pm \eta_s$ directions due to the excitation of sound modes (see Fig.~\ref{fig:f_eta}). This is in contrast to the hybrid model assumption, that considers the energy and momentum perturbation is located at a fixed $\eta_s$ throughout.
This difference in rapidity distribution is also behind the difference in yield between these two sets of calculations. Since we normalize the spectrum by conserving the total energy (\ref{eqn:normalize}), which involves the integration of $m_T\cosh\ma{y}$, the height of the $p_T$ spectrum from the linearized hydrodynamic calculation without any effects of flow  is smaller than that from the hybrid model, see Fig.~\ref{fig:spectrum}, because of the broader momentum rapidity distribution in the former, see Fig.~\ref{fig:distribution}. However, while the distribution obtained from the linearized analysis is rather wide in the absence of flow, incorporating the effects of radial flow reduces its rapidity extent, as particle production from the wake becomes more strongly correlated with the direction of motion of the high energy parton.  
Finally, we note by comparing panels (b) and (d) of Fig.~\ref{fig:distribution} that introducing radial flow substantially increases the number of particles produced with $p_T>1$~GeV, confirming what we already saw in Fig.~\ref{fig:spectrum}.

\begin{figure}[t]
        \centering
        \includegraphics[height=3.5in]{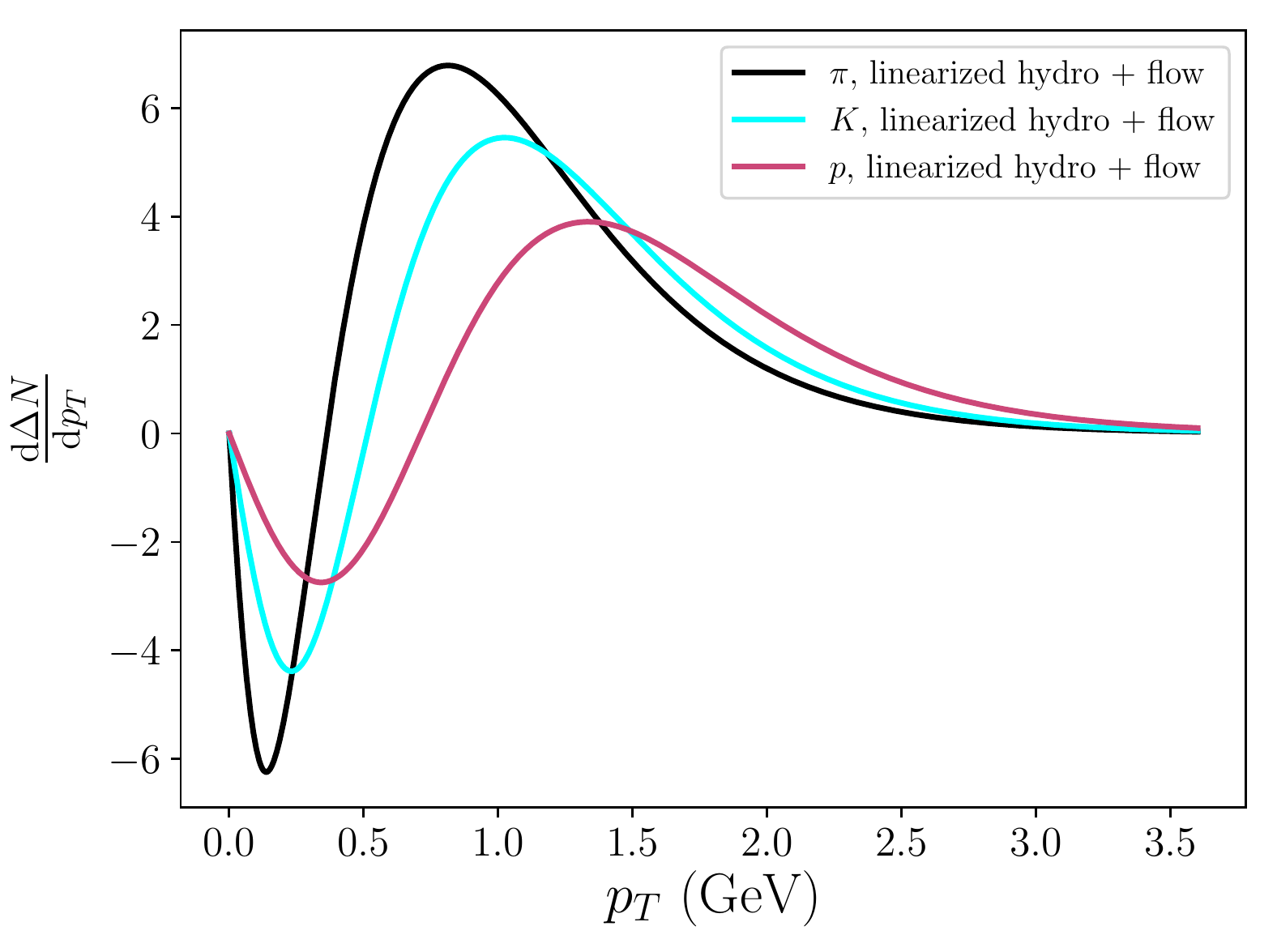}
\caption{Spectra of pions, kaons and protons produced at freezeout from the wake, with the effects of radial flow added. The curves are each independently normalized.}
\label{fig:mass}
\end{figure}

To conclude this Section, we explore a natural consequence of transverse flow. In Fig.~\ref{fig:mass}, we compare the spectra of pions, kaons ($m=495$ MeV) and protons ($m=938$ MeV) from our linearized hydrodynamic calculations with the effects of radial flow added. Each curve is independently normalized, i.e.~the integral of the distribution multiplied by the energy over the whole phase space gives the total energy lost by the jet, as if all of the energy in the wake went into a single particle species. We have plotted the curves this way to make it easy to compare their shapes. It is well known that a characteristic property of radial flow is that it makes the $p_T$ spectrum of hadrons increasingly harder as the mass of the produced hadron increases. This is a standard tell for hydrodynamic behavior. We see in Fig.~\ref{fig:mass} that the same phenomenon occurs for the particles coming from the wake.  
For hadrons with a higher mass, the location of the positive peak in the $p_T$-spectrum is pushed to a higher value, which is expected since the momentum gained by the particle from the transverse flow grows with the particle mass. 
This is not a surprise since we are treating the wake via linearized hydrodynamics, but it is good to see.
This may lead to interesting observables in the fragmentation function. For example, the ratio of the pion fragmentation functions in $pp$ and AA collisions may have a different shape from that of the proton fragmentation functions, since more protons at $2$ GeV are produced than pions at $2$ GeV from the jet wake.

\section{Towards Jet Observables}
\label{sect:observable}

As has been already demonstrated by several theoretical analyses~\cite{He:2015pra,Casalderrey-Solana:2016jvj,Tachibana:2017syd,KunnawalkamElayavalli:2017hxo,Chen:2017zte,Milhano:2017nzm,He:2018xjv,Park:2018acg,Tachibana:2018yae,Chang:2019sae,Casalderrey-Solana:2019ubu,Pablos:2019ngg,Chen:2020tbl,Brewer:2020chg,Pablos:2020wnp}, some jet observables can be significantly modified by medium response effects. 
By momentum conservation, the wake that a jet leaves behind in the droplet of QGP through which it has passed must carry momentum in the jet direction, since it carries the momentum lost by the jet.  That means that after hadronization the particles originating from the wake also must have net momentum in the jet direction.  And in turn this means that when experimentalists define what they observe as a jet -- via a reconstruction algorithm and a background subtraction procedure -- the jets that they reconstruct must include some particles coming from the wake since background subtraction procedures can at best only subtract effects of fluctuations that are uncorrelated with the jet direction. These effects~\cite{Floerchinger:2013rya,Floerchinger:2013hza} are likely comparable to or larger in magnitude than the effects originating from the wake of the jet, highlighting the importance of background subtraction in any experimental analysis. In an experimentally reconstructed jet, the harder particles likely originate directly from the jet shower whereas some of the soft particles originate from the wake in the plasma. So, jet observables that are sensitive to the softer particles within jets are likely to be significantly influenced by the wake in the plasma. 

The contribution of particles in jets that originate from the wake in the plasma to jet observables has been particularly well studied  in the hybrid strong/weak coupling model. In this model, the inclusion of these particles coming from the backreaction of the medium is essential to the description of many basic jet observables, like jet suppression and the medium-modification of  fragmentation functions~\cite{Casalderrey-Solana:2016jvj}. The calculations of the previous Section, in particular those illustrated in Figs.~\ref{fig:spectrum} and \ref{fig:distribution} (b), show that with our better description of medium response built upon linearized hydrodynamics and incorporating the effects of radial flow, when the wake of a jet hadronizes it yields more semi-hard particles ($1~{\rm GeV}\lesssim p_T \lesssim 3~{\rm GeV}$) that are more correlated along the jet direction than in the simplified implementation of the wake used in the hybrid model. 
At a qualitative level, both these features are exactly what is indicated by the discrepancies between the hybrid model calculations of so-called missing-$p_T$ observables in Ref.~\cite{Casalderrey-Solana:2016jvj} and the experimental measurements of these observables in heavy ion collisions at the LHC by CMS~\cite{Khachatryan:2015lha}. This provides strong motivation for a future quantitative study of the consequences
of this improved description of jet wakes for jet measurements at the LHC. 
To make quantitative comparison with experiments, however, this improved description of jet wakes will need to be incorporated into a Monte Carlo study of a realistic ensemble of jet showers with a distribution of points of origin, direction, and shower structure, each propagating through a realistic model for the droplet of QGP, for example as in the hybrid model.  We leave this to future work.

Nevertheless, to gauge the influence that this improved description could have on actual jet observables, in this work we will study two examples which share features with some of those jet observables. In each case, we will compute only the contribution to the observable that comes from the hadronization of the jet wake sourced by our single high energy parton, as in Section IV. In particular, we will compute the energy carried by particles coming from the backreaction of the medium that lie within a cone of a specified opening angle centered on the direction of the high energy parton; and the energy distribution of those particles. For simplicity, as in Section IV we will only consider a setup in which the high energy parton that loses energy propagates at zero rapidity; we shall measure azimuthal angles with respect to its transverse direction.


\subsection{Energy Recovered inside a Cone}

\begin{figure}
\centering
\includegraphics[height=3.5in]{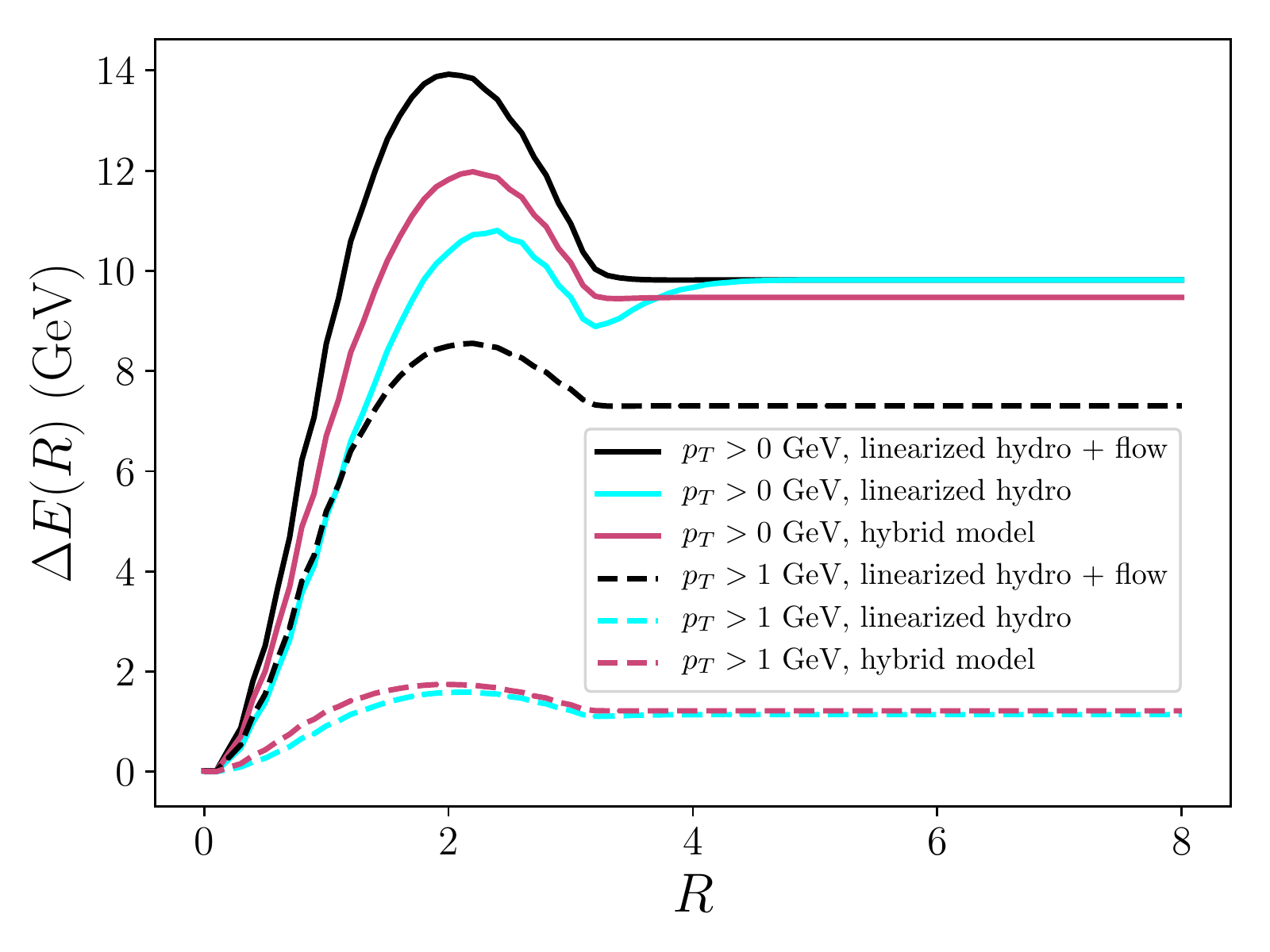}
\caption{``Lost'' energy from the high energy parton that, after propagation as a wake in the QGP and subsequent hadronization, is recovered within a cone with opening angle $R$ around the direction of the high energy parton. Results are shown for the oversimplified hybrid model treatment of the wake and for our linearized hydrodynamics analysis with and without the effects of radial flow, in each case for two different $p_T$ cuts.}
\label{fig:Econe}
\end{figure}

Our first example is the amount of energy lost by the high energy parton carried by particles originating from the wake in the plasma
recovered 
within a cone of opening angle $R=\sqrt{\phi^2+\ma{y}^2}$, defined by
\be
\label{eq:erecov}
\Delta E(R)\equiv \int_{\sqrt{\phi^2+\ma{y}^2}<R}\diff\phi\diff\ma{y}\int \diff p_T \, \sqrt{p_T^2+m^2} \cosh\ma{y}\, \frac{\diff\Delta N}{\diff p_T \diff\phi \diff\ma{y}}\,.
\ee
In any realistic simulation, any jet reconstruction algorithm will collect particles around the jet direction and incorporate them into the definition of the jet. If backreaction occurs, particles from the medium response will also be added to the jet. As a consequence, even if there is a transfer of energy from hard jet fragments to soft medium particles, those medium response particles will put energy back into the reconstructed jet. Therefore, Eq.~(\ref{eq:erecov}) gives us a way to gauge how much of the ``lost'' energy is ``recovered'', namely included as a part of the reconstructed jet coming from the backreaction of the medium.  
Our results for this quantity as a function of $R$ are shown in Fig.~\ref{fig:Econe}. The plots extend out to large enough values of $R$~\footnote{
The angular separation between two particles is given by
$R^2=2(\cosh{\ma{y}}-\cos\phi)$, which coincides with the standard approximation $R=\sqrt{\phi^2+\ma{y}^2}$ for $R\ll1$. For large $R>1$, while only the former gives the correct angular separation, the latter can still be used to define a distance in the $(\phi,\ma{y})$ plane. This definition, which is conventional, has a simpler representation in the $(\phi,\ma{y})$ plane.} that the total amount of energy lost by the high energy parton is fully recovered.

We note from the solid magenta curve in Fig.~\ref{fig:Econe} 
that while the hydrodynamic calculations recover the full deposited energy, a very small fraction of the energy is missed in the hybrid model calculation, even at very large angles. This is just a consequence of the approximations used to derive the simplified expression, Eq.~(\ref{eqn:spectrum_hb}), which assumes the production of massless final particles, but is used to study the productions of pions. 
%
%

To gauge the importance of semi-hard medium response particles in $\Delta E(R)$, in Fig.~\ref{fig:Econe} we also show results when we restrict the transverse momentum of the recovered particles to $p_T>1$ GeV. 
Doing so highlights an important comparison, namely the very different momentum composition of the medium response particles within a given angular region. For both the calculations with no transverse flow, energy is mostly recovered in soft particles, and the contribution from semi-hard particles, $p_T>1$, to $\Delta E(R)$ in Fig.~\ref{fig:Econe} is very small. However, when the effects of flow are added this distribution completely changes and a large fraction of the energy within the cone is recovered in the form of semi-hard particles, which is consistent with what we saw in the spectra of Fig.~\ref{fig:spectrum}. We will further investigate this in the next Subsection.


\begin{figure}
    \begin{subfigure}[t]{0.49\textwidth}
        \centering
        \includegraphics[height=2.5in]{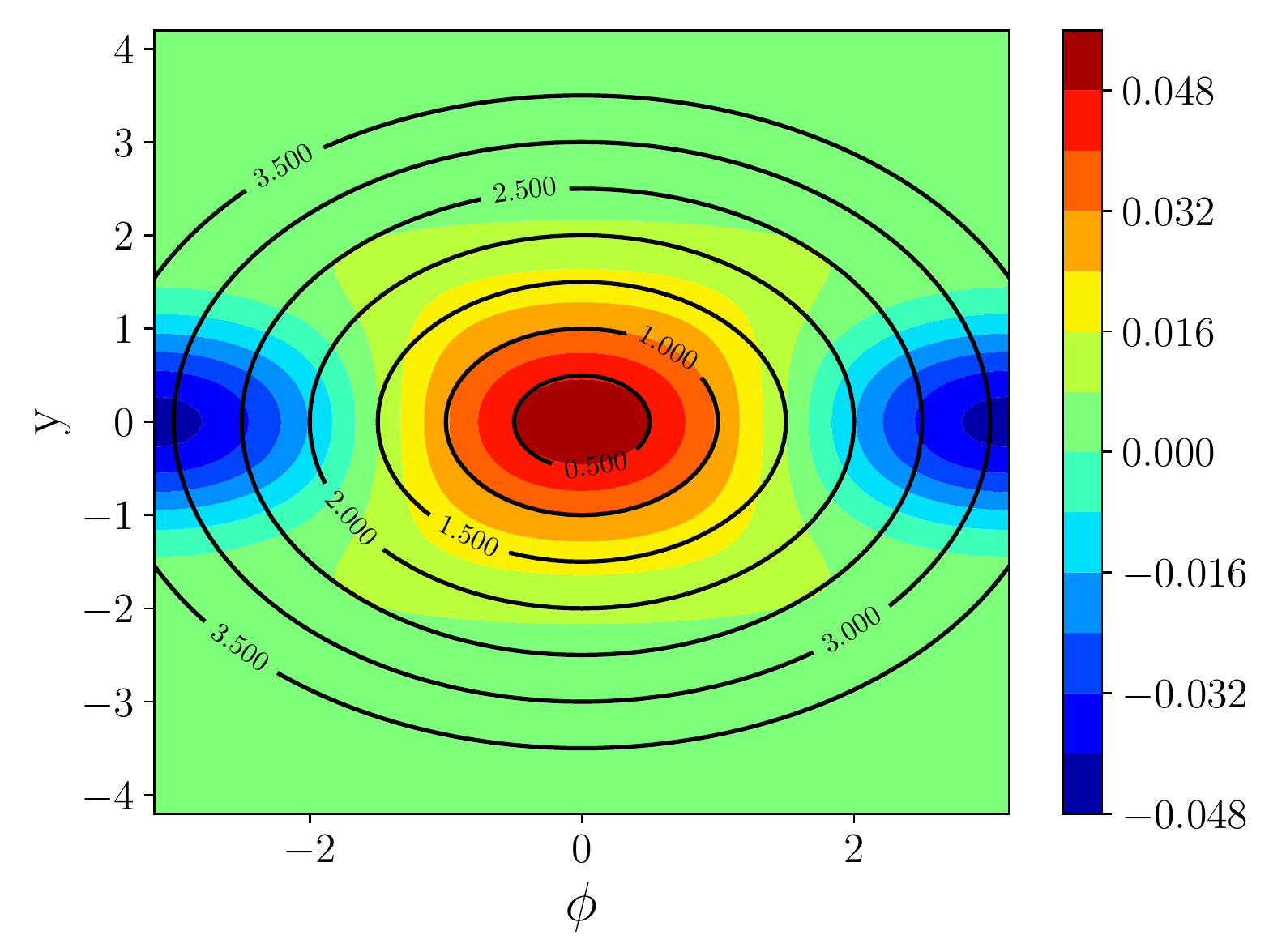}
        \caption{Linearized hydrodynamics without transverse flow.}
    \end{subfigure}%
    ~
    \begin{subfigure}[t]{0.49\textwidth}
        \centering
        \includegraphics[height=2.5in]{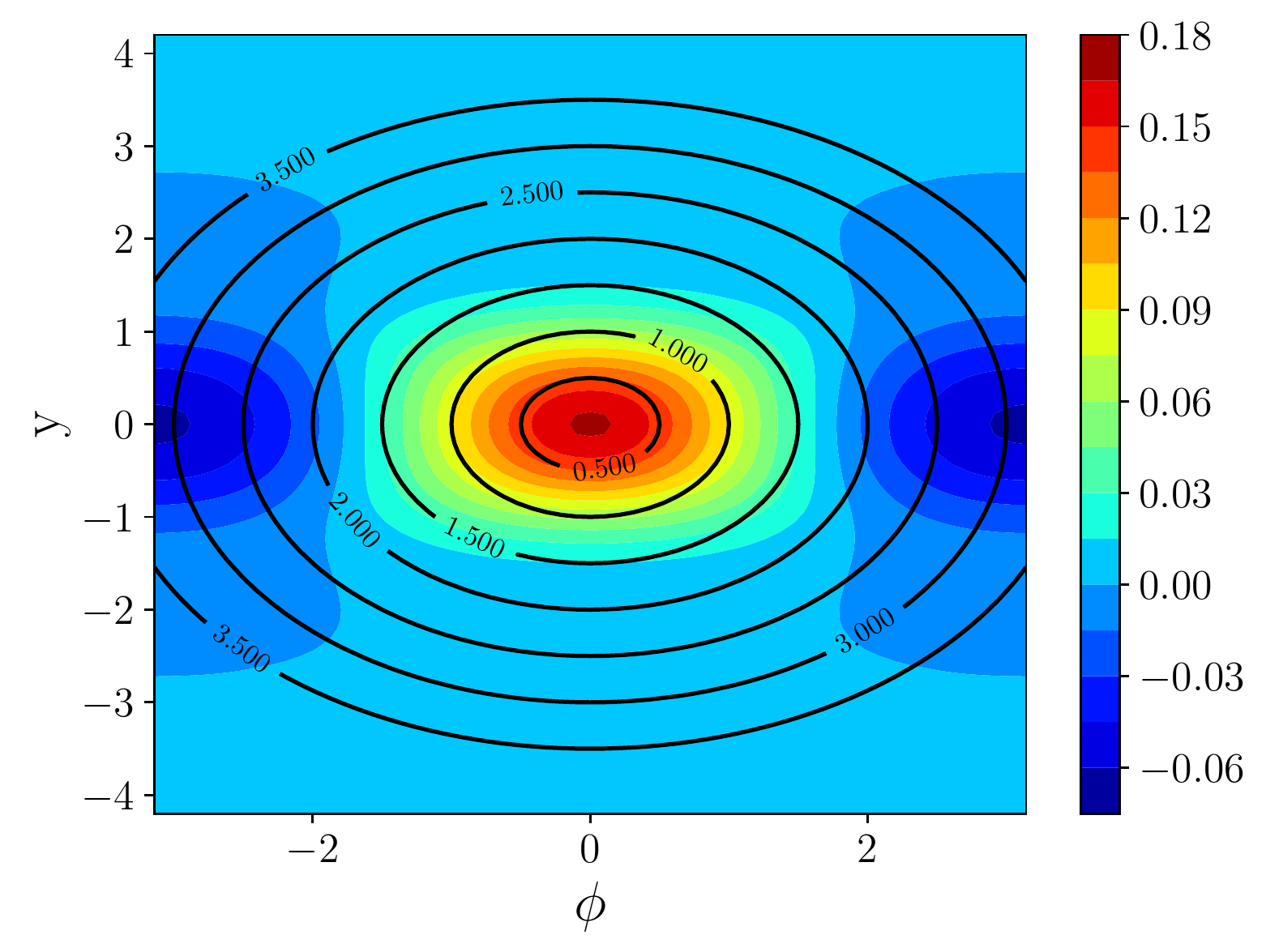}
        \caption{Linearized hydrodynamics with the effects of transverse flow added.}
    \end{subfigure}%
    
    \begin{subfigure}[t]{0.49\textwidth}
        \centering
        \includegraphics[height=2.5in]{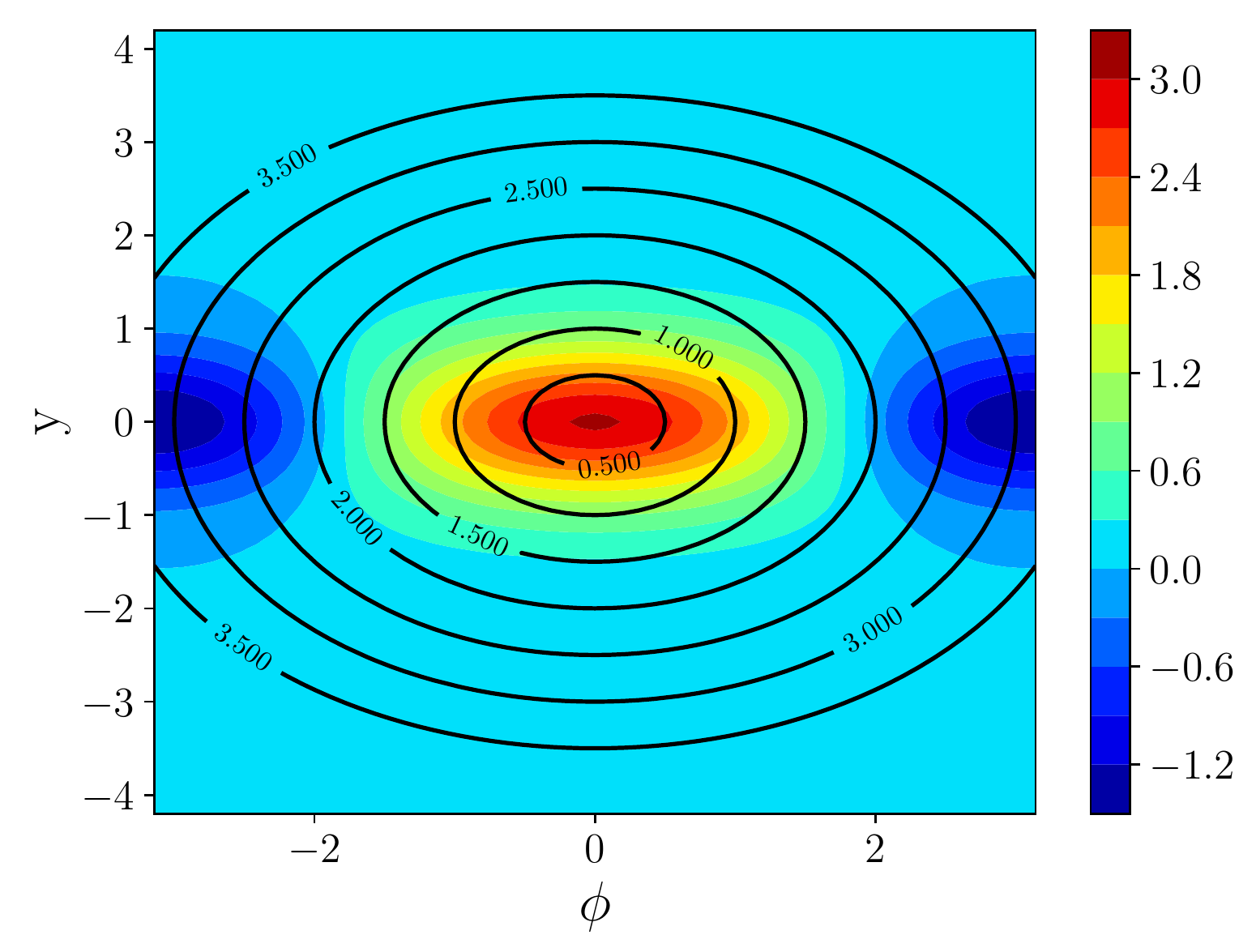}
        \caption{Hybrid model.}
    \end{subfigure}%
\caption{Energy density $\frac{\diff\Delta E}{\diff\phi \diff \ma{y}}$ in the $(\phi,\ma{y})$ plane, together with contours of constant $R$, for the same three calculations as in Fig.~\ref{fig:Econe}.}
\label{fig:dE}
\end{figure}

First, though, we should understand one of the most striking features of the energy recovery plots in Fig.~\ref{fig:Econe}, namely the non-monotonic behavior observed in all three calculations.
The fact that $\Delta E(R)$ is above its large-$R$ value for cones with opening angle $R\sim 2$ means that after hadronization the wake deposits more than its total energy, more energy than was lost by the high energy parton, within a cone with this radius.  How can this be?
This is a consequence of the 
fact that the spectrum in Fig.~\ref{fig:spectrum} and the azimuthal and spacetime rapidity distributions in Fig.~\ref{fig:distribution} are all negative in some kinematic regions. Recall that this means that in such regions the presence of the wake depletes the energy/momentum distribution relative to what it would be in the absence of the wake, which we can understand as a result of a physical effect: the high energy parton drags the fluid along its direction of propagation, creating a wake, and in so doing reduces the number of particles produced opposite to the jet direction.  Fig.~\ref{fig:distribution} does not display this well, because in that figure in some panels we have integrated over $\phi$ and in others we have integrated over $\ma{y}$.
To gain a better understanding of this phenomenon, in
Fig.~\ref{fig:dE} we show 
the energy density profile in the $(\phi,\ma{y})$ plane,
defined as 
\be
\frac{\diff\Delta E}{\diff\phi \diff \ma{y}} \equiv \int \diff p_T\, \sqrt{p_T^2+m^2} \cosh\ma{y}\, \frac{\diff\Delta N}{\diff p_T \diff\phi \diff\ma{y}}\,.
\ee
For convenience, in the same Figure we show contours corresponding to cones centered on the direction of the high energy parton with fixed cone sizes $R$.
 When the cone size is small $R<1$, only positive energy from the wake is recovered within the cone. When $R$ increases further and reaches about $R\sim 2$, the cone starts to touch the region 
 where the effect of the wake is to deplete the background, represented by a negative energy density change.
 As a consequence, the accumulated energy recovered by the cone starts to decrease as $R$ increases further in this regime, consistent with what we have seen in the solid curves of Fig.~\ref{fig:Econe}. 
 
 The details of the shape of the energy $\Delta E(R)$ recovered as a function of the cone opening angle $R$, and in particular the differences between the three solid curves in Fig.~\ref{fig:Econe}, reflect details of and differences between the energy density distribution in the three different panels of Fig.~\ref{fig:dE}, one for each of the three calculations. Both for the linearized hydrodynamic model with transverse flow added and for the hybrid model, the positive energy density distribution in Fig.~\ref{fig:dE} is narrow in rapidity $\ma{y}$, and it is recovered first as $R$ increases. The negative energy density contribution is only collected at larger $R\gtrsim 2$.
This is why both the black and magenta solid curves in Fig.~\ref{fig:Econe} drop after $R\sim 2$ without further increase. In contrast, the linearized hydrodynamic calculation leads to a wider $\ma{y}$ distribution, which implies that not all the positive energy contribution from the wake has been recovered at the angular distance when the effect of the background depletion starts to affect the energy balance. 
 Therefore, the recovered energy in the linearized hydrodynamic calculation without the transverse flow effect added increases first as increasing the cone angle $R$ catches more and more positive energy, then drops with increasing $R$ as negative energy is included, and finally increases again as $R$ increases further, as more positive energy is included.

\subsection{Energy Distribution of the Recovered Energy: a Proxy for Soft Contributions to In-medium  Fragmentation Functions}

The differences in the role of semi-hard particles in the energy recovered described in the last Subsection prompt us to study the spectrum of particles coming from the backreaction of the medium in further detail. In a realistic simulation, depending on the jet algorithm employed, many of these particles will be a part of what an experimentalist reconstructs as a jet, which will affect the measured fragmentation functions at very low momentum. Inspired by the definition of fragmentation functions, here we will study the number density of particles from the wake produced within a cone with opening angle $R$ with a given fraction $z$ of the longitudinal momentum of the jet along the jet direction. But, what shall we take as the longitudinal momentum of the jet after it has been quenched?
As we have done throughout, for us our ``jet'' is actually a single high energy parton, not a jet shower. We will assume, again as throughout, that our high energy parton has an initial energy $E_{\ma{in}}=100$ GeV. After it passes through the medium, it has lost $\Delta E_\ma{tot}$. However, we take into account the energy coming from the wake recovered within the cone of opening angle $R$, namely the quantity $\Delta E(R)$ that we plotted in Fig.~\ref{fig:Econe}, and define the energy of the ``reconstructed jet'' as $E(R) = E_{\ma{in}}-\Delta E_\ma{tot} + \Delta E(R)$. For this ``reconstructed jet'', we can define the longitudinal momentum fraction distribution as 
\be
\Delta f(z)\equiv \int \diff p_T \int_{\sqrt{\phi^2+\ma{y}^2}<R} \diff\phi \diff\ma{y} \frac{\diff \Delta N}{\diff p_T \diff\phi \diff\ma{y}} \delta\Big(z-\frac{p_T\cos\phi}{E(R)} \Big)\,.
\ee
$\Delta f(z)$ serves as a proxy for the contribution to the in-medium fragmentation function originating from the wake in the plasma.

\begin{figure}
    \begin{subfigure}[t]{0.49\textwidth}
        \centering
        \includegraphics[height=2.3in]{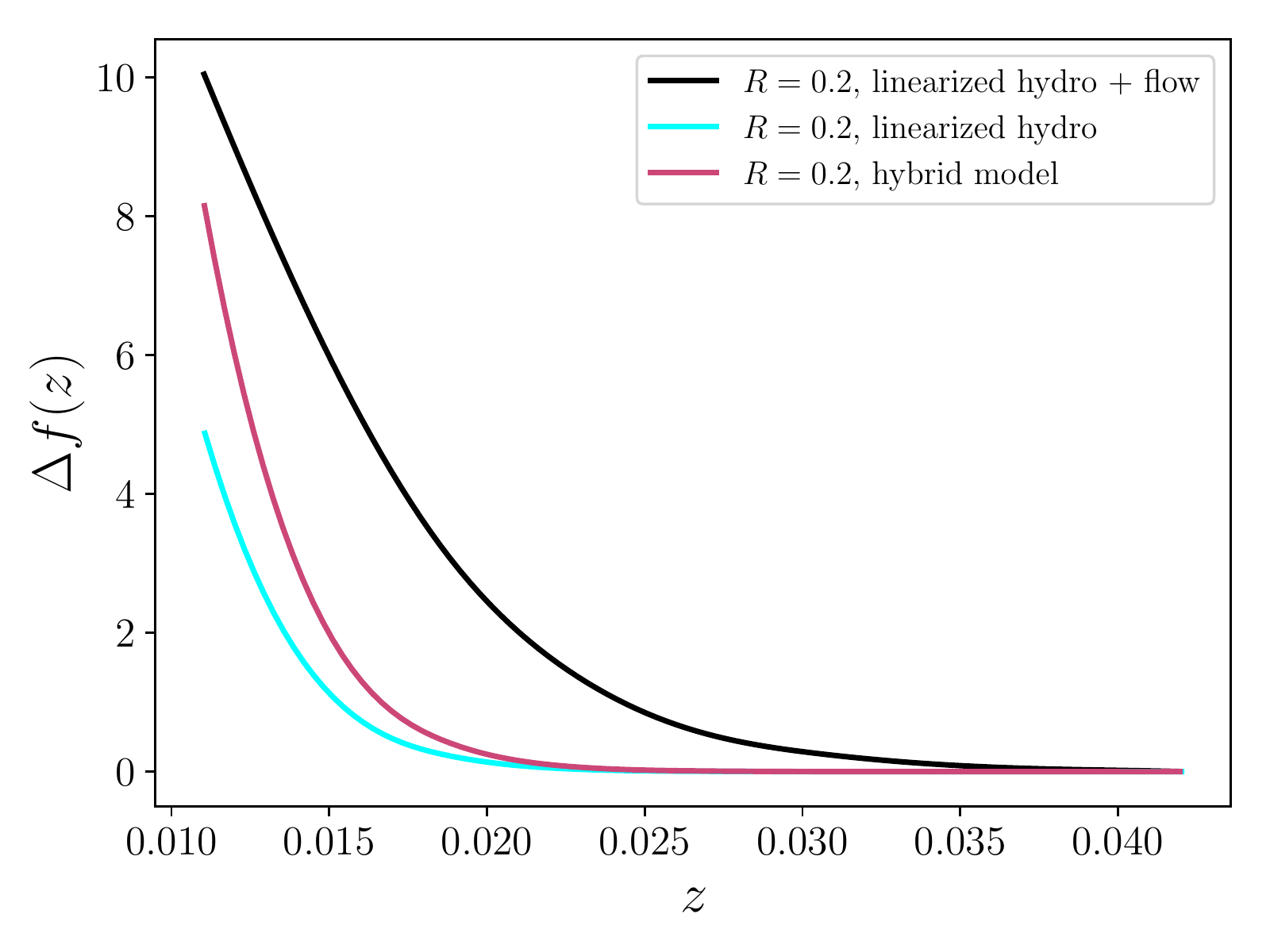}
        \caption{$R=0.2$.}
    \end{subfigure}%
    ~
    \begin{subfigure}[t]{0.49\textwidth}
        \centering
        \includegraphics[height=2.3in]{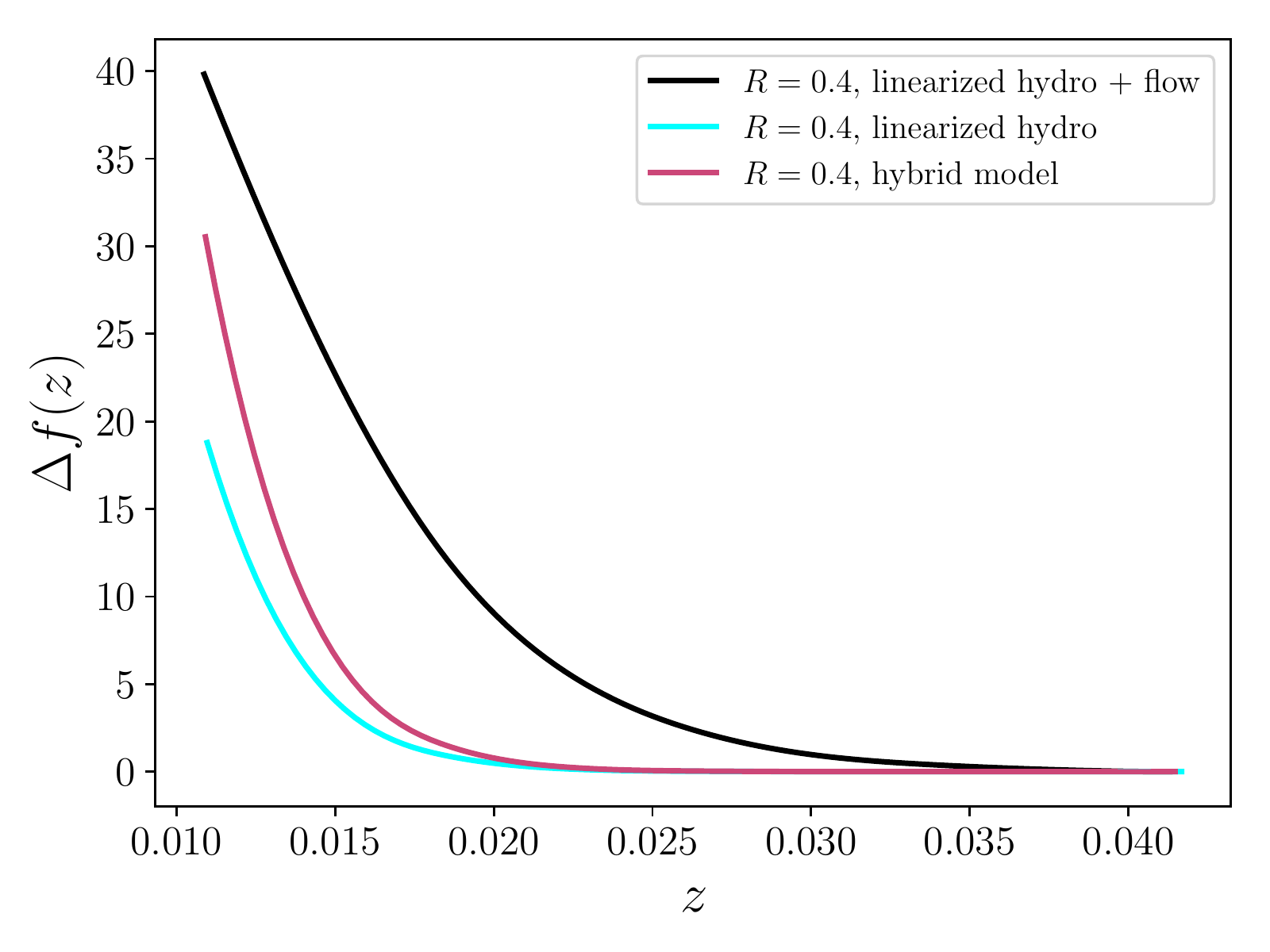}
        \caption{$R=0.4$.}
    \end{subfigure}%
\caption{Results from all three calculations of the proxy $\Delta f(z)$ for the contribution to the in-medium fragmentation function originating from the wake in the plasma for the ``reconstructed jet'' with cone angle $R=0.2$ and $R=0.4$.}
\label{fig:frag}
\end{figure}

Results from our calculations of $\Delta f(z)$ from all three calculations are shown in Fig.~\ref{fig:frag} for two different values of $R$. We have plotted $\Delta f(z)$ for $z>0.01$, corresponding to particles with $p_T\gtrsim 1$~GeV since the initial energy of the jet is 100 GeV. As we discussed in the Introduction, the oversimplified wake in the hybrid model lacks particles with momentum $p_T>2$ GeV, which corresponds to $z\gtrsim 0.02$ here. 
We see in Fig.~\ref{fig:frag} that the linearized hydrodynamic calculation of the wake without radial flow yields similar results, which is to say a similar paucity of semi-hard particles. And, we see that when we include the effects of radial flow in addition to the full hydrodynamic response that we have calculated using linearized hydrodynamics, as we have already inferred the hardening effect of the radial transverse flow is dramatic. In the black curves in Fig.~\ref{fig:frag} we see a tremendous increase in the contribution of semi-hard fragments with respect to the other two calculations. These results highlight the importance of the effects of radial transverse flow on the spectrum and distribution of particles obtained from the backreaction of the medium, the wake that the high energy parton deposits in the droplet of QGP.

\section{Concluding Remarks}
\label{sect:conclusion}

In this paper, we study the wake that a passing high energy parton leaves in the droplet of QGP through which it passes  by using linearized hydrodynamics on a background of Bjorken flow. We treat the jet energy and momentum loss as external currents to the hydrodynamic equations and expand the perturbation to these equations describing the jet wake to linear order. By solving the linearized hydrodynamic equations numerically, we have studied the evolution of, and the energy-momentum conservation in, the wake. We observed that viscous effects, even for viscosities as small as $\eta_0/s=1/4\pi$, have a significant effect on the evolution of the wake. These effects quickly spread out the hydrodynamic perturbation in the fluid and make the fluid backreaction to an energetic probe small in magnitude. This justifies the linearized hydrodynamics approach we have employed in this paper. By studying particle production from the perturbed hydrodynamic fields at freezeout, we conclude that most of the qualitative features of the backreaction of the medium seen in the complete linear analysis of this paper are well-captured by the simplified approach taken in the hybrid strong/weak coupling model.

We have also studied how radial transverse flow affects the spectrum, azimuthal distribution and momentum rapidity distribution of the particles produced when the medium, including the wake in it, hadronizes at freezeout. We have done so not by solving the full hydrodynamic problem, but rather by taking into account the boost that radial flow exerts on the particles coming from the perturbation of the medium at the time of freezeout. We have seen that this boost has a very significant effect on the particles originating from the backreaction of the medium. As a consequence of the radial transverse flow, these particles originating from the wake in the fluid have a harder spectrum and are more tightly correlated with the direction of the high energy parton than was the case in the simplified hybrid model calculation from previous work. This change in the distribution of the particles coming from the backreaction will have important consequences for jet observables. To get an initial qualitative sense of the extent to which this is the case, we have shown that radial flow effects increase the fraction of the energy lost by the initial high energy parton that ends up (after propagating as a hydrodynamic wake and then freezing out into particles) within a cone of opening angle $R$ around the direction of the high energy parton. This is so for cones with the small values of $R$ used in experimentalists' jet reconstruction algorithms and in fact is so for any $R \lesssim 2$.
We have also seen that radial flow effects
significantly increase the fraction of semi-hard particles with $p_T>1$ GeV coming from the perturbed medium.


The studies we have performed in this paper cannot be compared directly to experimental measurements, since they lack many important ingredients. For example, we have considered a single high energy parton produced at a specified place with a specified direction, rather than looking at an ensemble of different branching parton showers, with their initial position and direction of origin as well as their pattern of branching drawn from appropriate probability distributions in a Monte Carlo calculation, for example as in the hybrid strong/weak coupling model.  Our calculation could also be extended by computing the perturbation on top of a more realistic hydrodynamic background.
This notwithstanding, contrasting the results that we have obtained from our linearized hydrodynamic analysis including the effects of radial flow with the much more simplified analysis of particle production from the wake incorporated in the hybrid model is encouraging. Although the hybrid model incorporates the complexities of jet showers and a realistic hydrodynamic background, its oversimplified treatment of the wake has consequences for many jet observables that are sensitive to soft particles within jets~\cite{Casalderrey-Solana:2016jvj}.
In the detailed analyses of Ref.~\cite{Casalderrey-Solana:2016jvj}, the hybrid model implementation of particle production from the backreaction of the medium 
yields a spectrum that is too soft and particle production spread out over too wide an angle to explain experimental measurements of several jet observables in heavy ion collisions at the LHC.
We have seen in our analysis that incorporating the effects of radial flow on our linearized hydrodynamic calculation changes the spectrum and distribution of particle production in the direction indicated by the experimental measurements as it yields an increase in the number of semi-hard particles which are better correlated with the jet direction. While a Monte Carlo study is needed to assess whether the effect introduced by radial flow yields results in quantitative agreement with experimental measurements, the magnitude of the effect we have observed points in the right direction and encourages us to implement these findings in a future Monte Carlo analysis of medium response. 


\begin{acknowledgements}
We acknowledge helpful conversations with Yasuki Tachibana.
 The work of JCS is supported by grants SGR-2017-754, FPA2016-76005-C2-1-P, and PID2019-105614GB-C21. JGM is supported by European Research Council project ERC-2018-ADG-835105 YoctoLHC and Funda\c c\~ao para a Ci\^encia e a Tecnologia (Portugal) under project CERN/FIS-PAR/0024/2019, and he gratefully acknowledges the hospitality of the CERN theory group. DP is supported by the Trond Mohn Foundation under Grant No.  BFS2018REK01.  The work of KR and XY was supported by the U.S. Department of Energy, Office of Science, Office of Nuclear Physics grant DE-SC0011090. 
\end{acknowledgements}

\appendix
\section{Killing Vectors and Conservation Laws}
\label{sect:killing}
The Gauss law states that for a (contravariant) vector $v^\mu$ in a manifold $M$, the volume integral of $\nabla_\mu v^\mu$ is related to an integral on the boundary by
\be
\int_M \nabla_\mu v^\mu \sqrt{g} \diff^n x = \int_{\partial M} \hat{n}_\mu v^\mu \sqrt{h} \diff^{n-1} x\,,
\ee
in which $\hat{n}_\mu$ is the normal vector of the boundary, $g = |\det g_{\mu\nu}|$ and $h_{\mu\nu} = g_{\mu\nu} - \hat{n}_\mu\hat{n}_\nu$ is the induced metric on the boundary manifold $\partial M$. We cannot apply the Gauss law directly to the hydrodynamic equation $\nabla_\mu T^{\mu\nu} = 0$ since $T^{\mu\nu}$ is a rank $(2,0)$ tensor. But we can construct a conserved quantity with a Killing vector $\xi^\mu$:
\be
\nabla_\nu ( \xi_\mu T^{\mu\nu} )  = 0 \,.
\ee
Then applying the Gauss law gives
\be
\label{eqn:gausslaw1}
\int_M \nabla_\nu ( \xi_\mu T^{\mu\nu} ) \sqrt{g} \diff^n x = \int_{\partial M} \hat{n}_\nu \xi_\mu T^{\mu\nu}  \sqrt{h} \diff^{n-1} x =0 \,.
\ee

The existence of Killing vectors reflects symmetries of the spacetime. For Minkowski spacetime, the Killing vectors $\frac{\partial}{\partial t}$, $\frac{\partial}{\partial x}$, $\frac{\partial}{\partial y}$ and $\frac{\partial}{\partial z}$ are associated with the translational symmetries along $t$, $x$, $y$ and $z$. Thus they are related to energy and momentum conservation.

We consider the Killing vector $\frac{\partial}{\partial t}$ here. We choose the manifold to be the slice of spacetime between two times $\tau_1$ and $\tau_2$ ($\tau_1>\tau_2$), the normal vectors at $\tau_1$ and $\tau_2$ point to $\pm\tau$ direction respectively. If the stress-energy tensor vanishes sufficiently quickly at large $x$, $y$ and $\eta_s$, we find from Eq.~(\ref{eqn:gausslaw1})
\be
\label{eqn:gausslaw2}
\int \tau_1 \diff x \diff y \diff \eta_s\, T^{t\tau}(\tau_1,x,y,\eta_s)  - \int \tau_2 \diff x \diff y \diff \eta_s\, T^{t\tau}(\tau_2,x,y,\eta_s) = 0\,,
\ee
i.e., $\int \tau \diff x \diff y \diff \eta_s \,T^{t\tau}(\tau,x,y,\eta_s) $ is conserved as the total energy. By similar construction, the integration of $T^{z\tau}$ is conserved as the total momentum along the $z$-direction.

When the external currents of the hydrodynamic equations are nonvanishing, a similar construction with a Killing vector $\xi^\mu$ leads to
\be
\label{eqn:killing_J}
\nabla_\nu ( \xi_\mu T^{\mu\nu} )  = \xi_\mu J^\mu \,.
\ee
Then Eq.~(\ref{eqn:gausslaw2}) becomes
\be
\label{eqn:gausslaw3}
&&\int \tau_1 \diff x \diff y \diff \eta_s\, T^{t\tau}(\tau_1,x,y,\eta_s)  - \int \tau_2 \diff x \diff y \diff \eta_s\, T^{t\tau}(\tau_2,x,y,\eta_s) \nn\\ &=& \int_{\tau_2}^{\tau_1} \diff\tau \int \tau \diff x \diff y \diff \eta_s\, J^t(\tau,x,y,\eta_s) \,.
\ee
As explained earlier, the physical meaning of the first line in Eq.~(\ref{eqn:gausslaw3}) is the total energy $\Delta E$ deposited into the hydrodynamic system between $\tau_2$ and $\tau_1$. Writing $\tau_2=\tau$ and $\tau_1=\tau+ \Delta \tau$ and taking the limit $\Delta\tau\to0$, we find
\be
\label{eqn:gausslaw4}
\int \tau \diff x \diff y \diff \eta_s\, J^t(\tau,x,y,\eta_s) = \frac{\diff}{\diff\tau} \int \tau \diff x \diff y \diff \eta_s\, T^{t\tau}(\tau,x,y,\eta_s) = \frac{\diff E}{\diff \tau} \,,
\ee
which is Eq.~(\ref{eqn:conserve_ef}).
Similar equations for the momentum deposition can be worked out and lead to Eq.~(\ref{eqn:conserve_pf}).

We want to emphasize that the condition that the stress-energy tensor vanishes sufficiently quickly at large $x$, $y$ and $\eta_s$ is important via an example. In the Bjorken flow, the hydrodynamics is boost-invariant, which means the stress-energy tensor is independent of $\eta_s$ and is nonvanishing at $\eta_s = \pm \infty$. So when applying Eq.~(\ref{eqn:gausslaw1}), there are two more boundaries that contribute to Eq.~(\ref{eqn:gausslaw2}). The left hand side of Eq.~(\ref{eqn:gausslaw2}) in this case becomes
\be
\label{eqn:gausslaw5}
&& \int \tau_1 \diff x \diff y \int_{\eta_2}^{\eta_1} \diff \eta_s T^{t\tau}(\tau_1,x,y,\eta_s)  
- \int \tau_2 \diff x \diff y \int_{\eta_2}^{\eta_1}  \diff \eta_s T^{t\tau}(\tau_2,x,y,\eta_s)  \\ \nn
&+&\int \diff x \diff y \int_{\tau_2}^{\tau_1} \tau \diff\tau\, T^{t\eta_s}(\tau,x,y,\eta_1) 
- \int \diff x \diff y \int_{\tau_2}^{\tau_1} \tau \diff\tau\, T^{t\eta_s}(\tau,x,y,\eta_2)   \,.
\ee
For ideal Bjorken flow, we have
\be
T^{\tau\tau} &=& \varepsilon_0 \Big(\frac{\tau_0}{\tau}\Big)^{1+c_s^2} \\
T^{\tau\eta_s} &=& T^{\eta_s\tau} = 0\\
T^{\eta_s\eta_s} &=& \varepsilon_0 \frac{c_s^2}{\tau^2}  \Big(\frac{\tau_0}{\tau}\Big)^{1+c_s^2}\,.
\ee
Then we can calculate Eq.~(\ref{eqn:gausslaw5}) explicitly, which, up to an irrelevant constant factor, leads to
\be
&&\int_{\eta_2}^{\eta_1} \tau_1 \diff \eta_s\, \cosh\eta_s \Big(\frac{1}{\tau_1}\Big)^{1+c_s^2}  - \int_{\eta_2}^{\eta_1} \tau_2 \diff \eta_s\, \cosh\eta_s \Big(\frac{1}{\tau_2}\Big)^{1+c_s^2}  \\ \nn
&+& \int_{\tau_2}^{\tau_1} \tau^2 \diff \tau\, \sinh\eta_1 \, \frac{c_s^2}{\tau^2}  \Big(\frac{1}{\tau}\Big)^{1+c_s^2} 
- \int_{\tau_2}^{\tau_1} \tau^2 \diff \tau \,\sinh\eta_2 \, \frac{c_s^2}{\tau^2}  \Big(\frac{1}{\tau}\Big)^{1+c_s^2} = 0\,.
\ee
In a nutshell, energy is conserved in Bjorken flow. Momentum conservation can be worked out similarly.

\bibliography{main.bib}
\end{document}